\documentclass[aps,prl,superscriptaddress,twocolumn,longbibliography]{revtex4-2}

\usepackage{braket}
\usepackage{float}
\usepackage{graphicx}
\usepackage{amsmath}    
\usepackage{comment}
\usepackage{units}
\usepackage{amsthm}
\usepackage{amssymb}
\usepackage{graphicx}
\usepackage{color}
\usepackage{bbold}
\usepackage[colorlinks=true, urlcolor=blue, citecolor=blue,linkcolor=blue,citebordercolor={1 0 0},linkbordercolor={0 0 1}]{hyperref}
\usepackage{graphicx}
\graphicspath{{./images/},{./imagesAppendix/}}

\renewcommand{\eqref}[1]{Eq.~(\ref{#1})} % Reference to equation
\theoremstyle{plain}

\theoremstyle{plain}

\ifx\proof\undefined
\newenvironment{proof}[1][\protect\proofname]{\par
	\normalfont\topsep6\p@\@plus6\p@\relax
	\trivlist
	\itemindent\parindent
	\item[\hskip\labelsep\scshape #1]\ignorespaces
}{%
	\endtrivlist\@endpefalse
}
\providecommand{\proofname}{Proof}
\fi
\theoremstyle{plain}

\theoremstyle{remark}

\makeatother
\newcommand{\idg}[1]{{\bfseries #1)}}

\newcommand\numberthis{\addtocounter{equation}{1}\tag{\theequation}}

\providecommand{\factname}{Fact}
\providecommand{\theoremname}{Theorem}
\providecommand{\claimname}{Claim}
\providecommand{\lemmaname}{Lemma}
\providecommand{\definitionname}{Definition}

\definecolor{THc}{rgb}{0.9,0.3,0.2}

\newcommand{\revA}[1]{{#1}}
\newcommand{\revB}[1]{#1}

\theoremstyle{definition}

\newcommand{\subfigimg}[3][,]{%
	\setbox1=\hbox{\includegraphics[#1]{#3}}% Store image in box
	\leavevmode\rlap{\usebox1}% Print image
	\rlap{\hspace*{2pt}\raisebox{\dimexpr\ht1-0.5\baselineskip}{{\bfseries \large\textsf{#2}}}}% Print label
	\phantom{\usebox1}% Insert appropriate spcing
}

\usepackage{amssymb}
\usepackage{color}
\usepackage{color}
\definecolor{KB}{rgb}{0.4,0.3,0.9}

\begin{document}

\title{Atomtronic multi-terminal Aharonov-Bohm interferometer}

\author{Jonathan Wei Zhong Lau}	
\email{e0032323@u.nus.edu}
\affiliation{Centre for Quantum Technologies, National University of Singapore 117543, Singapore}

\author{Koon Siang Gan}	
\affiliation{Centre for Quantum Technologies, National University of Singapore 117543, Singapore}

\author{Rainer Dumke}
\affiliation{Centre for Quantum Technologies, National University of Singapore 117543, Singapore}
\affiliation{School of Physical and Mathematical Sciences, Nanyang Technological University 637371, Singapore}

\author{Luigi Amico}
\thanks{On leave from the Dipartimento di Fisica e Astronomia ``Ettore Majorana'', University of Catania.}
\affiliation{Centre for Quantum Technologies, National University of Singapore 117543, Singapore}
\affiliation{Quantum Research Centre, Technology Innovation Institute, Abu Dhabi, UAE}
\affiliation{INFN-Sezione di Catania, Via S. Sofia 64, 95127 Catania, Italy}
\affiliation{LANEF ’Chaire d’excellence’, Universitè Grenoble-Alpes \& CNRS, F-38000 Grenoble, France}

\author{Leong-Chuan Kwek}
\affiliation{Centre for Quantum Technologies, National University of Singapore 117543, Singapore}
\affiliation{MajuLab, CNRS-UNS-NUS-NTU International Joint Research Unit, UMI 3654, Singapore}
\affiliation{National Institute of Education,
Nanyang Technological University, 1 Nanyang Walk, Singapore 637616}
\affiliation{School of Electrical and Electronic Engineering
Block S2.1, 50 Nanyang Avenue, 
Singapore 639798 }

\author{Tobias Haug}
\email{tobias.haug@u.nus.edu}
\affiliation{QOLS, Blackett Laboratory, Imperial College London SW7 2AZ, UK}

\def\kwek#1{\textcolor{red}{#1}}
% \collaboration{CLEO Collaboration}%\noaffiliation

\date{\today}% It is always \today, today,

\begin{abstract}
We study a multi-functional device for cold atoms consisting of a three-terminal ring circuit pierced by a synthetic magnetic flux, where the ring can be continuous or discretized.  The flux controls the atomic current through the ring via the Aharonov-Bohm effect. 
Our device shows a flux-induced transition of reflections from an Andreev-like negative density to positive density.
Further, the flux can direct the atomic current into specific output ports, realizing a flexible non-reciprocal switch to connect multiple atomic systems or sense rotations. 
By changing the flux linearly in time, we convert constant matter wave currents into an AC modulated current. This effect can be used to realize an atomic frequency generator and study fundamental problems related to the Aharonov-Bohm effect. 
We experimentally demonstrate Bose-Einstein condensation into the light-shaped optical potential of the three-terminal ring.
Our work opens up the possibility of novel atomtronic devices for practical applications in quantum technologies. 
\end{abstract}

%\tableofcontents

\maketitle

% \paragraph{Introduction.---}
%\section{Introduction}
Precise control over quantum systems has led to the rapid development of quantum technologies for applications in quantum simulation~\cite{bloch2012quantum}, quantum communication~\cite{keil2016fifteen} and metrology~\cite{amico2021atomtronic}.
These latter fields are fundamental to atomtronics~\cite{amico2021roadmap}, an emerging quantum technology of propagating cold atoms in matter-wave circuits~\cite{amico2017focus,amico2021roadmap,amico2021atomtronic}.
Inspired originally by electronics, atomtronics exploits the advancement in optical traps and cooling to precisely move ultracold atoms to realize novel and practical quantum devices~\cite{mcgloin2003applications,ryu2015integrated,gauthier2016direct,barredo2018synthetic}.
Indeed, simple atomtronic circuits with Bose-Einstein condensates (BECs) or degenerate fermions that mimic classical transport have already exhibited interesting physics with potential applications ~\cite{seaman2007atomtronics,pepino2009atomtronic,pepino2021advances,anderson2021matter,caliga2016principles,caliga2016transport,caliga2017experimental,wilsmann2018control, ryu2020quantum,eckel2014hysteresis,wright2013driving,perez2021coherent,kiehn2022implementation}.

The construction of atomtronic circuits requires an in-depth understanding of cold-atom transport both theoretically and experimentally~\cite{bloch2012quantum,chien2015quantum,heeger1988solitons,atala2013direct,miyake2013realizing,aidelsburger2013realization,jotzu2014experimental,caliga2017experimental}. 
Analogues of one-dimensional mesoscopic conductors also have been investigated~\cite{brantut2012conduction,krinner2015observation,husmann2015connecting,lebrat2018band,gauthier2019quantitative}, with transport now possible over macroscopic distances~\cite{pandey2019hypersonic}.
In particular, BECs trapped in ring shaped potentials~\cite{haug2018mesoscopic,haug2019topological,haug2019aharonov,safaei2019monitoring} and Y-shaped junctions~\cite{ryu2015integrated,haug2019andreev,haug2021quantum} augurs potential practical applications due to its subtle similarity to integrated photonic chips. 

\begin{figure*}
    \centering
    \subfigimg[width=0.8\textwidth]{}{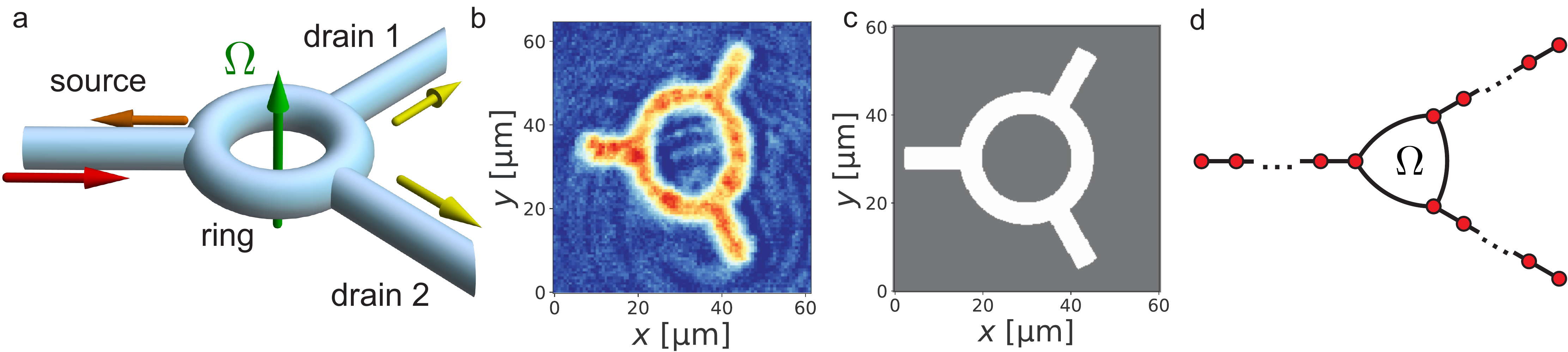} 
    \caption{\idg{a} Three-terminal Aharonov-Bohm circuit with source lead (left) attached to a ring with two drain leads (right). A synthetic Aharonov-Bohm flux $\Omega$ through the ring controls the current flowing from source to the drains, which can be used for multiple functionalities.
    \idg{b} An initial experimental demonstration of the setup with a BEC. $6 \cdot 10^{4}$ Rubidium atoms are cooled to 50~nK by atom evaporation into an optical potential created by a DMD. We show the {\it in-situ} atomic density measured with absorption imaging.
    \idg{c} Potential of two-dimensional ring-lead system simulated with GPE with ring diameter $R=30\mu m$.
    \idg{d} Sketch of lattice ring-lead system for Bose-Hubbard simulations. Source and drain leads consist of an extensive number of lattice sites, connected to a $L=3$ site ring. 
    }
    \label{fig:diagram_device}
\end{figure*}

A promising geometry for cold atom devices is the ring-shaped circuit. Such systems can exhibit superfluid current flows~\cite{ramanathan2011superflow,wright2013driving,ryu2013experimental,eller2020producing} and can realize effective two-level dynamics for a potential atomtronic qubit~\cite{aghamalyan2015coherent,haug2018readout}. Here, the transport can be controlled by the Aharonov-Bohm effect where the magnetic flux \revA{through the ring} changes the interference of matter~\cite{aharonov1959significance}. 
The static Aharonov-Bohm effect controls the conductance in mesoscopic electronic rings~\cite{gefen1984quantum,buttiker1984quantum,webb1985observation,imry2002introduction}, while the nature of the time-dependent Aharonov-Bohm effect is still controversially discussed~\cite{singleton2013covariant,macdougall2015revisiting,jing2017time,choudhury2019direct}.
Through the application of suitable synthetic fields~\cite{jaksch2003creation,lin2009synthetic,dalibard2011colloquium,haug2021machine, wright2013driving,eckel2014hysteresis,del2022imprinting}, cold atoms can harness the Aharonov-Bohm effect with a high degree of control and coherence that is difficult to reach in other systems. An important example is the transport through two-terminal Aharonov-Bohm rings with bosonic atoms~\cite{haug2019aharonov,haug2019andreev,haug2021quantum}.

\begin{figure}
    \centering
    \subfigimg[width=0.49\textwidth]{}{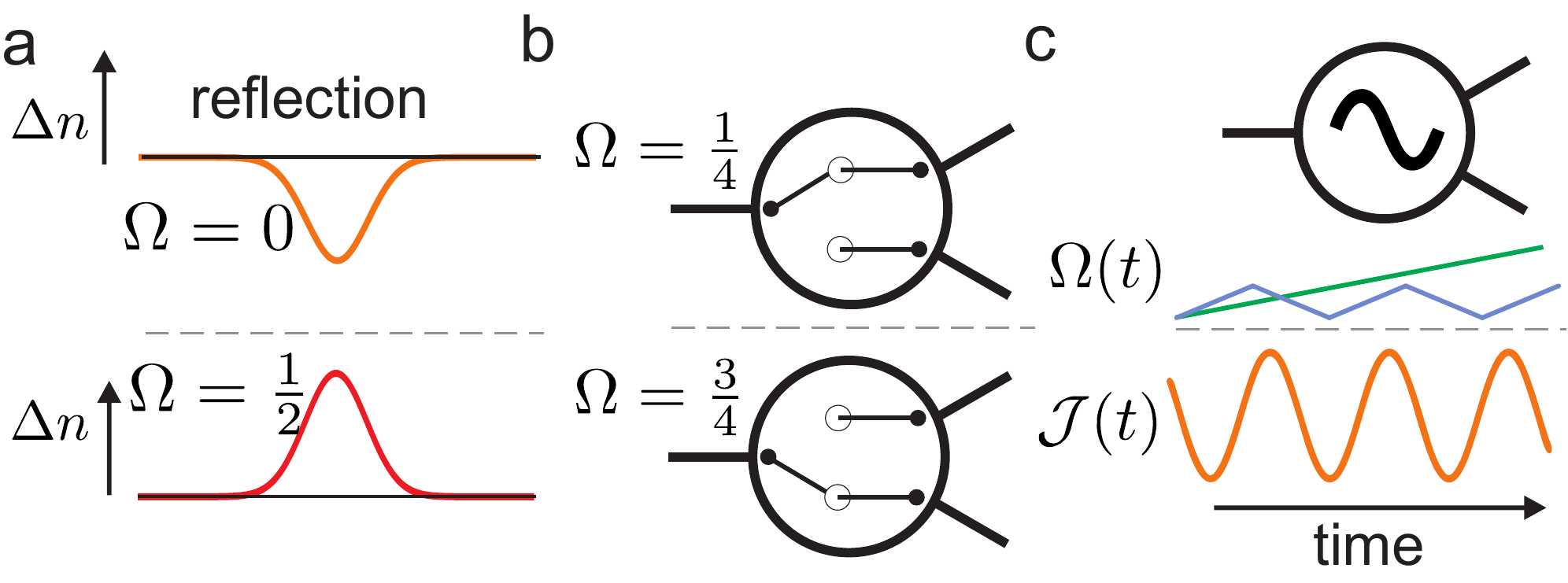} 
    \caption{
    Applications of three-terminal cold atom Aharonov-Bohm circuit.
    \idg{a} Control reflections of density waves from negative (Andreev-like) to positive with flux $\Omega$. 
    \idg{b} Directional switch of the current to one of the output terminals by adjusting $\Omega$, which can also be used as rotation sensor.
    \idg{c} Atomic frequency generator with a sinusoidal output current $\mathcal{J}(t)$ of period $T$ by linearly increasing $\Omega(t)=t/T$ in time or via a periodic ramp.
    }
    \label{fig:diagram_application}
\end{figure}

Here, we study the transport in a three terminal circuit in which a bosonic condensate is guided from a source lead through a Aharonov-Bohm ring attached to two drains - see Fig.\ref{fig:diagram_device}a. 
\revA{We simulate this system with extensive leads coupled to a continuous 2D ring or a discretized ring of three sites.}
We show an experimental demonstration of the continuous setup by loading a Bose-Einstein condensate (BEC) of $^{87}$Rb atoms in a Digital Micromirror Device (DMD) generated optical potential - see Fig.\ref{fig:diagram_device}b.
We demonstrate  that our scheme  provides a novel concept for a  multi-functional  device. \revB{The applications of our work are summarized in Fig \ref{fig:diagram_application}}. \revB{We show that our} system can (i) control density waves; (ii) realize a non-reciprocal switch and sense rotations; (iii) convert a direct current (DC) matter-wave into an alternating current (AC) modulation of the DC matter-wave. 

We first introduce the system together with an experimental demonstration of the setup. We then analyze the low energy and highly non-equilibrium dynamics of the system as well as the dynamics under a time-dependent driving of the flux. \revB{We finally discuss the applications of our work. }

\paragraph{Model.---}
%\section{Bose-Hubbard model}
A sketch of the three-terminal ring pierced by flux $\Omega$ is shown in Fig.~\ref{fig:diagram_device}a. 
We experimentally demonstrate the feasibility of this setup by loading a BEC into a static optical potential generated by a Digital Micromirror Device  (see Fig.~\ref{fig:diagram_device}b and the Supplemental Material (SM)~\ref{sec:experiment} for details).
\revA{In the dilute limit with weak interactions, we simulate the setup with the 2D Gross-Pitaevskii equation (GPE)
\begin{align*}
i\hbar\partial_t\psi=[-\frac{\hbar}{2m}(\partial_x^2+\partial_y^2)+V(x,y)+g_\text{2D}N\vert\psi\vert^2+\omega L_\text{z}]\psi
\end{align*}
where $\psi\equiv\psi(x,y,t)$ is the wavefunction, $m$ is the mass of the atoms, $g_\text{2D}$ the atom-atom interaction strength for Rubidium atoms in two dimensions~\cite{bao2003numerical}, $\omega$ is the rotation of the system to induce flux, $L_\text{z}=-i(x\partial_y-y\partial_x)$ the angular momentum operator and $N$ the number of atoms. The potential $V(x,y)$ is shown in Fig.\ref{fig:diagram_device}c.}

In the limit where leads and ring are strongly confined, we can treat the system as effectively one-dimensional. 
Here, we simulate this system numerically for \revA{different interaction strengths} with the Bose-Hubbard model by dividing the system into the source lead, two drain leads and the ring with $L$ sites (see Fig.\ref{fig:diagram_device}d). The source lead $s$ and the two drains $b,c$ are connected to the ring in a symmetric manner with $x_s=1$, $x_b=L/3$ and $x_c=2L/3$. We choose an extensive number of source and drain sites, while the ring is assumed to be small with $L=3$ sites.
The system Hamiltonian $H = H_{r} + H_{\ell}+ H_{r\ell}$ with $N$ bosons is given by 
\begin{align*}
    &H_{r} = \sum_{j=1}^L \left[\frac{U}{2}\hat{n}_j(\hat{n}_j-1) - J(e^{-i 2 \pi\Omega(t)/L}\hat{a}_{j+1}^\dagger \hat{a}_j + H.C.)\right]\\
    &H_{r\ell} = -K\sum_{\alpha=\{b,c,s\}}\left(\hat{\alpha}_1^\dagger \hat{a}_{x_\alpha} + H.C. \right)\label{eq:Hamilton}\numberthis\\
    &H_{\ell} = \sum_{\alpha=\{b,c,s\}}\sum_{j=1}^{L_\alpha} \left[\frac{U_\alpha}{2}\hat{n}_j^\alpha(\hat{n}_j^\alpha-1)
    - J_\alpha( \hat{\alpha}_{j+1}^\dagger \hat{\alpha}_j + H.C.)\right]\,
\end{align*}
where $\hat{a}_j (\hat{a}_j^\dagger)$ is the bosonic annihilation (creation) operator at site $j$ on the ring, 
$\hat{n}_j = \hat{a}_j^\dagger \hat{a}_j$ is the corresponding number operator, 
$J$ is the intra-ring coupling strength and $U$ the interaction strength of the ring. We impose periodic boundary conditions in the ring $\hat{a}_{L+1} = \hat{a}_1$. For the leads, $\hat{\alpha}_j$ is the annihilation operator, 
$\hat{n}_j^\alpha$ the number operator, 
$L_\alpha$ the number of sites, $U_\alpha$ the interaction strength and $J_\alpha$ the intra-reservoir couplings for the source and drains with $\alpha\in\{s,b,c\}$. 

$\Omega(t)$ represents the flux through the ring which can be dependent on time $t$. 
\revA{This flux can be generated for neutral atoms via rotation, where the Coriolis flux mimics the effect of the magnetic field~\cite{engels2003observation,wright2013driving}. 
A suitable approach is to rotate the whole potential with rotational frequency $\omega=\frac{\Omega\hbar}{mR^2}$, where $R$ is the radius of the ring, yielding $\omega=5.1\Omega\text{Hz}$ for the parameters of Fig.\ref{fig:diagram_device}b~\cite{wright2013driving,haug2018mesoscopic}. The undesired centrifugal potential can be removed with the correction potential $V(r)=\frac{1}{2}\omega^2r^2$, where $r$ is the distance to the rotation center.
As an alternative approach, synthetic magnetic fields can be achieved by counter-propagating Raman beams~\cite{dalibard2011colloquium,lin2009synthetic} or driving the optical potential in time~\cite{weitenberg2021tailoring}, which has been demonstrated for lattice systems~\cite{wintersperger2020realization}. }
Due to flux quantization in the ring, the spectrum of $H(\Omega)$ is periodic with $\Omega\rightarrow\Omega+k$, $k$ being integer, with the flux quantum set to 1.
The current operator between lead $\alpha$ and ring is given by 
%\begin{equation}\label{eq:current}
$\mathcal{J}_{\alpha} = -iK (\hat{\alpha}_1^\dagger \hat{a}_{x_\alpha} - \text{H.C.}).$%\,.
%\end{equation}

\begin{figure}[htbp]
	\centering	
	\subfigimg[width=0.49\textwidth]{}{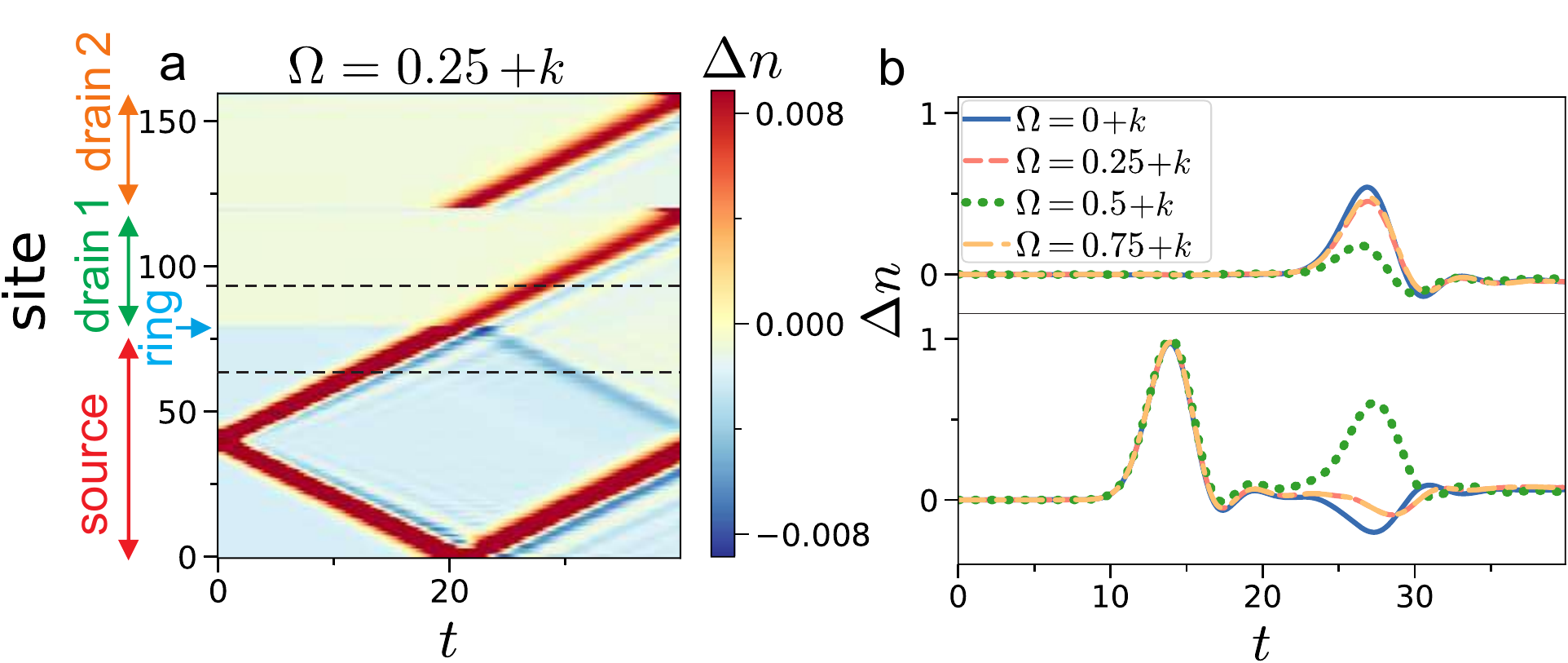}
	\caption{Dynamics of low-energy excitations in three-terminal ring device for $L=3$ ring sites. \idg{a} Change of density relative to average density $\Delta n(t)=\langle n(t)\rangle-n_0$ as function of time and sites of source, ring, drain 1 and 2 for $\Omega=\frac{1}{4}+k$, where $k$ is an integer. The forward propagating wave is transmitted into the drains, as well as reflected back into the source. 
	%\idg{b} Current $\mathcal{J}$ into the first drain against time and $\Omega$. 
	\idg{b} $\Delta n$ in source (bottom) and first drain (top) measured at positions shown as dashed lines in a). %At $t=27$ in the bottom curve, the reflected density wave is seen with positive or negative amplitude depending on $\Omega$. 
 We show $N=80$ hard-core bosons with $J=1$, $K=0.5$ and in total $L=160$ lattice sites ($L=3$, $L_s=79$, $L_b=L_c=39$).
	}
	\label{fig:fulldynamics}
\end{figure}

%\section{Low-excitation dynamics}
\paragraph{Low-energy dynamics.---}
First, we study the dynamics close to the ground state using the Bose-Hubbard model. We perturb the local potential in the source
$H_\text{e}=-\epsilon_s\sum_{j=1}^{L_s}\exp[-(j-D)^2/2\sigma^2]\hat{n}^s_j$
with $D=L_s/2$, $\sigma=2$ and $\epsilon_s=0.3$, We prepare the ground state of the Hamiltonian $H+H_\text{e}$, where now $H_\text{e}$ leads to a locally raised density in the source. At $t>0$, we evolve the system with $H$ only, resulting in two density waves traveling in positive and negative direction (the negative direction can be ignored for a sufficiently large source). We investigate the change in density $\Delta n(t)=\langle n(t)\rangle-n_0$, where $n_0$ is the average density. The dynamics is calculated using matrix product states with the ITensor library~\cite{itensor}. 
In Fig.\ref{fig:fulldynamics}a we show $\Delta n(t)$ for $\Omega=\frac{1}{4}+k$ as function of time $t$ and the sites of source, ring and drain (see SM~\ref{sec:gs_sup} for other values of $\Omega$). The forward propagating density wave moves from the source to the ring, then is transmitted into the drains as well as reflected back to the source.  For any value of $\Omega$ the transmission into drain 1 and drain 2 is nearly the same.
We show the density $\Delta n$ at a fixed site in source and drain in Fig.\ref{fig:fulldynamics}b. The transmission is maximal for $\Omega=k$ and minimal for $\Omega=\frac{1}{2}+k$. We observe identical results for $\Omega=\frac{1}{4}+k$ and $\Omega=\frac{3}{4}+k$, which is the result of an emergent reflection symmetry $\Omega\rightarrow-\Omega$.
We find that the reflection into the source at a specific time $t_r$ ($t_r\sim 27$ in Fig.\ref{fig:fulldynamics}b) changes in nature with $\Omega$. For $\Omega=k$ we find a clear negative reflection, which is a hallmark of Andreev reflections. With increasing $\Omega$, the Andreev reflections turn into positive reflections.

\paragraph{Dynamics far from ground state.---}
%\section{Open System Steady State}
We investigate the dynamics when the system is far from the ground state via a quench protocol. At $t>0$, the filled source lead injects atoms into the initially empty ring and drains. We study this highly non-equilibrium setting for zero, weak and infinite interaction.

\begin{figure*}[htbp]
	\centering	
\subfigimg[width=0.26\textwidth]{a}{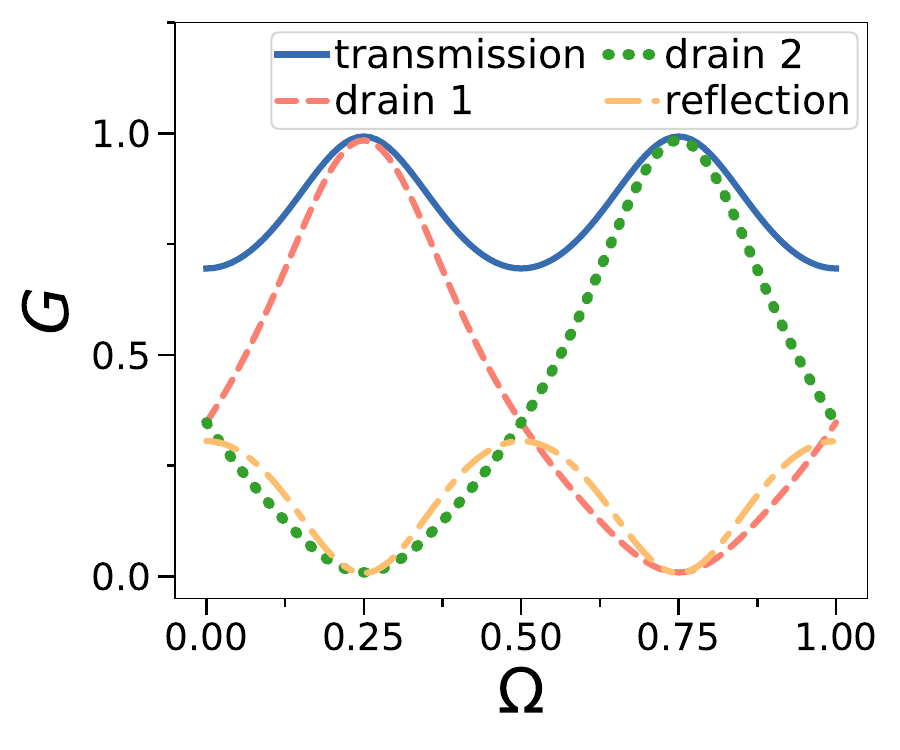}
  \subfigimg[width=0.26\textwidth]{b}{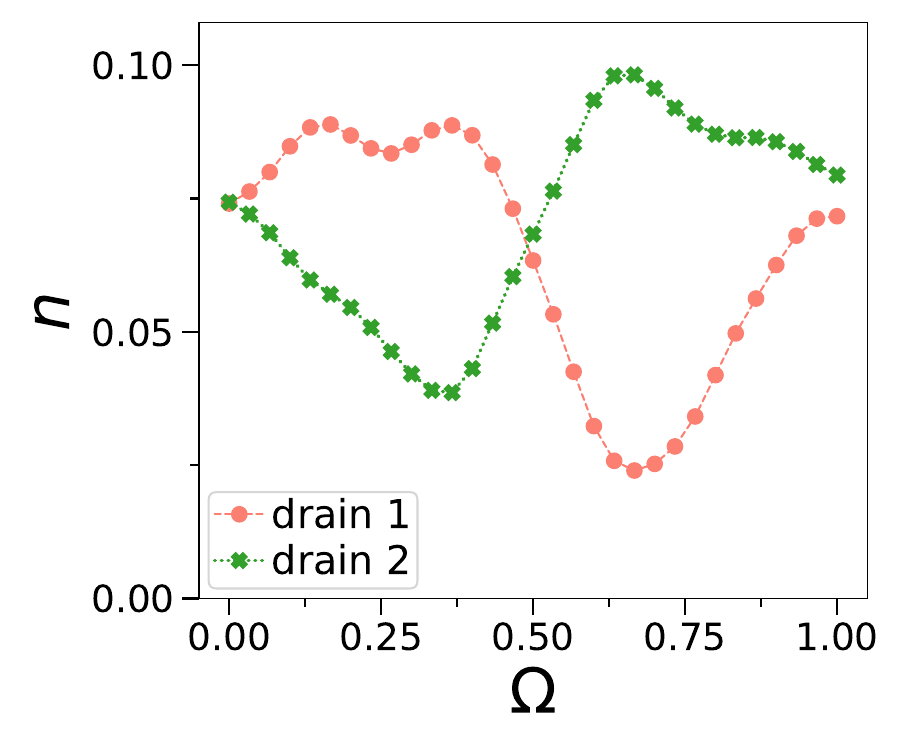}
  \subfigimg[width=0.26\textwidth]{c}{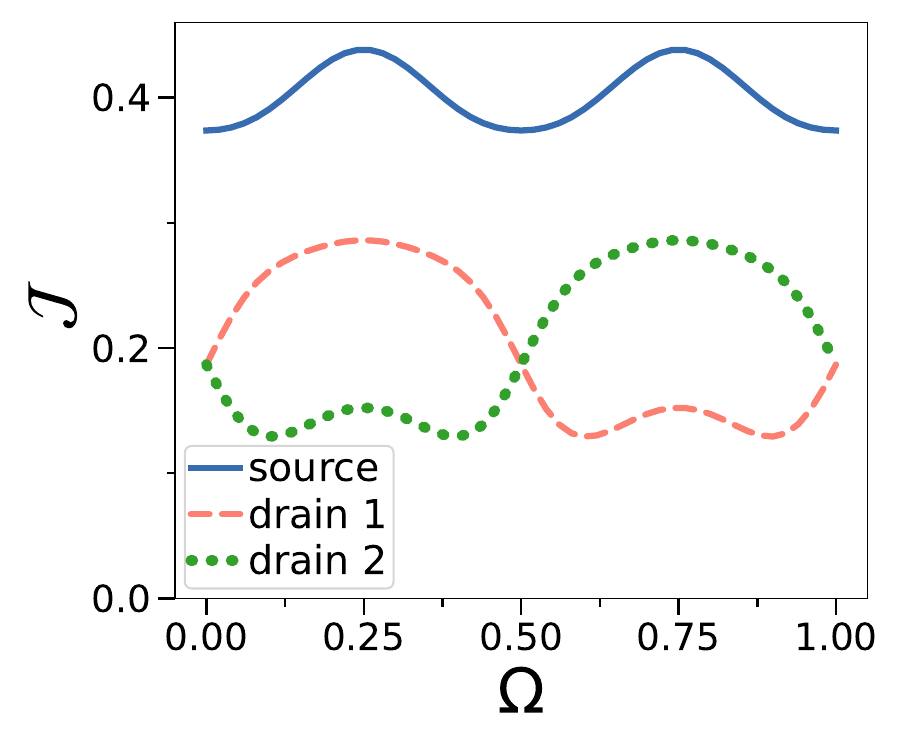}
	\caption{Transport through three-terminal device for non-equilibrium setting and different interaction strengths. 
	\idg{a} Landauer formalism for non-interacting bosons. We show the total transmission and reflection of the system, as well as the transmission into drain 1 and 2 as function of flux $\Omega$. 
  \idg{b} Atom fraction $n$ in the drain 
 averaged over time $t=400\text{ms}$ for continuous ring-lead system with the 2D GPE. $N=2000$ atoms are initially prepared in the source and evolved in the potential shown in Fig.\ref{fig:diagram_device}c, \revB{where the physical rotation frequency is given by $\omega=\frac{\Omega\hbar}{mR^2}$.}
\idg{c} Steady-state current $\mathcal{J}$ for hard-core bosons and $L=3$ ring sites as function of flux $\Omega$ for $J=1$, $L=3$, $K=1$ and $B_\text{s}=B_\text{d}=1$.
	}
	\label{fig:current_neq}
\end{figure*}

In the limit of zero interaction $U=0$, we describe the dynamics with the Landauer formalism as explained in SM~\ref{sec:landauer}, which yields a transmission of 
\begin{equation}
G_\alpha=16\left\vert\frac{1-\sqrt{2}+2i(2\sqrt{2}-3)\exp(-i\pi(2\Omega+\alpha))}{62-46\sqrt{2}+2i\cos(\pi(2\Omega+\alpha))}\right\vert^2
\end{equation}
into the respective drains $\alpha\in\{1,2\}$. The resulting transmission and reflection of the system is shown in Fig.\ref{fig:current_neq}a.  For $\Omega=\frac{1}{4}$ we have unit transmission into drain 1, and zero transmission into drain 2, while for $\Omega=\frac{3}{4}$ the dynamics of the drains is interchanged. Thus, tuning $\Omega$ can direct the current either into drain 1 or drain 2, realizing a perfect non-reciprocal switch with zero back-reflection.

Next, we investigate the system in the dilute limit with the continuous 2D GPE~\cite{wittek2015extended}. In Fig.\ref{fig:current_neq}b, we find that physically rotating the setup with $\Omega$ modulates the average fraction of atoms in the drains.   The flux dependence of the current arises from interference pattern in the ring, which are modulated by $\Omega$. Increasing interaction leads to smaller interference patterns, which reduces the flux sensitivity.
\revB{Due to the finite width of the ring and the radial dependence of the flux, we find that the system is not perfectly periodic with $\Omega$ in contrast to the one-dimensional case. By reducing the width of the ring and leads we expect that the symmetry can be restored}. Further details are shown in SM~\ref{sec:2dgpe}.

Now, we investigate the limit of strong interaction with hard-core bosons, where each lattice site occupies at most one boson.
We use $L=3$ ring sites, and simplify the source and drain leads by tracing out all of their sites except the very first one coupled to the ring ($\hat{s}_1$, $\hat{b}_1$ and $\hat{c}_1$). The 
dynamics of the reduced density matrix $\rho(t)$ within the Born-Markov approximation is described by~\cite{breuer2002theory,haug2019andreev}
\begin{gather}
    \frac{\partial \rho}{\partial t} = - \frac{i}{\hbar}\left[H,\rho\right] - \frac{1}{2}\sum_m \left\{ L_m^\dagger L_m , \rho \right\} +\sum_m L_m \rho L_m^\dagger
\end{gather}
with the Lindblad operators $L_1 = B_\text{s}\hat{s}_1^\dagger$ , $L_2=B_\text{b}\hat{b}_1$, $L_3=B_\text{c}\hat{c}_1$ and coupling strength $B_\alpha$. $L_1$ describes bosons entering the system at the source site, and $L_2$, $L_3$ atoms leaving to the respective drains. 
We solve for the steady-state $\rho_\text{ss}$ via $\partial \rho_\text{ss}/\partial t = 0$~\cite{guo2017dissipatively}. 
In Fig.\ref{fig:current_neq}c we show the steady-state current $\mathcal{J}(\Omega)$. The current in drain 1 and 2 varies strongly with $\Omega$, allowing for directional control into either drains. The source current shows a transition from being flux-independent to flux-dependent with intra-ring coupling $J$ (see SM~\ref{sec:nonequ}).

\begin{figure}[htbp]
	\centering	
		%\subfigimg[width=0.3\textwidth]{a}{CurrentSDT1.pdf}
	\subfigimg[width=0.24\textwidth]{a}{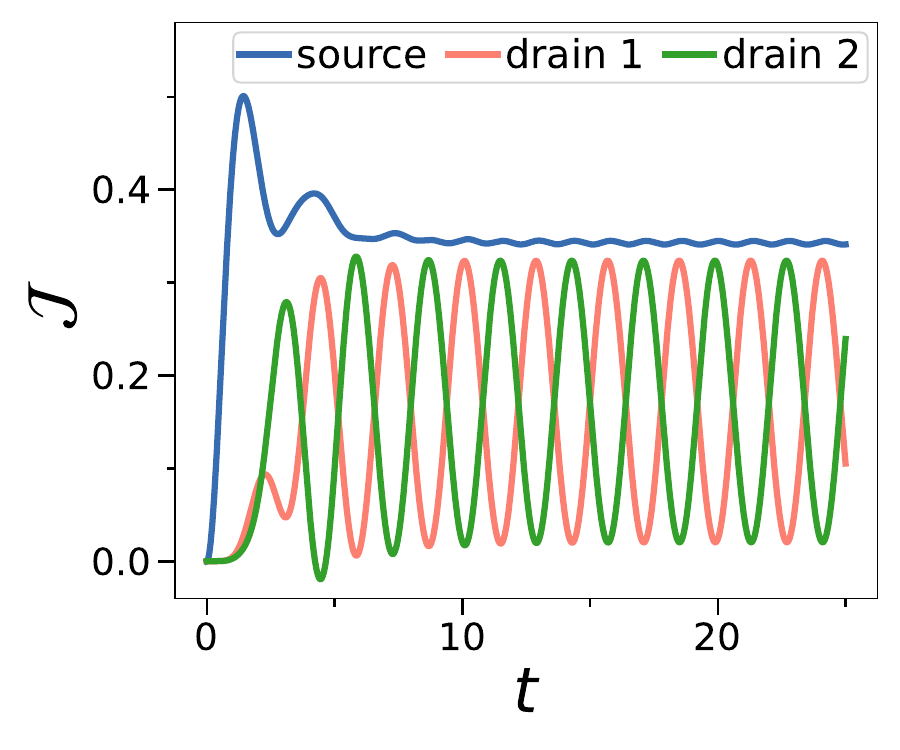}\hfill
		\subfigimg[width=0.24\textwidth]{b}{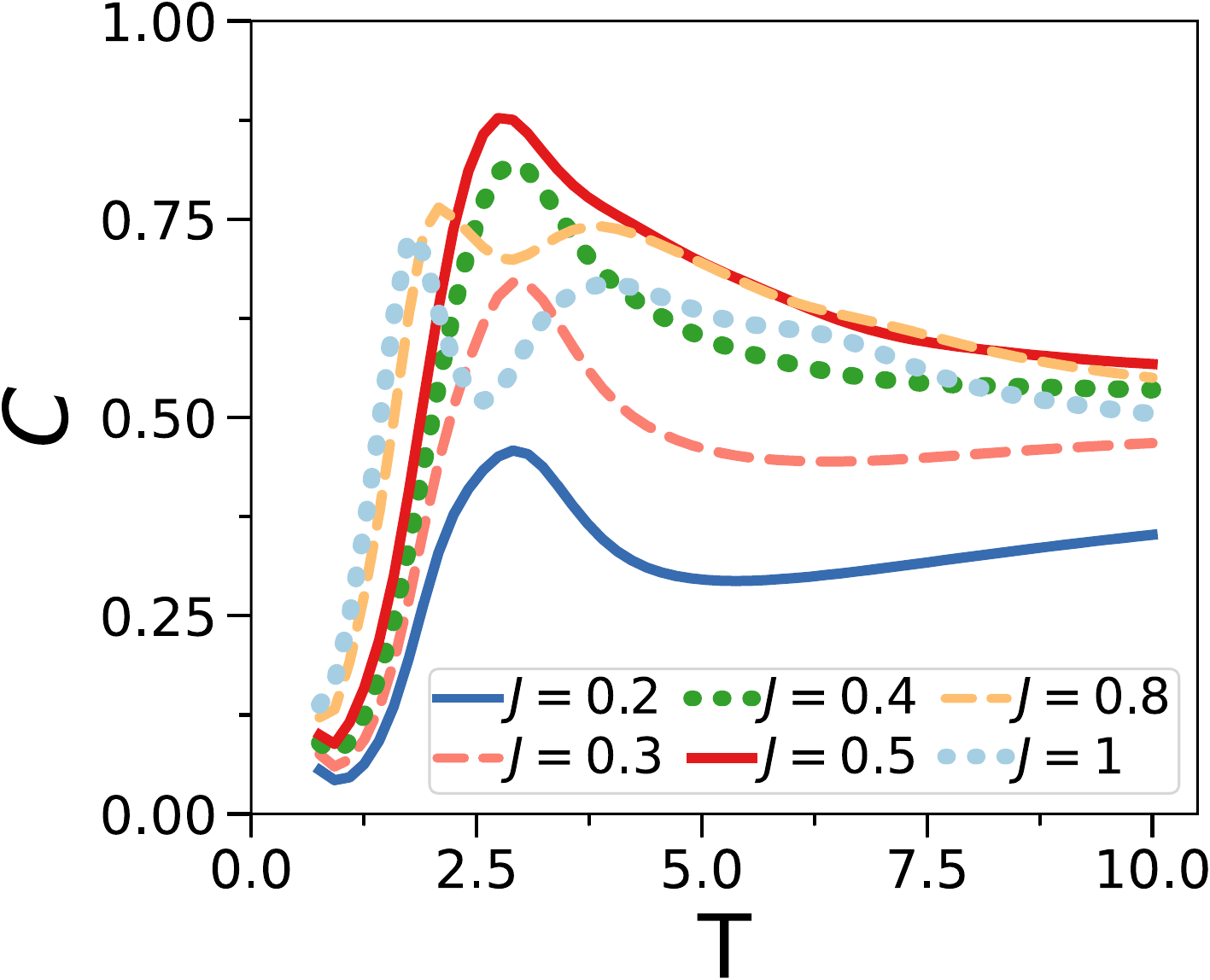}
	\caption{Time-dependent flux $\Omega(t)$ for $L=3$ ring sites. \idg{a} Current $\mathcal{J}(t)$ in time $t$ for a linearly increasing flux $\Omega(t)=t/T$ with  driving period $T=2.8$ and $J=\frac{1}{2}$. 
	\idg{b} DC/AC conversion efficiency $C=\Delta\mathcal{J}_\text{drain}/\langle\mathcal{J}_\text{source}\rangle$ measured as the drain current amplitude relative to average source current against driving period $T$ of the flux $\Omega(t)=t/T$.
	We have hard-core bosons with $L=3$, $K=1$ and $B_\alpha=1$.
	}
	\label{fig:dynamicflux}
\end{figure}

%\section{Atomic frequency generator}
\paragraph{Time-dependent flux.---}
The flux $\Omega(t)=t/T$ is now linearly increased in time by one flux quantum for one period $T$. As a result, the current undergoes a periodic modulation. 
For $t>0$, we inject atoms via the source into the initially empty ring and drains for the lattice Bose-Hubbard model. 
We show the case of $T=2.8$ in Fig.\ref{fig:dynamicflux}a. The current undergoes \revA{initial transient dynamics} until it settles into periodic sinusoidal oscillations, where the currents in the two drains are shifted by $T/2$. The drain current oscillates between close to 0 and nearly the magnitude of the source current.
Thus, this device realizes a \revA{form of} atomic DC/AC converter where a constant source current \revA{converts into AC modulated currents.}

We investigate the conversion efficiency $C=\Delta \mathcal{J}_\text{drain}/\text{max}(\mathcal{J}_\text{source})$ of the DC/AC converter as a function of $T$ and $J$ in Fig.\ref{fig:dynamicflux}b. $C=1$ \revA{indicates that} the amplitude of the drain current oscillation matches the source current.
We define the drain current amplitude $\Delta \mathcal{J}_\text{drain}=\text{max}_{t/T\gg1}\mathcal{J}_\text{drain}(t)-\text{min}_{t/T\gg1}\mathcal{J}_\text{drain}(t)$ and the maximal source current $\text{max}(\mathcal{J}_\text{source})=\text{max}_{t/T\gg1}\mathcal{J}_\text{source}(t)$, where take the maximum over the long-time behavior $t/T\gg1$. 
For small $T$, the driving is much faster than the system dynamics which suppresses \revA{large} oscillations.
For large $T$, \revA{the driving is much slower than the system dynamics, causing it to be dominated by the properties of the instantaneous steady state as a function of $\Omega$ (see SM~\ref{sec:drive_sup})}.
We find a sweet spot in the regime of intermediate $T\approx2.8$ with $C\approx0.88$ for $J=\frac{1}{2}$. \revA{While increasing the flux linearly might be experimentally difficult, similar results can achieved by a simple periodic modulation between $\Omega=0$ and $\Omega=1$ as shown in SM~\ref{sec:drive_osc_sup}} 

\paragraph{Discussion.---}
We propose a multi-functional atomtronic device with a three-terminal ring circuit. \revA{We study the setup with a continuous potential as well as a lattice with $L=3$ ring sites and extensive leads. }
In the low energy regime for the lattice setup, transport through the device is realized with density waves. The flux controls the conductance of the source-ring interface 
yielding a maximal current for $\Omega=0$ and minimal for $\Omega=\frac{1}{2}$. {\it By tuning $\Omega$, our  setup controls  the type of reflection with a crossover from negative Andreev-like to a normal one.} 
\revA{This effect opens up a new way to control the transport of density waves, as well as detect flux in the system. The value of flux can be obtained by measuring the reflection, with positive reflection indicating a value of flux close to half-integer. 
While we studied Andreev-reflections of relatively small density waves, given its wave-origin and persistence in the GPE regime~\cite{haug2019andreev}, we expect larger density waves to show similar behavior~\cite{daley2008andreev}. The density wave in lattice systems can be read out via the {\it in-situ} atom density with state-of-the-art atom microscopes~\cite{sherson2010single}.
For the low-energy transport, the transport is carried by collective density wave excitations of the atomic condensate. In contrast to the non-equilibrium regime, the flux controls only the magnitude, but not the direction of the current into the leads. 
For a 1D Bose-Liquid ring coupled to two leads, the low-energy current is known to be independent of flux~\cite{tokuno2008dynamics}. 
Here, only the persistent current within the ring couples to the flux, while the transmission is flux independent. For low-energy currents, we believe this is also true for three leads systems, yielding the observed non-directional currents.
However, finite size effects can change the effective coupling strength between ring and leads~\cite{tokuno2008dynamics}.
Flux causes substantial shifts of the energy levels in small rings, likely leading to a decreased effective ring-lead coupling and transmission for half-flux. }

Far from the ground state, the system is characterized by a substantially different dynamics. 
We analyse the dynamics for \revA{continuous} non-interacting (Landauer formalism), dilute (continuous 2D GPE) and \revA{discrete} strongly interacting limit (Bose-Hubbard lattice with Lindblad). \revA{The flux controls the direction of the current in all three regimes}. \revB{Our work shows that control over the directionality is robust, appearing for zero, weak and strong interactions, as well as for exactly one-dimensional and finite two-dimensional systems. }
{\it By choosing the flux around $\Omega=(2k+1)/4$, $k$ being integer, we find a non-reciprocal behavior where we can direct the flow into either of the drains}. With this effect we can switch the matter-wave  between different output terminals to realize a transistor or a rotation sensor.

By sweeping the flux in  time, the matter-wave experiences a time-dependent Aharonov-Bohm effect. This could be experimentally achieved by a constant acceleration of the rotation affecting the ring, \revA{or ramping the flux periodically up and down.}
This driving generates a sinusoidal modulation of the current in time. 
{\it As a result, a constant source current is converted into an \revA{AC modulated current} in the drains.} We control the frequency \revA{and amplitude} of the \revA{modulation} via the change of the flux, with maximal conversion  efficiency $C\approx0.88$. 
In a reverse operation, this remarkable feature can be used  as a sensor for time-dependent rotations $\Omega(t)$ by measuring the frequency of the current.

The time-dependent flux also allows us to study the time-dependent Aharonov-Bohm effect with cold atoms in a controlled environment, which has remained an open problem in other systems~\cite{singleton2013covariant,macdougall2015revisiting,jing2017time,choudhury2019direct}.
\revB{Depending on the cold-atom implementation, additional terms  can appear in the effective Hamiltonian for the time-dependent driving of the flux, which have to be carefully studied in future work.}

To our knowledge, we provide the first cold atom system that integrates switch-like and frequency generating capabilities. 
%Our work relies on current experimental capabilities of the field, where we experimentally demonstrate the initial setup with a BEC in a light-shaped potential.
\revB{Our work relies on current experimental capabilities of the field, where we experimentally demonstrate the feasibility of the parameters of our simulations. }
\revA{Note that our setup is a special case of a much larger class of possible atomtronic setups. Inspired by vast applications of classical ring wave guides for electromagnetic fields~\cite{pon1961hybrid,yariv2002critical,dong2010ghz}, analogous atomtronic devices for directional couplers, frequency filters or wave splitters could be designed. We hope to draw attention to this rich field for application in cold atom technologies, which to our knowledge has been barely explored.}

\begin{acknowledgements}
\paragraph{Acknowledgements.---}
%\section{Acknowledgements}
We thank many participants of the Atomtronics conference 2022 in Benasque for fruitful discussions. This work is supported by the Singapore Ministry of Education (MOE) and the Singapore National Research Foundation (NRF).
\end{acknowledgements}

\bibliography{paper}% Produces the bibliography via BibTeX.

\begin{thebibliography}{78}%
\makeatletter
\providecommand \@ifxundefined [1]{%
 \@ifx{#1\undefined}
}%
\providecommand \@ifnum [1]{%
 \ifnum #1\expandafter \@firstoftwo
 \else \expandafter \@secondoftwo
 \fi
}%
\providecommand \@ifx [1]{%
 \ifx #1\expandafter \@firstoftwo
 \else \expandafter \@secondoftwo
 \fi
}%
\providecommand \natexlab [1]{#1}%
\providecommand \enquote  [1]{``#1''}%
\providecommand \bibnamefont  [1]{#1}%
\providecommand \bibfnamefont [1]{#1}%
\providecommand \citenamefont [1]{#1}%
\providecommand \href@noop [0]{\@secondoftwo}%
\providecommand \href [0]{\begingroup \@sanitize@url \@href}%
\providecommand \@href[1]{\@@startlink{#1}\@@href}%
\providecommand \@@href[1]{\endgroup#1\@@endlink}%
\providecommand \@sanitize@url [0]{\catcode `\\12\catcode `\$12\catcode
  `\&12\catcode `\#12\catcode `\^12\catcode `\_12\catcode `\%12\relax}%
\providecommand \@@startlink[1]{}%
\providecommand \@@endlink[0]{}%
\providecommand \url  [0]{\begingroup\@sanitize@url \@url }%
\providecommand \@url [1]{\endgroup\@href {#1}{\urlprefix }}%
\providecommand \urlprefix  [0]{URL }%
\providecommand \Eprint [0]{\href }%
\providecommand \doibase [0]{https://doi.org/}%
\providecommand \selectlanguage [0]{\@gobble}%
\providecommand \bibinfo  [0]{\@secondoftwo}%
\providecommand \bibfield  [0]{\@secondoftwo}%
\providecommand \translation [1]{[#1]}%
\providecommand \BibitemOpen [0]{}%
\providecommand \bibitemStop [0]{}%
\providecommand \bibitemNoStop [0]{.\EOS\space}%
\providecommand \EOS [0]{\spacefactor3000\relax}%
\providecommand \BibitemShut  [1]{\csname bibitem#1\endcsname}%
\let\auto@bib@innerbib\@empty
%</preamble>
\bibitem [{\citenamefont {Bloch}\ \emph {et~al.}(2012)\citenamefont {Bloch},
  \citenamefont {Dalibard},\ and\ \citenamefont
  {Nascimbene}}]{bloch2012quantum}%
  \BibitemOpen
  \bibfield  {author} {\bibinfo {author} {\bibfnamefont {I.}~\bibnamefont
  {Bloch}}, \bibinfo {author} {\bibfnamefont {J.}~\bibnamefont {Dalibard}},\
  and\ \bibinfo {author} {\bibfnamefont {S.}~\bibnamefont {Nascimbene}},\
  }\bibfield  {title} {\bibinfo {title} {Quantum simulations with ultracold
  quantum gases},\ }\href@noop {} {\bibfield  {journal} {\bibinfo  {journal}
  {Nature Physics}\ }\textbf {\bibinfo {volume} {8}},\ \bibinfo {pages} {267}
  (\bibinfo {year} {2012})}\BibitemShut {NoStop}%
\bibitem [{\citenamefont {Keil}\ \emph {et~al.}(2016)\citenamefont {Keil},
  \citenamefont {Amit}, \citenamefont {Zhou}, \citenamefont {Groswasser},
  \citenamefont {Japha},\ and\ \citenamefont {Folman}}]{keil2016fifteen}%
  \BibitemOpen
  \bibfield  {author} {\bibinfo {author} {\bibfnamefont {M.}~\bibnamefont
  {Keil}}, \bibinfo {author} {\bibfnamefont {O.}~\bibnamefont {Amit}}, \bibinfo
  {author} {\bibfnamefont {S.}~\bibnamefont {Zhou}}, \bibinfo {author}
  {\bibfnamefont {D.}~\bibnamefont {Groswasser}}, \bibinfo {author}
  {\bibfnamefont {Y.}~\bibnamefont {Japha}},\ and\ \bibinfo {author}
  {\bibfnamefont {R.}~\bibnamefont {Folman}},\ }\bibfield  {title} {\bibinfo
  {title} {Fifteen years of cold matter on the atom chip: promise,
  realizations, and prospects},\ }\href@noop {} {\bibfield  {journal} {\bibinfo
   {journal} {Journal of modern optics}\ }\textbf {\bibinfo {volume} {63}},\
  \bibinfo {pages} {1840} (\bibinfo {year} {2016})}\BibitemShut {NoStop}%
\bibitem [{\citenamefont {Amico}\ \emph {et~al.}(2022)\citenamefont {Amico},
  \citenamefont {Anderson}, \citenamefont {Boshier}, \citenamefont {Brantut},
  \citenamefont {Kwek}, \citenamefont {Minguzzi},\ and\ \citenamefont {von
  Klitzing}}]{amico2021atomtronic}%
  \BibitemOpen
  \bibfield  {author} {\bibinfo {author} {\bibfnamefont {L.}~\bibnamefont
  {Amico}}, \bibinfo {author} {\bibfnamefont {D.}~\bibnamefont {Anderson}},
  \bibinfo {author} {\bibfnamefont {M.}~\bibnamefont {Boshier}}, \bibinfo
  {author} {\bibfnamefont {J.-P.}\ \bibnamefont {Brantut}}, \bibinfo {author}
  {\bibfnamefont {L.-C.}\ \bibnamefont {Kwek}}, \bibinfo {author}
  {\bibfnamefont {A.}~\bibnamefont {Minguzzi}},\ and\ \bibinfo {author}
  {\bibfnamefont {W.}~\bibnamefont {von Klitzing}},\ }\bibfield  {title}
  {\bibinfo {title} {Colloquium: Atomtronic circuits: From many-body physics to
  quantum technologies},\ }\href {https://doi.org/10.1103/RevModPhys.94.041001}
  {\bibfield  {journal} {\bibinfo  {journal} {Rev. Mod. Phys.}\ }\textbf
  {\bibinfo {volume} {94}},\ \bibinfo {pages} {041001} (\bibinfo {year}
  {2022})}\BibitemShut {NoStop}%
\bibitem [{\citenamefont {Amico}\ \emph {et~al.}(2021)\citenamefont {Amico},
  \citenamefont {Boshier}, \citenamefont {Birkl}, \citenamefont {Minguzzi},
  \citenamefont {Miniatura}, \citenamefont {Kwek}, \citenamefont {Aghamalyan},
  \citenamefont {Ahufinger}, \citenamefont {Anderson}, \citenamefont {Andrei}
  \emph {et~al.}}]{amico2021roadmap}%
  \BibitemOpen
  \bibfield  {author} {\bibinfo {author} {\bibfnamefont {L.}~\bibnamefont
  {Amico}}, \bibinfo {author} {\bibfnamefont {M.}~\bibnamefont {Boshier}},
  \bibinfo {author} {\bibfnamefont {G.}~\bibnamefont {Birkl}}, \bibinfo
  {author} {\bibfnamefont {A.}~\bibnamefont {Minguzzi}}, \bibinfo {author}
  {\bibfnamefont {C.}~\bibnamefont {Miniatura}}, \bibinfo {author}
  {\bibfnamefont {L.-C.}\ \bibnamefont {Kwek}}, \bibinfo {author}
  {\bibfnamefont {D.}~\bibnamefont {Aghamalyan}}, \bibinfo {author}
  {\bibfnamefont {V.}~\bibnamefont {Ahufinger}}, \bibinfo {author}
  {\bibfnamefont {D.}~\bibnamefont {Anderson}}, \bibinfo {author}
  {\bibfnamefont {N.}~\bibnamefont {Andrei}}, \emph {et~al.},\ }\bibfield
  {title} {\bibinfo {title} {Roadmap on atomtronics: State of the art and
  perspective},\ }\href@noop {} {\bibfield  {journal} {\bibinfo  {journal} {AVS
  Quantum Science}\ }\textbf {\bibinfo {volume} {3}},\ \bibinfo {pages}
  {039201} (\bibinfo {year} {2021})}\BibitemShut {NoStop}%
\bibitem [{\citenamefont {Amico}\ \emph {et~al.}(2017)\citenamefont {Amico},
  \citenamefont {Birkl}, \citenamefont {Boshier},\ and\ \citenamefont
  {Kwek}}]{amico2017focus}%
  \BibitemOpen
  \bibfield  {author} {\bibinfo {author} {\bibfnamefont {L.}~\bibnamefont
  {Amico}}, \bibinfo {author} {\bibfnamefont {G.}~\bibnamefont {Birkl}},
  \bibinfo {author} {\bibfnamefont {M.}~\bibnamefont {Boshier}},\ and\ \bibinfo
  {author} {\bibfnamefont {L.-C.}\ \bibnamefont {Kwek}},\ }\bibfield  {title}
  {\bibinfo {title} {Focus on atomtronics-enabled quantum technologies},\
  }\href@noop {} {\bibfield  {journal} {\bibinfo  {journal} {New Journal of
  Physics}\ }\textbf {\bibinfo {volume} {19}},\ \bibinfo {pages} {020201}
  (\bibinfo {year} {2017})}\BibitemShut {NoStop}%
\bibitem [{\citenamefont {McGloin}\ \emph {et~al.}(2003)\citenamefont
  {McGloin}, \citenamefont {Spalding}, \citenamefont {Melville}, \citenamefont
  {Sibbett},\ and\ \citenamefont {Dholakia}}]{mcgloin2003applications}%
  \BibitemOpen
  \bibfield  {author} {\bibinfo {author} {\bibfnamefont {D.}~\bibnamefont
  {McGloin}}, \bibinfo {author} {\bibfnamefont {G.~C.}\ \bibnamefont
  {Spalding}}, \bibinfo {author} {\bibfnamefont {H.}~\bibnamefont {Melville}},
  \bibinfo {author} {\bibfnamefont {W.}~\bibnamefont {Sibbett}},\ and\ \bibinfo
  {author} {\bibfnamefont {K.}~\bibnamefont {Dholakia}},\ }\bibfield  {title}
  {\bibinfo {title} {Applications of spatial light modulators in atom optics},\
  }\href@noop {} {\bibfield  {journal} {\bibinfo  {journal} {Optics Express}\
  }\textbf {\bibinfo {volume} {11}},\ \bibinfo {pages} {158} (\bibinfo {year}
  {2003})}\BibitemShut {NoStop}%
\bibitem [{\citenamefont {Ryu}\ and\ \citenamefont
  {Boshier}(2015)}]{ryu2015integrated}%
  \BibitemOpen
  \bibfield  {author} {\bibinfo {author} {\bibfnamefont {C.}~\bibnamefont
  {Ryu}}\ and\ \bibinfo {author} {\bibfnamefont {M.~G.}\ \bibnamefont
  {Boshier}},\ }\bibfield  {title} {\bibinfo {title} {Integrated coherent
  matter wave circuits},\ }\href@noop {} {\bibfield  {journal} {\bibinfo
  {journal} {New Journal of Physics}\ }\textbf {\bibinfo {volume} {17}},\
  \bibinfo {pages} {092002} (\bibinfo {year} {2015})}\BibitemShut {NoStop}%
\bibitem [{\citenamefont {Gauthier}\ \emph {et~al.}(2016)\citenamefont
  {Gauthier}, \citenamefont {Lenton}, \citenamefont {Parry}, \citenamefont
  {Baker}, \citenamefont {Davis}, \citenamefont {Rubinsztein-Dunlop},\ and\
  \citenamefont {Neely}}]{gauthier2016direct}%
  \BibitemOpen
  \bibfield  {author} {\bibinfo {author} {\bibfnamefont {G.}~\bibnamefont
  {Gauthier}}, \bibinfo {author} {\bibfnamefont {I.}~\bibnamefont {Lenton}},
  \bibinfo {author} {\bibfnamefont {N.~M.}\ \bibnamefont {Parry}}, \bibinfo
  {author} {\bibfnamefont {M.}~\bibnamefont {Baker}}, \bibinfo {author}
  {\bibfnamefont {M.}~\bibnamefont {Davis}}, \bibinfo {author} {\bibfnamefont
  {H.}~\bibnamefont {Rubinsztein-Dunlop}},\ and\ \bibinfo {author}
  {\bibfnamefont {T.}~\bibnamefont {Neely}},\ }\bibfield  {title} {\bibinfo
  {title} {Direct imaging of a digital-micromirror device for configurable
  microscopic optical potentials},\ }\href@noop {} {\bibfield  {journal}
  {\bibinfo  {journal} {Optica}\ }\textbf {\bibinfo {volume} {3}},\ \bibinfo
  {pages} {1136} (\bibinfo {year} {2016})}\BibitemShut {NoStop}%
\bibitem [{\citenamefont {Barredo}\ \emph {et~al.}(2018)\citenamefont
  {Barredo}, \citenamefont {Lienhard}, \citenamefont {De~Leseleuc},
  \citenamefont {Lahaye},\ and\ \citenamefont
  {Browaeys}}]{barredo2018synthetic}%
  \BibitemOpen
  \bibfield  {author} {\bibinfo {author} {\bibfnamefont {D.}~\bibnamefont
  {Barredo}}, \bibinfo {author} {\bibfnamefont {V.}~\bibnamefont {Lienhard}},
  \bibinfo {author} {\bibfnamefont {S.}~\bibnamefont {De~Leseleuc}}, \bibinfo
  {author} {\bibfnamefont {T.}~\bibnamefont {Lahaye}},\ and\ \bibinfo {author}
  {\bibfnamefont {A.}~\bibnamefont {Browaeys}},\ }\bibfield  {title} {\bibinfo
  {title} {Synthetic three-dimensional atomic structures assembled atom by
  atom},\ }\href@noop {} {\bibfield  {journal} {\bibinfo  {journal} {Nature}\
  }\textbf {\bibinfo {volume} {561}},\ \bibinfo {pages} {79} (\bibinfo {year}
  {2018})}\BibitemShut {NoStop}%
\bibitem [{\citenamefont {Seaman}\ \emph {et~al.}(2007)\citenamefont {Seaman},
  \citenamefont {Kr{\"a}mer}, \citenamefont {Anderson},\ and\ \citenamefont
  {Holland}}]{seaman2007atomtronics}%
  \BibitemOpen
  \bibfield  {author} {\bibinfo {author} {\bibfnamefont {B.}~\bibnamefont
  {Seaman}}, \bibinfo {author} {\bibfnamefont {M.}~\bibnamefont {Kr{\"a}mer}},
  \bibinfo {author} {\bibfnamefont {D.}~\bibnamefont {Anderson}},\ and\
  \bibinfo {author} {\bibfnamefont {M.}~\bibnamefont {Holland}},\ }\bibfield
  {title} {\bibinfo {title} {Atomtronics: Ultracold-atom analogs of electronic
  devices},\ }\href@noop {} {\bibfield  {journal} {\bibinfo  {journal}
  {Physical Review A}\ }\textbf {\bibinfo {volume} {75}},\ \bibinfo {pages}
  {023615} (\bibinfo {year} {2007})}\BibitemShut {NoStop}%
\bibitem [{\citenamefont {Pepino}\ \emph {et~al.}(2009)\citenamefont {Pepino},
  \citenamefont {Cooper}, \citenamefont {Anderson},\ and\ \citenamefont
  {Holland}}]{pepino2009atomtronic}%
  \BibitemOpen
  \bibfield  {author} {\bibinfo {author} {\bibfnamefont {R.}~\bibnamefont
  {Pepino}}, \bibinfo {author} {\bibfnamefont {J.}~\bibnamefont {Cooper}},
  \bibinfo {author} {\bibfnamefont {D.}~\bibnamefont {Anderson}},\ and\
  \bibinfo {author} {\bibfnamefont {M.}~\bibnamefont {Holland}},\ }\bibfield
  {title} {\bibinfo {title} {Atomtronic circuits of diodes and transistors},\
  }\href@noop {} {\bibfield  {journal} {\bibinfo  {journal} {Physical review
  letters}\ }\textbf {\bibinfo {volume} {103}},\ \bibinfo {pages} {140405}
  (\bibinfo {year} {2009})}\BibitemShut {NoStop}%
\bibitem [{\citenamefont {Pepino}(2021)}]{pepino2021advances}%
  \BibitemOpen
  \bibfield  {author} {\bibinfo {author} {\bibfnamefont {R.~A.}\ \bibnamefont
  {Pepino}},\ }\bibfield  {title} {\bibinfo {title} {Advances in atomtronics},\
  }\href@noop {} {\bibfield  {journal} {\bibinfo  {journal} {Entropy}\ }\textbf
  {\bibinfo {volume} {23}},\ \bibinfo {pages} {534} (\bibinfo {year}
  {2021})}\BibitemShut {NoStop}%
\bibitem [{\citenamefont {Anderson}(2021)}]{anderson2021matter}%
  \BibitemOpen
  \bibfield  {author} {\bibinfo {author} {\bibfnamefont {D.~Z.}\ \bibnamefont
  {Anderson}},\ }\bibfield  {title} {\bibinfo {title} {Matter waves,
  single-mode excitations of the matter-wave field, and the atomtronic
  transistor oscillator},\ }\href@noop {} {\bibfield  {journal} {\bibinfo
  {journal} {Physical Review A}\ }\textbf {\bibinfo {volume} {104}},\ \bibinfo
  {pages} {033311} (\bibinfo {year} {2021})}\BibitemShut {NoStop}%
\bibitem [{\citenamefont {Caliga}\ \emph
  {et~al.}(2016{\natexlab{a}})\citenamefont {Caliga}, \citenamefont
  {Straatsma}, \citenamefont {Zozulya},\ and\ \citenamefont
  {Anderson}}]{caliga2016principles}%
  \BibitemOpen
  \bibfield  {author} {\bibinfo {author} {\bibfnamefont {S.~C.}\ \bibnamefont
  {Caliga}}, \bibinfo {author} {\bibfnamefont {C.~J.}\ \bibnamefont
  {Straatsma}}, \bibinfo {author} {\bibfnamefont {A.~A.}\ \bibnamefont
  {Zozulya}},\ and\ \bibinfo {author} {\bibfnamefont {D.~Z.}\ \bibnamefont
  {Anderson}},\ }\bibfield  {title} {\bibinfo {title} {Principles of an
  atomtronic transistor},\ }\href@noop {} {\bibfield  {journal} {\bibinfo
  {journal} {New Journal of Physics}\ }\textbf {\bibinfo {volume} {18}},\
  \bibinfo {pages} {015012} (\bibinfo {year} {2016}{\natexlab{a}})}\BibitemShut
  {NoStop}%
\bibitem [{\citenamefont {Caliga}\ \emph
  {et~al.}(2016{\natexlab{b}})\citenamefont {Caliga}, \citenamefont
  {Straatsma},\ and\ \citenamefont {Anderson}}]{caliga2016transport}%
  \BibitemOpen
  \bibfield  {author} {\bibinfo {author} {\bibfnamefont {S.~C.}\ \bibnamefont
  {Caliga}}, \bibinfo {author} {\bibfnamefont {C.~J.}\ \bibnamefont
  {Straatsma}},\ and\ \bibinfo {author} {\bibfnamefont {D.~Z.}\ \bibnamefont
  {Anderson}},\ }\bibfield  {title} {\bibinfo {title} {Transport dynamics of
  ultracold atoms in a triple-well transistor-like potential},\ }\href@noop {}
  {\bibfield  {journal} {\bibinfo  {journal} {New Journal of Physics}\ }\textbf
  {\bibinfo {volume} {18}},\ \bibinfo {pages} {025010} (\bibinfo {year}
  {2016}{\natexlab{b}})}\BibitemShut {NoStop}%
\bibitem [{\citenamefont {Caliga}\ \emph {et~al.}(2017)\citenamefont {Caliga},
  \citenamefont {Straatsma},\ and\ \citenamefont
  {Anderson}}]{caliga2017experimental}%
  \BibitemOpen
  \bibfield  {author} {\bibinfo {author} {\bibfnamefont {S.~C.}\ \bibnamefont
  {Caliga}}, \bibinfo {author} {\bibfnamefont {C.~J.}\ \bibnamefont
  {Straatsma}},\ and\ \bibinfo {author} {\bibfnamefont {D.~Z.}\ \bibnamefont
  {Anderson}},\ }\bibfield  {title} {\bibinfo {title} {Experimental
  demonstration of an atomtronic battery},\ }\href@noop {} {\bibfield
  {journal} {\bibinfo  {journal} {New Journal of Physics}\ }\textbf {\bibinfo
  {volume} {19}},\ \bibinfo {pages} {013036} (\bibinfo {year}
  {2017})}\BibitemShut {NoStop}%
\bibitem [{\citenamefont {Wilsmann}\ \emph {et~al.}(2018)\citenamefont
  {Wilsmann}, \citenamefont {Ymai}, \citenamefont {Tonel}, \citenamefont
  {Links},\ and\ \citenamefont {Foerster}}]{wilsmann2018control}%
  \BibitemOpen
  \bibfield  {author} {\bibinfo {author} {\bibfnamefont {K.~W.}\ \bibnamefont
  {Wilsmann}}, \bibinfo {author} {\bibfnamefont {L.~H.}\ \bibnamefont {Ymai}},
  \bibinfo {author} {\bibfnamefont {A.~P.}\ \bibnamefont {Tonel}}, \bibinfo
  {author} {\bibfnamefont {J.}~\bibnamefont {Links}},\ and\ \bibinfo {author}
  {\bibfnamefont {A.}~\bibnamefont {Foerster}},\ }\bibfield  {title} {\bibinfo
  {title} {Control of tunneling in an atomtronic switching device},\
  }\href@noop {} {\bibfield  {journal} {\bibinfo  {journal} {Communications
  physics}\ }\textbf {\bibinfo {volume} {1}},\ \bibinfo {pages} {1} (\bibinfo
  {year} {2018})}\BibitemShut {NoStop}%
\bibitem [{\citenamefont {Ryu}\ \emph {et~al.}(2020)\citenamefont {Ryu},
  \citenamefont {Samson},\ and\ \citenamefont {Boshier}}]{ryu2020quantum}%
  \BibitemOpen
  \bibfield  {author} {\bibinfo {author} {\bibfnamefont {C.}~\bibnamefont
  {Ryu}}, \bibinfo {author} {\bibfnamefont {E.}~\bibnamefont {Samson}},\ and\
  \bibinfo {author} {\bibfnamefont {M.~G.}\ \bibnamefont {Boshier}},\
  }\bibfield  {title} {\bibinfo {title} {Quantum interference of currents in an
  atomtronic squid},\ }\href@noop {} {\bibfield  {journal} {\bibinfo  {journal}
  {Nature communications}\ }\textbf {\bibinfo {volume} {11}},\ \bibinfo {pages}
  {1} (\bibinfo {year} {2020})}\BibitemShut {NoStop}%
\bibitem [{\citenamefont {Eckel}\ \emph {et~al.}(2014)\citenamefont {Eckel},
  \citenamefont {Lee}, \citenamefont {Jendrzejewski}, \citenamefont {Murray},
  \citenamefont {Clark}, \citenamefont {Lobb}, \citenamefont {Phillips},
  \citenamefont {Edwards},\ and\ \citenamefont
  {Campbell}}]{eckel2014hysteresis}%
  \BibitemOpen
  \bibfield  {author} {\bibinfo {author} {\bibfnamefont {S.}~\bibnamefont
  {Eckel}}, \bibinfo {author} {\bibfnamefont {J.~G.}\ \bibnamefont {Lee}},
  \bibinfo {author} {\bibfnamefont {F.}~\bibnamefont {Jendrzejewski}}, \bibinfo
  {author} {\bibfnamefont {N.}~\bibnamefont {Murray}}, \bibinfo {author}
  {\bibfnamefont {C.~W.}\ \bibnamefont {Clark}}, \bibinfo {author}
  {\bibfnamefont {C.~J.}\ \bibnamefont {Lobb}}, \bibinfo {author}
  {\bibfnamefont {W.~D.}\ \bibnamefont {Phillips}}, \bibinfo {author}
  {\bibfnamefont {M.}~\bibnamefont {Edwards}},\ and\ \bibinfo {author}
  {\bibfnamefont {G.~K.}\ \bibnamefont {Campbell}},\ }\bibfield  {title}
  {\bibinfo {title} {Hysteresis in a quantized superfluid
  ‘atomtronic’circuit},\ }\href@noop {} {\bibfield  {journal} {\bibinfo
  {journal} {Nature}\ }\textbf {\bibinfo {volume} {506}},\ \bibinfo {pages}
  {200} (\bibinfo {year} {2014})}\BibitemShut {NoStop}%
\bibitem [{\citenamefont {Wright}\ \emph {et~al.}(2013)\citenamefont {Wright},
  \citenamefont {Blakestad}, \citenamefont {Lobb}, \citenamefont {Phillips},\
  and\ \citenamefont {Campbell}}]{wright2013driving}%
  \BibitemOpen
  \bibfield  {author} {\bibinfo {author} {\bibfnamefont {K.~C.}\ \bibnamefont
  {Wright}}, \bibinfo {author} {\bibfnamefont {R.}~\bibnamefont {Blakestad}},
  \bibinfo {author} {\bibfnamefont {C.}~\bibnamefont {Lobb}}, \bibinfo {author}
  {\bibfnamefont {W.}~\bibnamefont {Phillips}},\ and\ \bibinfo {author}
  {\bibfnamefont {G.}~\bibnamefont {Campbell}},\ }\bibfield  {title} {\bibinfo
  {title} {Driving phase slips in a superfluid atom circuit with a rotating
  weak link},\ }\href@noop {} {\bibfield  {journal} {\bibinfo  {journal}
  {Physical review letters}\ }\textbf {\bibinfo {volume} {110}},\ \bibinfo
  {pages} {025302} (\bibinfo {year} {2013})}\BibitemShut {NoStop}%
\bibitem [{\citenamefont {P{\'e}rez-Obiol}\ \emph {et~al.}(2022)\citenamefont
  {P{\'e}rez-Obiol}, \citenamefont {Polo},\ and\ \citenamefont
  {Amico}}]{perez2021coherent}%
  \BibitemOpen
  \bibfield  {author} {\bibinfo {author} {\bibfnamefont {A.}~\bibnamefont
  {P{\'e}rez-Obiol}}, \bibinfo {author} {\bibfnamefont {J.}~\bibnamefont
  {Polo}},\ and\ \bibinfo {author} {\bibfnamefont {L.}~\bibnamefont {Amico}},\
  }\bibfield  {title} {\bibinfo {title} {Coherent phase slips in coupled
  matter-wave circuits},\ }\href@noop {} {\bibfield  {journal} {\bibinfo
  {journal} {Physical Review Research}\ }\textbf {\bibinfo {volume} {4}},\
  \bibinfo {pages} {L022038} (\bibinfo {year} {2022})}\BibitemShut {NoStop}%
\bibitem [{\citenamefont {Kiehn}\ \emph {et~al.}(2022)\citenamefont {Kiehn},
  \citenamefont {Singh},\ and\ \citenamefont
  {Mathey}}]{kiehn2022implementation}%
  \BibitemOpen
  \bibfield  {author} {\bibinfo {author} {\bibfnamefont {H.}~\bibnamefont
  {Kiehn}}, \bibinfo {author} {\bibfnamefont {V.~P.}\ \bibnamefont {Singh}},\
  and\ \bibinfo {author} {\bibfnamefont {L.}~\bibnamefont {Mathey}},\
  }\bibfield  {title} {\bibinfo {title} {Implementation of an atomtronic squid
  in a strongly confined toroidal condensate},\ }\href@noop {} {\bibfield
  {journal} {\bibinfo  {journal} {Physical Review Research}\ }\textbf {\bibinfo
  {volume} {4}},\ \bibinfo {pages} {033024} (\bibinfo {year}
  {2022})}\BibitemShut {NoStop}%
\bibitem [{\citenamefont {Chien}\ \emph {et~al.}(2015)\citenamefont {Chien},
  \citenamefont {Peotta},\ and\ \citenamefont {Di~Ventra}}]{chien2015quantum}%
  \BibitemOpen
  \bibfield  {author} {\bibinfo {author} {\bibfnamefont {C.-C.}\ \bibnamefont
  {Chien}}, \bibinfo {author} {\bibfnamefont {S.}~\bibnamefont {Peotta}},\ and\
  \bibinfo {author} {\bibfnamefont {M.}~\bibnamefont {Di~Ventra}},\ }\bibfield
  {title} {\bibinfo {title} {Quantum transport in ultracold atoms},\
  }\href@noop {} {\bibfield  {journal} {\bibinfo  {journal} {Nature Physics}\
  }\textbf {\bibinfo {volume} {11}},\ \bibinfo {pages} {998} (\bibinfo {year}
  {2015})}\BibitemShut {NoStop}%
\bibitem [{\citenamefont {Heeger}\ \emph {et~al.}(1988)\citenamefont {Heeger},
  \citenamefont {Kivelson}, \citenamefont {Schrieffer},\ and\ \citenamefont
  {Su}}]{heeger1988solitons}%
  \BibitemOpen
  \bibfield  {author} {\bibinfo {author} {\bibfnamefont {A.~J.}\ \bibnamefont
  {Heeger}}, \bibinfo {author} {\bibfnamefont {S.}~\bibnamefont {Kivelson}},
  \bibinfo {author} {\bibfnamefont {J.}~\bibnamefont {Schrieffer}},\ and\
  \bibinfo {author} {\bibfnamefont {W.-P.}\ \bibnamefont {Su}},\ }\bibfield
  {title} {\bibinfo {title} {Solitons in conducting polymers},\ }\href@noop {}
  {\bibfield  {journal} {\bibinfo  {journal} {Reviews of Modern Physics}\
  }\textbf {\bibinfo {volume} {60}},\ \bibinfo {pages} {781} (\bibinfo {year}
  {1988})}\BibitemShut {NoStop}%
\bibitem [{\citenamefont {Atala}\ \emph {et~al.}(2013)\citenamefont {Atala},
  \citenamefont {Aidelsburger}, \citenamefont {Barreiro}, \citenamefont
  {Abanin}, \citenamefont {Kitagawa}, \citenamefont {Demler},\ and\
  \citenamefont {Bloch}}]{atala2013direct}%
  \BibitemOpen
  \bibfield  {author} {\bibinfo {author} {\bibfnamefont {M.}~\bibnamefont
  {Atala}}, \bibinfo {author} {\bibfnamefont {M.}~\bibnamefont {Aidelsburger}},
  \bibinfo {author} {\bibfnamefont {J.~T.}\ \bibnamefont {Barreiro}}, \bibinfo
  {author} {\bibfnamefont {D.}~\bibnamefont {Abanin}}, \bibinfo {author}
  {\bibfnamefont {T.}~\bibnamefont {Kitagawa}}, \bibinfo {author}
  {\bibfnamefont {E.}~\bibnamefont {Demler}},\ and\ \bibinfo {author}
  {\bibfnamefont {I.}~\bibnamefont {Bloch}},\ }\bibfield  {title} {\bibinfo
  {title} {Direct measurement of the zak phase in topological bloch bands},\
  }\href@noop {} {\bibfield  {journal} {\bibinfo  {journal} {Nature Physics}\
  }\textbf {\bibinfo {volume} {9}},\ \bibinfo {pages} {795} (\bibinfo {year}
  {2013})}\BibitemShut {NoStop}%
\bibitem [{\citenamefont {Miyake}\ \emph {et~al.}(2013)\citenamefont {Miyake},
  \citenamefont {Siviloglou}, \citenamefont {Kennedy}, \citenamefont {Burton},\
  and\ \citenamefont {Ketterle}}]{miyake2013realizing}%
  \BibitemOpen
  \bibfield  {author} {\bibinfo {author} {\bibfnamefont {H.}~\bibnamefont
  {Miyake}}, \bibinfo {author} {\bibfnamefont {G.~A.}\ \bibnamefont
  {Siviloglou}}, \bibinfo {author} {\bibfnamefont {C.~J.}\ \bibnamefont
  {Kennedy}}, \bibinfo {author} {\bibfnamefont {W.~C.}\ \bibnamefont
  {Burton}},\ and\ \bibinfo {author} {\bibfnamefont {W.}~\bibnamefont
  {Ketterle}},\ }\bibfield  {title} {\bibinfo {title} {Realizing the harper
  hamiltonian with laser-assisted tunneling in optical lattices},\ }\href@noop
  {} {\bibfield  {journal} {\bibinfo  {journal} {Physical review letters}\
  }\textbf {\bibinfo {volume} {111}},\ \bibinfo {pages} {185302} (\bibinfo
  {year} {2013})}\BibitemShut {NoStop}%
\bibitem [{\citenamefont {Aidelsburger}\ \emph {et~al.}(2013)\citenamefont
  {Aidelsburger}, \citenamefont {Atala}, \citenamefont {Lohse}, \citenamefont
  {Barreiro}, \citenamefont {Paredes},\ and\ \citenamefont
  {Bloch}}]{aidelsburger2013realization}%
  \BibitemOpen
  \bibfield  {author} {\bibinfo {author} {\bibfnamefont {M.}~\bibnamefont
  {Aidelsburger}}, \bibinfo {author} {\bibfnamefont {M.}~\bibnamefont {Atala}},
  \bibinfo {author} {\bibfnamefont {M.}~\bibnamefont {Lohse}}, \bibinfo
  {author} {\bibfnamefont {J.~T.}\ \bibnamefont {Barreiro}}, \bibinfo {author}
  {\bibfnamefont {B.}~\bibnamefont {Paredes}},\ and\ \bibinfo {author}
  {\bibfnamefont {I.}~\bibnamefont {Bloch}},\ }\bibfield  {title} {\bibinfo
  {title} {Realization of the hofstadter hamiltonian with ultracold atoms in
  optical lattices},\ }\href@noop {} {\bibfield  {journal} {\bibinfo  {journal}
  {Physical review letters}\ }\textbf {\bibinfo {volume} {111}},\ \bibinfo
  {pages} {185301} (\bibinfo {year} {2013})}\BibitemShut {NoStop}%
\bibitem [{\citenamefont {Jotzu}\ \emph {et~al.}(2014)\citenamefont {Jotzu},
  \citenamefont {Messer}, \citenamefont {Desbuquois}, \citenamefont {Lebrat},
  \citenamefont {Uehlinger}, \citenamefont {Greif},\ and\ \citenamefont
  {Esslinger}}]{jotzu2014experimental}%
  \BibitemOpen
  \bibfield  {author} {\bibinfo {author} {\bibfnamefont {G.}~\bibnamefont
  {Jotzu}}, \bibinfo {author} {\bibfnamefont {M.}~\bibnamefont {Messer}},
  \bibinfo {author} {\bibfnamefont {R.}~\bibnamefont {Desbuquois}}, \bibinfo
  {author} {\bibfnamefont {M.}~\bibnamefont {Lebrat}}, \bibinfo {author}
  {\bibfnamefont {T.}~\bibnamefont {Uehlinger}}, \bibinfo {author}
  {\bibfnamefont {D.}~\bibnamefont {Greif}},\ and\ \bibinfo {author}
  {\bibfnamefont {T.}~\bibnamefont {Esslinger}},\ }\bibfield  {title} {\bibinfo
  {title} {Experimental realization of the topological haldane model with
  ultracold fermions},\ }\href@noop {} {\bibfield  {journal} {\bibinfo
  {journal} {Nature}\ }\textbf {\bibinfo {volume} {515}},\ \bibinfo {pages}
  {237} (\bibinfo {year} {2014})}\BibitemShut {NoStop}%
\bibitem [{\citenamefont {Brantut}\ \emph {et~al.}(2012)\citenamefont
  {Brantut}, \citenamefont {Meineke}, \citenamefont {Stadler}, \citenamefont
  {Krinner},\ and\ \citenamefont {Esslinger}}]{brantut2012conduction}%
  \BibitemOpen
  \bibfield  {author} {\bibinfo {author} {\bibfnamefont {J.-P.}\ \bibnamefont
  {Brantut}}, \bibinfo {author} {\bibfnamefont {J.}~\bibnamefont {Meineke}},
  \bibinfo {author} {\bibfnamefont {D.}~\bibnamefont {Stadler}}, \bibinfo
  {author} {\bibfnamefont {S.}~\bibnamefont {Krinner}},\ and\ \bibinfo {author}
  {\bibfnamefont {T.}~\bibnamefont {Esslinger}},\ }\bibfield  {title} {\bibinfo
  {title} {Conduction of ultracold fermions through a mesoscopic channel},\
  }\href@noop {} {\bibfield  {journal} {\bibinfo  {journal} {Science}\ }\textbf
  {\bibinfo {volume} {337}},\ \bibinfo {pages} {1069} (\bibinfo {year}
  {2012})}\BibitemShut {NoStop}%
\bibitem [{\citenamefont {Krinner}\ \emph {et~al.}(2015)\citenamefont
  {Krinner}, \citenamefont {Stadler}, \citenamefont {Husmann}, \citenamefont
  {Brantut},\ and\ \citenamefont {Esslinger}}]{krinner2015observation}%
  \BibitemOpen
  \bibfield  {author} {\bibinfo {author} {\bibfnamefont {S.}~\bibnamefont
  {Krinner}}, \bibinfo {author} {\bibfnamefont {D.}~\bibnamefont {Stadler}},
  \bibinfo {author} {\bibfnamefont {D.}~\bibnamefont {Husmann}}, \bibinfo
  {author} {\bibfnamefont {J.-P.}\ \bibnamefont {Brantut}},\ and\ \bibinfo
  {author} {\bibfnamefont {T.}~\bibnamefont {Esslinger}},\ }\bibfield  {title}
  {\bibinfo {title} {Observation of quantized conductance in neutral matter},\
  }\href@noop {} {\bibfield  {journal} {\bibinfo  {journal} {Nature}\ }\textbf
  {\bibinfo {volume} {517}},\ \bibinfo {pages} {64} (\bibinfo {year}
  {2015})}\BibitemShut {NoStop}%
\bibitem [{\citenamefont {Husmann}\ \emph {et~al.}(2015)\citenamefont
  {Husmann}, \citenamefont {Uchino}, \citenamefont {Krinner}, \citenamefont
  {Lebrat}, \citenamefont {Giamarchi}, \citenamefont {Esslinger},\ and\
  \citenamefont {Brantut}}]{husmann2015connecting}%
  \BibitemOpen
  \bibfield  {author} {\bibinfo {author} {\bibfnamefont {D.}~\bibnamefont
  {Husmann}}, \bibinfo {author} {\bibfnamefont {S.}~\bibnamefont {Uchino}},
  \bibinfo {author} {\bibfnamefont {S.}~\bibnamefont {Krinner}}, \bibinfo
  {author} {\bibfnamefont {M.}~\bibnamefont {Lebrat}}, \bibinfo {author}
  {\bibfnamefont {T.}~\bibnamefont {Giamarchi}}, \bibinfo {author}
  {\bibfnamefont {T.}~\bibnamefont {Esslinger}},\ and\ \bibinfo {author}
  {\bibfnamefont {J.-P.}\ \bibnamefont {Brantut}},\ }\bibfield  {title}
  {\bibinfo {title} {Connecting strongly correlated superfluids by a quantum
  point contact},\ }\href@noop {} {\bibfield  {journal} {\bibinfo  {journal}
  {Science}\ }\textbf {\bibinfo {volume} {350}},\ \bibinfo {pages} {1498}
  (\bibinfo {year} {2015})}\BibitemShut {NoStop}%
\bibitem [{\citenamefont {Lebrat}\ \emph {et~al.}(2018)\citenamefont {Lebrat},
  \citenamefont {Gri{\v{s}}ins}, \citenamefont {Husmann}, \citenamefont
  {H{\"a}usler}, \citenamefont {Corman}, \citenamefont {Giamarchi},
  \citenamefont {Brantut},\ and\ \citenamefont {Esslinger}}]{lebrat2018band}%
  \BibitemOpen
  \bibfield  {author} {\bibinfo {author} {\bibfnamefont {M.}~\bibnamefont
  {Lebrat}}, \bibinfo {author} {\bibfnamefont {P.}~\bibnamefont
  {Gri{\v{s}}ins}}, \bibinfo {author} {\bibfnamefont {D.}~\bibnamefont
  {Husmann}}, \bibinfo {author} {\bibfnamefont {S.}~\bibnamefont
  {H{\"a}usler}}, \bibinfo {author} {\bibfnamefont {L.}~\bibnamefont {Corman}},
  \bibinfo {author} {\bibfnamefont {T.}~\bibnamefont {Giamarchi}}, \bibinfo
  {author} {\bibfnamefont {J.-P.}\ \bibnamefont {Brantut}},\ and\ \bibinfo
  {author} {\bibfnamefont {T.}~\bibnamefont {Esslinger}},\ }\bibfield  {title}
  {\bibinfo {title} {Band and correlated insulators of cold fermions in a
  mesoscopic lattice},\ }\href@noop {} {\bibfield  {journal} {\bibinfo
  {journal} {Physical Review X}\ }\textbf {\bibinfo {volume} {8}},\ \bibinfo
  {pages} {011053} (\bibinfo {year} {2018})}\BibitemShut {NoStop}%
\bibitem [{\citenamefont {Gauthier}\ \emph {et~al.}(2019)\citenamefont
  {Gauthier}, \citenamefont {Szigeti}, \citenamefont {Reeves}, \citenamefont
  {Baker}, \citenamefont {Bell}, \citenamefont {Rubinsztein-Dunlop},
  \citenamefont {Davis},\ and\ \citenamefont
  {Neely}}]{gauthier2019quantitative}%
  \BibitemOpen
  \bibfield  {author} {\bibinfo {author} {\bibfnamefont {G.}~\bibnamefont
  {Gauthier}}, \bibinfo {author} {\bibfnamefont {S.~S.}\ \bibnamefont
  {Szigeti}}, \bibinfo {author} {\bibfnamefont {M.~T.}\ \bibnamefont {Reeves}},
  \bibinfo {author} {\bibfnamefont {M.}~\bibnamefont {Baker}}, \bibinfo
  {author} {\bibfnamefont {T.~A.}\ \bibnamefont {Bell}}, \bibinfo {author}
  {\bibfnamefont {H.}~\bibnamefont {Rubinsztein-Dunlop}}, \bibinfo {author}
  {\bibfnamefont {M.~J.}\ \bibnamefont {Davis}},\ and\ \bibinfo {author}
  {\bibfnamefont {T.~W.}\ \bibnamefont {Neely}},\ }\bibfield  {title} {\bibinfo
  {title} {Quantitative acoustic models for superfluid circuits},\ }\href@noop
  {} {\bibfield  {journal} {\bibinfo  {journal} {Phys. Rev. Lett.}\ }\textbf
  {\bibinfo {volume} {123}},\ \bibinfo {pages} {260402} (\bibinfo {year}
  {2019})}\BibitemShut {NoStop}%
\bibitem [{\citenamefont {Pandey}\ \emph {et~al.}(2019)\citenamefont {Pandey},
  \citenamefont {Mas}, \citenamefont {Drougakis}, \citenamefont {Thekkeppatt},
  \citenamefont {Bolpasi}, \citenamefont {Vasilakis}, \citenamefont {Poulios},\
  and\ \citenamefont {von Klitzing}}]{pandey2019hypersonic}%
  \BibitemOpen
  \bibfield  {author} {\bibinfo {author} {\bibfnamefont {S.}~\bibnamefont
  {Pandey}}, \bibinfo {author} {\bibfnamefont {H.}~\bibnamefont {Mas}},
  \bibinfo {author} {\bibfnamefont {G.}~\bibnamefont {Drougakis}}, \bibinfo
  {author} {\bibfnamefont {P.}~\bibnamefont {Thekkeppatt}}, \bibinfo {author}
  {\bibfnamefont {V.}~\bibnamefont {Bolpasi}}, \bibinfo {author} {\bibfnamefont
  {G.}~\bibnamefont {Vasilakis}}, \bibinfo {author} {\bibfnamefont
  {K.}~\bibnamefont {Poulios}},\ and\ \bibinfo {author} {\bibfnamefont
  {W.}~\bibnamefont {von Klitzing}},\ }\bibfield  {title} {\bibinfo {title}
  {Hypersonic bose--einstein condensates in accelerator rings},\ }\href@noop {}
  {\bibfield  {journal} {\bibinfo  {journal} {Nature}\ }\textbf {\bibinfo
  {volume} {570}},\ \bibinfo {pages} {205} (\bibinfo {year}
  {2019})}\BibitemShut {NoStop}%
\bibitem [{\citenamefont {Haug}\ \emph
  {et~al.}(2018{\natexlab{a}})\citenamefont {Haug}, \citenamefont {Amico},
  \citenamefont {Dumke},\ and\ \citenamefont {Kwek}}]{haug2018mesoscopic}%
  \BibitemOpen
  \bibfield  {author} {\bibinfo {author} {\bibfnamefont {T.}~\bibnamefont
  {Haug}}, \bibinfo {author} {\bibfnamefont {L.}~\bibnamefont {Amico}},
  \bibinfo {author} {\bibfnamefont {R.}~\bibnamefont {Dumke}},\ and\ \bibinfo
  {author} {\bibfnamefont {L.-C.}\ \bibnamefont {Kwek}},\ }\bibfield  {title}
  {\bibinfo {title} {Mesoscopic vortex--meissner currents in ring ladders},\
  }\href@noop {} {\bibfield  {journal} {\bibinfo  {journal} {Quantum Science
  and Technology}\ }\textbf {\bibinfo {volume} {3}},\ \bibinfo {pages} {035006}
  (\bibinfo {year} {2018}{\natexlab{a}})}\BibitemShut {NoStop}%
\bibitem [{\citenamefont {Haug}\ \emph
  {et~al.}(2019{\natexlab{a}})\citenamefont {Haug}, \citenamefont {Dumke},
  \citenamefont {Kwek},\ and\ \citenamefont {Amico}}]{haug2019topological}%
  \BibitemOpen
  \bibfield  {author} {\bibinfo {author} {\bibfnamefont {T.}~\bibnamefont
  {Haug}}, \bibinfo {author} {\bibfnamefont {R.}~\bibnamefont {Dumke}},
  \bibinfo {author} {\bibfnamefont {L.-C.}\ \bibnamefont {Kwek}},\ and\
  \bibinfo {author} {\bibfnamefont {L.}~\bibnamefont {Amico}},\ }\bibfield
  {title} {\bibinfo {title} {Topological pumping in aharonov--bohm rings},\
  }\href@noop {} {\bibfield  {journal} {\bibinfo  {journal} {Communications
  Physics}\ }\textbf {\bibinfo {volume} {2}},\ \bibinfo {pages} {1} (\bibinfo
  {year} {2019}{\natexlab{a}})}\BibitemShut {NoStop}%
\bibitem [{\citenamefont {Haug}\ \emph
  {et~al.}(2019{\natexlab{b}})\citenamefont {Haug}, \citenamefont {Heimonen},
  \citenamefont {Dumke}, \citenamefont {Kwek},\ and\ \citenamefont
  {Amico}}]{haug2019aharonov}%
  \BibitemOpen
  \bibfield  {author} {\bibinfo {author} {\bibfnamefont {T.}~\bibnamefont
  {Haug}}, \bibinfo {author} {\bibfnamefont {H.}~\bibnamefont {Heimonen}},
  \bibinfo {author} {\bibfnamefont {R.}~\bibnamefont {Dumke}}, \bibinfo
  {author} {\bibfnamefont {L.-C.}\ \bibnamefont {Kwek}},\ and\ \bibinfo
  {author} {\bibfnamefont {L.}~\bibnamefont {Amico}},\ }\bibfield  {title}
  {\bibinfo {title} {Aharonov-bohm effect in mesoscopic bose-einstein
  condensates},\ }\href@noop {} {\bibfield  {journal} {\bibinfo  {journal}
  {Physical Review A}\ }\textbf {\bibinfo {volume} {100}},\ \bibinfo {pages}
  {041601} (\bibinfo {year} {2019}{\natexlab{b}})}\BibitemShut {NoStop}%
\bibitem [{\citenamefont {Safaei}\ \emph {et~al.}(2019)\citenamefont {Safaei},
  \citenamefont {Kwek}, \citenamefont {Dumke},\ and\ \citenamefont
  {Amico}}]{safaei2019monitoring}%
  \BibitemOpen
  \bibfield  {author} {\bibinfo {author} {\bibfnamefont {S.}~\bibnamefont
  {Safaei}}, \bibinfo {author} {\bibfnamefont {L.-C.}\ \bibnamefont {Kwek}},
  \bibinfo {author} {\bibfnamefont {R.}~\bibnamefont {Dumke}},\ and\ \bibinfo
  {author} {\bibfnamefont {L.}~\bibnamefont {Amico}},\ }\bibfield  {title}
  {\bibinfo {title} {Monitoring currents in cold-atom circuis},\ }\href@noop {}
  {\bibfield  {journal} {\bibinfo  {journal} {Physical Review A}\ }\textbf
  {\bibinfo {volume} {100}},\ \bibinfo {pages} {013621} (\bibinfo {year}
  {2019})}\BibitemShut {NoStop}%
\bibitem [{\citenamefont {Haug}\ \emph
  {et~al.}(2019{\natexlab{c}})\citenamefont {Haug}, \citenamefont {Dumke},
  \citenamefont {Kwek},\ and\ \citenamefont {Amico}}]{haug2019andreev}%
  \BibitemOpen
  \bibfield  {author} {\bibinfo {author} {\bibfnamefont {T.}~\bibnamefont
  {Haug}}, \bibinfo {author} {\bibfnamefont {R.}~\bibnamefont {Dumke}},
  \bibinfo {author} {\bibfnamefont {L.-C.}\ \bibnamefont {Kwek}},\ and\
  \bibinfo {author} {\bibfnamefont {L.}~\bibnamefont {Amico}},\ }\bibfield
  {title} {\bibinfo {title} {Andreev-reflection and aharonov--bohm dynamics in
  atomtronic circuits},\ }\href@noop {} {\bibfield  {journal} {\bibinfo
  {journal} {Quantum Science and Technology}\ }\textbf {\bibinfo {volume}
  {4}},\ \bibinfo {pages} {045001} (\bibinfo {year}
  {2019}{\natexlab{c}})}\BibitemShut {NoStop}%
\bibitem [{\citenamefont {Haug}(2021)}]{haug2021quantum}%
  \BibitemOpen
  \bibfield  {author} {\bibinfo {author} {\bibfnamefont {T.~F.}\ \bibnamefont
  {Haug}},\ }\emph {\bibinfo {title} {Quantum transport with cold atoms}},\
  \href@noop {} {Ph.D. thesis},\ \bibinfo  {school} {National University of
  Singapore (Singapore)} (\bibinfo {year} {2021})\BibitemShut {NoStop}%
\bibitem [{\citenamefont {Ramanathan}\ \emph {et~al.}(2011)\citenamefont
  {Ramanathan}, \citenamefont {Wright}, \citenamefont {Muniz}, \citenamefont
  {Zelan}, \citenamefont {Hill~III}, \citenamefont {Lobb}, \citenamefont
  {Helmerson}, \citenamefont {Phillips},\ and\ \citenamefont
  {Campbell}}]{ramanathan2011superflow}%
  \BibitemOpen
  \bibfield  {author} {\bibinfo {author} {\bibfnamefont {A.}~\bibnamefont
  {Ramanathan}}, \bibinfo {author} {\bibfnamefont {K.}~\bibnamefont {Wright}},
  \bibinfo {author} {\bibfnamefont {S.~R.}\ \bibnamefont {Muniz}}, \bibinfo
  {author} {\bibfnamefont {M.}~\bibnamefont {Zelan}}, \bibinfo {author}
  {\bibfnamefont {W.}~\bibnamefont {Hill~III}}, \bibinfo {author}
  {\bibfnamefont {C.}~\bibnamefont {Lobb}}, \bibinfo {author} {\bibfnamefont
  {K.}~\bibnamefont {Helmerson}}, \bibinfo {author} {\bibfnamefont
  {W.}~\bibnamefont {Phillips}},\ and\ \bibinfo {author} {\bibfnamefont
  {G.}~\bibnamefont {Campbell}},\ }\bibfield  {title} {\bibinfo {title}
  {Superflow in a toroidal bose-einstein condensate: an atom circuit with a
  tunable weak link},\ }\href@noop {} {\bibfield  {journal} {\bibinfo
  {journal} {Physical review letters}\ }\textbf {\bibinfo {volume} {106}},\
  \bibinfo {pages} {130401} (\bibinfo {year} {2011})}\BibitemShut {NoStop}%
\bibitem [{\citenamefont {Ryu}\ \emph {et~al.}(2013)\citenamefont {Ryu},
  \citenamefont {Blackburn}, \citenamefont {Blinova},\ and\ \citenamefont
  {Boshier}}]{ryu2013experimental}%
  \BibitemOpen
  \bibfield  {author} {\bibinfo {author} {\bibfnamefont {C.}~\bibnamefont
  {Ryu}}, \bibinfo {author} {\bibfnamefont {P.}~\bibnamefont {Blackburn}},
  \bibinfo {author} {\bibfnamefont {A.}~\bibnamefont {Blinova}},\ and\ \bibinfo
  {author} {\bibfnamefont {M.}~\bibnamefont {Boshier}},\ }\bibfield  {title}
  {\bibinfo {title} {Experimental realization of josephson junctions for an
  atom squid},\ }\href@noop {} {\bibfield  {journal} {\bibinfo  {journal}
  {Physical review letters}\ }\textbf {\bibinfo {volume} {111}},\ \bibinfo
  {pages} {205301} (\bibinfo {year} {2013})}\BibitemShut {NoStop}%
\bibitem [{\citenamefont {Eller}\ \emph {et~al.}(2020)\citenamefont {Eller},
  \citenamefont {Oladehin}, \citenamefont {Fogarty}, \citenamefont {Heller},
  \citenamefont {Clark},\ and\ \citenamefont {Edwards}}]{eller2020producing}%
  \BibitemOpen
  \bibfield  {author} {\bibinfo {author} {\bibfnamefont {B.}~\bibnamefont
  {Eller}}, \bibinfo {author} {\bibfnamefont {O.}~\bibnamefont {Oladehin}},
  \bibinfo {author} {\bibfnamefont {D.}~\bibnamefont {Fogarty}}, \bibinfo
  {author} {\bibfnamefont {C.}~\bibnamefont {Heller}}, \bibinfo {author}
  {\bibfnamefont {C.~W.}\ \bibnamefont {Clark}},\ and\ \bibinfo {author}
  {\bibfnamefont {M.}~\bibnamefont {Edwards}},\ }\bibfield  {title} {\bibinfo
  {title} {Producing flow in racetrack atom circuits by stirring},\ }\href@noop
  {} {\bibfield  {journal} {\bibinfo  {journal} {Physical Review A}\ }\textbf
  {\bibinfo {volume} {102}},\ \bibinfo {pages} {063324} (\bibinfo {year}
  {2020})}\BibitemShut {NoStop}%
\bibitem [{\citenamefont {Aghamalyan}\ \emph {et~al.}(2015)\citenamefont
  {Aghamalyan}, \citenamefont {Cominotti}, \citenamefont {Rizzi}, \citenamefont
  {Rossini}, \citenamefont {Hekking}, \citenamefont {Minguzzi}, \citenamefont
  {Kwek},\ and\ \citenamefont {Amico}}]{aghamalyan2015coherent}%
  \BibitemOpen
  \bibfield  {author} {\bibinfo {author} {\bibfnamefont {D.}~\bibnamefont
  {Aghamalyan}}, \bibinfo {author} {\bibfnamefont {M.}~\bibnamefont
  {Cominotti}}, \bibinfo {author} {\bibfnamefont {M.}~\bibnamefont {Rizzi}},
  \bibinfo {author} {\bibfnamefont {D.}~\bibnamefont {Rossini}}, \bibinfo
  {author} {\bibfnamefont {F.}~\bibnamefont {Hekking}}, \bibinfo {author}
  {\bibfnamefont {A.}~\bibnamefont {Minguzzi}}, \bibinfo {author}
  {\bibfnamefont {L.-C.}\ \bibnamefont {Kwek}},\ and\ \bibinfo {author}
  {\bibfnamefont {L.}~\bibnamefont {Amico}},\ }\bibfield  {title} {\bibinfo
  {title} {Coherent superposition of current flows in an atomtronic quantum
  interference device},\ }\href@noop {} {\bibfield  {journal} {\bibinfo
  {journal} {New journal of Physics}\ }\textbf {\bibinfo {volume} {17}},\
  \bibinfo {pages} {045023} (\bibinfo {year} {2015})}\BibitemShut {NoStop}%
\bibitem [{\citenamefont {Haug}\ \emph
  {et~al.}(2018{\natexlab{b}})\citenamefont {Haug}, \citenamefont {Tan},
  \citenamefont {Theng}, \citenamefont {Dumke}, \citenamefont {Kwek},\ and\
  \citenamefont {Amico}}]{haug2018readout}%
  \BibitemOpen
  \bibfield  {author} {\bibinfo {author} {\bibfnamefont {T.}~\bibnamefont
  {Haug}}, \bibinfo {author} {\bibfnamefont {J.}~\bibnamefont {Tan}}, \bibinfo
  {author} {\bibfnamefont {M.}~\bibnamefont {Theng}}, \bibinfo {author}
  {\bibfnamefont {R.}~\bibnamefont {Dumke}}, \bibinfo {author} {\bibfnamefont
  {L.-C.}\ \bibnamefont {Kwek}},\ and\ \bibinfo {author} {\bibfnamefont
  {L.}~\bibnamefont {Amico}},\ }\bibfield  {title} {\bibinfo {title} {Readout
  of the atomtronic quantum interference device},\ }\href@noop {} {\bibfield
  {journal} {\bibinfo  {journal} {Physical Review A}\ }\textbf {\bibinfo
  {volume} {97}},\ \bibinfo {pages} {013633} (\bibinfo {year}
  {2018}{\natexlab{b}})}\BibitemShut {NoStop}%
\bibitem [{\citenamefont {Aharonov}\ and\ \citenamefont
  {Bohm}(1959)}]{aharonov1959significance}%
  \BibitemOpen
  \bibfield  {author} {\bibinfo {author} {\bibfnamefont {Y.}~\bibnamefont
  {Aharonov}}\ and\ \bibinfo {author} {\bibfnamefont {D.}~\bibnamefont
  {Bohm}},\ }\bibfield  {title} {\bibinfo {title} {Significance of
  electromagnetic potentials in the quantum theory},\ }\href@noop {} {\bibfield
   {journal} {\bibinfo  {journal} {Physical Review}\ }\textbf {\bibinfo
  {volume} {115}},\ \bibinfo {pages} {485} (\bibinfo {year}
  {1959})}\BibitemShut {NoStop}%
\bibitem [{\citenamefont {Gefen}\ \emph {et~al.}(1984)\citenamefont {Gefen},
  \citenamefont {Imry},\ and\ \citenamefont {Azbel}}]{gefen1984quantum}%
  \BibitemOpen
  \bibfield  {author} {\bibinfo {author} {\bibfnamefont {Y.}~\bibnamefont
  {Gefen}}, \bibinfo {author} {\bibfnamefont {Y.}~\bibnamefont {Imry}},\ and\
  \bibinfo {author} {\bibfnamefont {M.~Y.}\ \bibnamefont {Azbel}},\ }\bibfield
  {title} {\bibinfo {title} {Quantum oscillations and the aharonov-bohm effect
  for parallel resistors},\ }\href@noop {} {\bibfield  {journal} {\bibinfo
  {journal} {Physical review letters}\ }\textbf {\bibinfo {volume} {52}},\
  \bibinfo {pages} {129} (\bibinfo {year} {1984})}\BibitemShut {NoStop}%
\bibitem [{\citenamefont {B{\"u}ttiker}\ \emph {et~al.}(1984)\citenamefont
  {B{\"u}ttiker}, \citenamefont {Imry},\ and\ \citenamefont
  {Azbel}}]{buttiker1984quantum}%
  \BibitemOpen
  \bibfield  {author} {\bibinfo {author} {\bibfnamefont {M.}~\bibnamefont
  {B{\"u}ttiker}}, \bibinfo {author} {\bibfnamefont {Y.}~\bibnamefont {Imry}},\
  and\ \bibinfo {author} {\bibfnamefont {M.~Y.}\ \bibnamefont {Azbel}},\
  }\bibfield  {title} {\bibinfo {title} {Quantum oscillations in
  one-dimensional normal-metal rings},\ }\href@noop {} {\bibfield  {journal}
  {\bibinfo  {journal} {Physical Review A}\ }\textbf {\bibinfo {volume} {30}},\
  \bibinfo {pages} {1982} (\bibinfo {year} {1984})}\BibitemShut {NoStop}%
\bibitem [{\citenamefont {Webb}\ \emph {et~al.}(1985)\citenamefont {Webb},
  \citenamefont {Washburn}, \citenamefont {Umbach},\ and\ \citenamefont
  {Laibowitz}}]{webb1985observation}%
  \BibitemOpen
  \bibfield  {author} {\bibinfo {author} {\bibfnamefont {R.~A.}\ \bibnamefont
  {Webb}}, \bibinfo {author} {\bibfnamefont {S.}~\bibnamefont {Washburn}},
  \bibinfo {author} {\bibfnamefont {C.}~\bibnamefont {Umbach}},\ and\ \bibinfo
  {author} {\bibfnamefont {R.}~\bibnamefont {Laibowitz}},\ }\bibfield  {title}
  {\bibinfo {title} {Observation of h e aharonov-bohm oscillations in
  normal-metal rings},\ }\href@noop {} {\bibfield  {journal} {\bibinfo
  {journal} {Physical Review Letters}\ }\textbf {\bibinfo {volume} {54}},\
  \bibinfo {pages} {2696} (\bibinfo {year} {1985})}\BibitemShut {NoStop}%
\bibitem [{\citenamefont {Imry}(2002)}]{imry2002introduction}%
  \BibitemOpen
  \bibfield  {author} {\bibinfo {author} {\bibfnamefont {Y.}~\bibnamefont
  {Imry}},\ }\href@noop {} {\emph {\bibinfo {title} {Introduction to mesoscopic
  physics}}},\ \bibinfo {number} {2}\ (\bibinfo  {publisher} {Oxford University
  Press on Demand},\ \bibinfo {year} {2002})\BibitemShut {NoStop}%
\bibitem [{\citenamefont {Singleton}\ and\ \citenamefont
  {Vagenas}(2013)}]{singleton2013covariant}%
  \BibitemOpen
  \bibfield  {author} {\bibinfo {author} {\bibfnamefont {D.}~\bibnamefont
  {Singleton}}\ and\ \bibinfo {author} {\bibfnamefont {E.~C.}\ \bibnamefont
  {Vagenas}},\ }\bibfield  {title} {\bibinfo {title} {The covariant,
  time-dependent aharonov--bohm effect},\ }\href@noop {} {\bibfield  {journal}
  {\bibinfo  {journal} {Physics Letters B}\ }\textbf {\bibinfo {volume}
  {723}},\ \bibinfo {pages} {241} (\bibinfo {year} {2013})}\BibitemShut
  {NoStop}%
\bibitem [{\citenamefont {Macdougall}\ \emph {et~al.}(2015)\citenamefont
  {Macdougall}, \citenamefont {Singleton},\ and\ \citenamefont
  {Vagenas}}]{macdougall2015revisiting}%
  \BibitemOpen
  \bibfield  {author} {\bibinfo {author} {\bibfnamefont {J.}~\bibnamefont
  {Macdougall}}, \bibinfo {author} {\bibfnamefont {D.}~\bibnamefont
  {Singleton}},\ and\ \bibinfo {author} {\bibfnamefont {E.~C.}\ \bibnamefont
  {Vagenas}},\ }\bibfield  {title} {\bibinfo {title} {Revisiting the marton,
  simpson, and suddeth experimental confirmation of the aharonov--bohm
  effect},\ }\href@noop {} {\bibfield  {journal} {\bibinfo  {journal} {Physics
  Letters A}\ }\textbf {\bibinfo {volume} {379}},\ \bibinfo {pages} {1689}
  (\bibinfo {year} {2015})}\BibitemShut {NoStop}%
\bibitem [{\citenamefont {Jing}\ \emph {et~al.}(2017)\citenamefont {Jing},
  \citenamefont {Zhang}, \citenamefont {Wang}, \citenamefont {Long},\ and\
  \citenamefont {Dong}}]{jing2017time}%
  \BibitemOpen
  \bibfield  {author} {\bibinfo {author} {\bibfnamefont {J.}~\bibnamefont
  {Jing}}, \bibinfo {author} {\bibfnamefont {Y.-F.}\ \bibnamefont {Zhang}},
  \bibinfo {author} {\bibfnamefont {K.}~\bibnamefont {Wang}}, \bibinfo {author}
  {\bibfnamefont {Z.-W.}\ \bibnamefont {Long}},\ and\ \bibinfo {author}
  {\bibfnamefont {S.-H.}\ \bibnamefont {Dong}},\ }\bibfield  {title} {\bibinfo
  {title} {On the time-dependent aharonov--bohm effect},\ }\href@noop {}
  {\bibfield  {journal} {\bibinfo  {journal} {Physics Letters B}\ }\textbf
  {\bibinfo {volume} {774}},\ \bibinfo {pages} {87} (\bibinfo {year}
  {2017})}\BibitemShut {NoStop}%
\bibitem [{\citenamefont {Choudhury}\ and\ \citenamefont
  {Mahajan}(2019)}]{choudhury2019direct}%
  \BibitemOpen
  \bibfield  {author} {\bibinfo {author} {\bibfnamefont {S.~R.}\ \bibnamefont
  {Choudhury}}\ and\ \bibinfo {author} {\bibfnamefont {S.}~\bibnamefont
  {Mahajan}},\ }\bibfield  {title} {\bibinfo {title} {Direct calculation of
  time varying aharonov-bohm effect},\ }\href@noop {} {\bibfield  {journal}
  {\bibinfo  {journal} {Physics Letters A}\ }\textbf {\bibinfo {volume}
  {383}},\ \bibinfo {pages} {2467} (\bibinfo {year} {2019})}\BibitemShut
  {NoStop}%
\bibitem [{\citenamefont {Jaksch}\ and\ \citenamefont
  {Zoller}(2003)}]{jaksch2003creation}%
  \BibitemOpen
  \bibfield  {author} {\bibinfo {author} {\bibfnamefont {D.}~\bibnamefont
  {Jaksch}}\ and\ \bibinfo {author} {\bibfnamefont {P.}~\bibnamefont
  {Zoller}},\ }\bibfield  {title} {\bibinfo {title} {Creation of effective
  magnetic fields in optical lattices: the hofstadter butterfly for cold
  neutral atoms},\ }\href@noop {} {\bibfield  {journal} {\bibinfo  {journal}
  {New Journal of Physics}\ }\textbf {\bibinfo {volume} {5}},\ \bibinfo {pages}
  {56} (\bibinfo {year} {2003})}\BibitemShut {NoStop}%
\bibitem [{\citenamefont {Lin}\ \emph {et~al.}(2009)\citenamefont {Lin},
  \citenamefont {Compton}, \citenamefont {Jim{\'e}nez-Garc{\'\i}a},
  \citenamefont {Porto},\ and\ \citenamefont {Spielman}}]{lin2009synthetic}%
  \BibitemOpen
  \bibfield  {author} {\bibinfo {author} {\bibfnamefont {Y.-J.}\ \bibnamefont
  {Lin}}, \bibinfo {author} {\bibfnamefont {R.~L.}\ \bibnamefont {Compton}},
  \bibinfo {author} {\bibfnamefont {K.}~\bibnamefont
  {Jim{\'e}nez-Garc{\'\i}a}}, \bibinfo {author} {\bibfnamefont {J.~V.}\
  \bibnamefont {Porto}},\ and\ \bibinfo {author} {\bibfnamefont {I.~B.}\
  \bibnamefont {Spielman}},\ }\bibfield  {title} {\bibinfo {title} {Synthetic
  magnetic fields for ultracold neutral atoms},\ }\href
  {https://pubmed.ncbi.nlm.nih.gov/19956256/} {\bibfield  {journal} {\bibinfo
  {journal} {Nature}\ }\textbf {\bibinfo {volume} {462}},\ \bibinfo {pages}
  {628} (\bibinfo {year} {2009})}\BibitemShut {NoStop}%
\bibitem [{\citenamefont {Dalibard}\ \emph {et~al.}(2011)\citenamefont
  {Dalibard}, \citenamefont {Gerbier}, \citenamefont {Juzeli{\=u}nas},\ and\
  \citenamefont {{\"O}hberg}}]{dalibard2011colloquium}%
  \BibitemOpen
  \bibfield  {author} {\bibinfo {author} {\bibfnamefont {J.}~\bibnamefont
  {Dalibard}}, \bibinfo {author} {\bibfnamefont {F.}~\bibnamefont {Gerbier}},
  \bibinfo {author} {\bibfnamefont {G.}~\bibnamefont {Juzeli{\=u}nas}},\ and\
  \bibinfo {author} {\bibfnamefont {P.}~\bibnamefont {{\"O}hberg}},\ }\bibfield
   {title} {\bibinfo {title} {Colloquium: Artificial gauge potentials for
  neutral atoms},\ }\href@noop {} {\bibfield  {journal} {\bibinfo  {journal}
  {Reviews of Modern Physics}\ }\textbf {\bibinfo {volume} {83}},\ \bibinfo
  {pages} {1523} (\bibinfo {year} {2011})}\BibitemShut {NoStop}%
\bibitem [{\citenamefont {Haug}\ \emph {et~al.}(2021)\citenamefont {Haug},
  \citenamefont {Dumke}, \citenamefont {Kwek}, \citenamefont {Miniatura},\ and\
  \citenamefont {Amico}}]{haug2021machine}%
  \BibitemOpen
  \bibfield  {author} {\bibinfo {author} {\bibfnamefont {T.}~\bibnamefont
  {Haug}}, \bibinfo {author} {\bibfnamefont {R.}~\bibnamefont {Dumke}},
  \bibinfo {author} {\bibfnamefont {L.-C.}\ \bibnamefont {Kwek}}, \bibinfo
  {author} {\bibfnamefont {C.}~\bibnamefont {Miniatura}},\ and\ \bibinfo
  {author} {\bibfnamefont {L.}~\bibnamefont {Amico}},\ }\bibfield  {title}
  {\bibinfo {title} {Machine-learning engineering of quantum currents},\
  }\href@noop {} {\bibfield  {journal} {\bibinfo  {journal} {Physical Review
  Research}\ }\textbf {\bibinfo {volume} {3}},\ \bibinfo {pages} {013034}
  (\bibinfo {year} {2021})}\BibitemShut {NoStop}%
\bibitem [{\citenamefont {Del~Pace}\ \emph {et~al.}(2022)\citenamefont
  {Del~Pace}, \citenamefont {Xhani}, \citenamefont {Falconi}, \citenamefont
  {Fedrizzi}, \citenamefont {Grani}, \citenamefont {Rajkov}, \citenamefont
  {Inguscio}, \citenamefont {Scazza}, \citenamefont {Kwon},\ and\ \citenamefont
  {Roati}}]{del2022imprinting}%
  \BibitemOpen
  \bibfield  {author} {\bibinfo {author} {\bibfnamefont {G.}~\bibnamefont
  {Del~Pace}}, \bibinfo {author} {\bibfnamefont {K.}~\bibnamefont {Xhani}},
  \bibinfo {author} {\bibfnamefont {A.~M.}\ \bibnamefont {Falconi}}, \bibinfo
  {author} {\bibfnamefont {M.}~\bibnamefont {Fedrizzi}}, \bibinfo {author}
  {\bibfnamefont {N.}~\bibnamefont {Grani}}, \bibinfo {author} {\bibfnamefont
  {D.~H.}\ \bibnamefont {Rajkov}}, \bibinfo {author} {\bibfnamefont
  {M.}~\bibnamefont {Inguscio}}, \bibinfo {author} {\bibfnamefont
  {F.}~\bibnamefont {Scazza}}, \bibinfo {author} {\bibfnamefont
  {W.}~\bibnamefont {Kwon}},\ and\ \bibinfo {author} {\bibfnamefont
  {G.}~\bibnamefont {Roati}},\ }\bibfield  {title} {\bibinfo {title}
  {Imprinting persistent currents in tunable fermionic rings},\ }\href@noop {}
  {\bibfield  {journal} {\bibinfo  {journal} {Physical Review X}\ }\textbf
  {\bibinfo {volume} {12}},\ \bibinfo {pages} {041037} (\bibinfo {year}
  {2022})}\BibitemShut {NoStop}%
\bibitem [{\citenamefont {Bao}\ \emph {et~al.}(2003)\citenamefont {Bao},
  \citenamefont {Jaksch},\ and\ \citenamefont {Markowich}}]{bao2003numerical}%
  \BibitemOpen
  \bibfield  {author} {\bibinfo {author} {\bibfnamefont {W.}~\bibnamefont
  {Bao}}, \bibinfo {author} {\bibfnamefont {D.}~\bibnamefont {Jaksch}},\ and\
  \bibinfo {author} {\bibfnamefont {P.~A.}\ \bibnamefont {Markowich}},\
  }\bibfield  {title} {\bibinfo {title} {Numerical solution of the
  gross--pitaevskii equation for bose--einstein condensation},\ }\href@noop {}
  {\bibfield  {journal} {\bibinfo  {journal} {Journal of Computational
  Physics}\ }\textbf {\bibinfo {volume} {187}},\ \bibinfo {pages} {318}
  (\bibinfo {year} {2003})}\BibitemShut {NoStop}%
\bibitem [{\citenamefont {Engels}\ \emph {et~al.}(2003)\citenamefont {Engels},
  \citenamefont {Coddington}, \citenamefont {Haljan}, \citenamefont
  {Schweikhard},\ and\ \citenamefont {Cornell}}]{engels2003observation}%
  \BibitemOpen
  \bibfield  {author} {\bibinfo {author} {\bibfnamefont {P.}~\bibnamefont
  {Engels}}, \bibinfo {author} {\bibfnamefont {I.}~\bibnamefont {Coddington}},
  \bibinfo {author} {\bibfnamefont {P.}~\bibnamefont {Haljan}}, \bibinfo
  {author} {\bibfnamefont {V.}~\bibnamefont {Schweikhard}},\ and\ \bibinfo
  {author} {\bibfnamefont {E.~A.}\ \bibnamefont {Cornell}},\ }\bibfield
  {title} {\bibinfo {title} {Observation of long-lived vortex aggregates in
  rapidly rotating bose-einstein condensates},\ }\href@noop {} {\bibfield
  {journal} {\bibinfo  {journal} {Physical review letters}\ }\textbf {\bibinfo
  {volume} {90}},\ \bibinfo {pages} {170405} (\bibinfo {year}
  {2003})}\BibitemShut {NoStop}%
\bibitem [{\citenamefont {Weitenberg}\ and\ \citenamefont
  {Simonet}(2021)}]{weitenberg2021tailoring}%
  \BibitemOpen
  \bibfield  {author} {\bibinfo {author} {\bibfnamefont {C.}~\bibnamefont
  {Weitenberg}}\ and\ \bibinfo {author} {\bibfnamefont {J.}~\bibnamefont
  {Simonet}},\ }\bibfield  {title} {\bibinfo {title} {Tailoring quantum gases
  by floquet engineering},\ }\href@noop {} {\bibfield  {journal} {\bibinfo
  {journal} {Nature Physics}\ }\textbf {\bibinfo {volume} {17}},\ \bibinfo
  {pages} {1342} (\bibinfo {year} {2021})}\BibitemShut {NoStop}%
\bibitem [{\citenamefont {Wintersperger}\ \emph {et~al.}(2020)\citenamefont
  {Wintersperger}, \citenamefont {Braun}, \citenamefont {{\"U}nal},
  \citenamefont {Eckardt}, \citenamefont {Liberto}, \citenamefont {Goldman},
  \citenamefont {Bloch},\ and\ \citenamefont
  {Aidelsburger}}]{wintersperger2020realization}%
  \BibitemOpen
  \bibfield  {author} {\bibinfo {author} {\bibfnamefont {K.}~\bibnamefont
  {Wintersperger}}, \bibinfo {author} {\bibfnamefont {C.}~\bibnamefont
  {Braun}}, \bibinfo {author} {\bibfnamefont {F.~N.}\ \bibnamefont {{\"U}nal}},
  \bibinfo {author} {\bibfnamefont {A.}~\bibnamefont {Eckardt}}, \bibinfo
  {author} {\bibfnamefont {M.~D.}\ \bibnamefont {Liberto}}, \bibinfo {author}
  {\bibfnamefont {N.}~\bibnamefont {Goldman}}, \bibinfo {author} {\bibfnamefont
  {I.}~\bibnamefont {Bloch}},\ and\ \bibinfo {author} {\bibfnamefont
  {M.}~\bibnamefont {Aidelsburger}},\ }\bibfield  {title} {\bibinfo {title}
  {Realization of an anomalous floquet topological system with ultracold
  atoms},\ }\href@noop {} {\bibfield  {journal} {\bibinfo  {journal} {Nature
  Physics}\ }\textbf {\bibinfo {volume} {16}},\ \bibinfo {pages} {1058}
  (\bibinfo {year} {2020})}\BibitemShut {NoStop}%
\bibitem [{\citenamefont {Fishman}\ \emph {et~al.}(2022)\citenamefont
  {Fishman}, \citenamefont {White},\ and\ \citenamefont
  {Stoudenmire}}]{itensor}%
  \BibitemOpen
  \bibfield  {author} {\bibinfo {author} {\bibfnamefont {M.}~\bibnamefont
  {Fishman}}, \bibinfo {author} {\bibfnamefont {S.~R.}\ \bibnamefont {White}},\
  and\ \bibinfo {author} {\bibfnamefont {E.~M.}\ \bibnamefont {Stoudenmire}},\
  }\bibfield  {title} {\bibinfo {title} {{The ITensor Software Library for
  Tensor Network Calculations}},\ }\href
  {https://doi.org/10.21468/SciPostPhysCodeb.4} {\bibfield  {journal} {\bibinfo
   {journal} {SciPost Phys. Codebases}\ ,\ \bibinfo {pages} {4}} (\bibinfo
  {year} {2022})}\BibitemShut {NoStop}%
\bibitem [{\citenamefont {Wittek}\ and\ \citenamefont
  {Calderaro}(2015)}]{wittek2015extended}%
  \BibitemOpen
  \bibfield  {author} {\bibinfo {author} {\bibfnamefont {P.}~\bibnamefont
  {Wittek}}\ and\ \bibinfo {author} {\bibfnamefont {L.}~\bibnamefont
  {Calderaro}},\ }\bibfield  {title} {\bibinfo {title} {Extended computational
  kernels in a massively parallel implementation of the trotter--suzuki
  approximation},\ }\href@noop {} {\bibfield  {journal} {\bibinfo  {journal}
  {Computer Physics Communications}\ }\textbf {\bibinfo {volume} {197}},\
  \bibinfo {pages} {339} (\bibinfo {year} {2015})}\BibitemShut {NoStop}%
\bibitem [{\citenamefont {Breuer}\ \emph {et~al.}(2002)\citenamefont {Breuer},
  \citenamefont {Petruccione} \emph {et~al.}}]{breuer2002theory}%
  \BibitemOpen
  \bibfield  {author} {\bibinfo {author} {\bibfnamefont {H.-P.}\ \bibnamefont
  {Breuer}}, \bibinfo {author} {\bibfnamefont {F.}~\bibnamefont {Petruccione}},
  \emph {et~al.},\ }\href@noop {} {\emph {\bibinfo {title} {The theory of open
  quantum systems}}}\ (\bibinfo  {publisher} {Oxford University Press on
  Demand},\ \bibinfo {year} {2002})\BibitemShut {NoStop}%
\bibitem [{\citenamefont {Guo}\ and\ \citenamefont
  {Poletti}(2017)}]{guo2017dissipatively}%
  \BibitemOpen
  \bibfield  {author} {\bibinfo {author} {\bibfnamefont {C.}~\bibnamefont
  {Guo}}\ and\ \bibinfo {author} {\bibfnamefont {D.}~\bibnamefont {Poletti}},\
  }\bibfield  {title} {\bibinfo {title} {Dissipatively driven hardcore bosons
  steered by a gauge field},\ }\href@noop {} {\bibfield  {journal} {\bibinfo
  {journal} {Physical Review B}\ }\textbf {\bibinfo {volume} {96}},\ \bibinfo
  {pages} {165409} (\bibinfo {year} {2017})}\BibitemShut {NoStop}%
\bibitem [{\citenamefont {Daley}\ \emph {et~al.}(2008)\citenamefont {Daley},
  \citenamefont {Zoller},\ and\ \citenamefont {Trauzettel}}]{daley2008andreev}%
  \BibitemOpen
  \bibfield  {author} {\bibinfo {author} {\bibfnamefont {A.}~\bibnamefont
  {Daley}}, \bibinfo {author} {\bibfnamefont {P.}~\bibnamefont {Zoller}},\ and\
  \bibinfo {author} {\bibfnamefont {B.}~\bibnamefont {Trauzettel}},\ }\bibfield
   {title} {\bibinfo {title} {Andreev-like reflections with cold atoms},\
  }\href@noop {} {\bibfield  {journal} {\bibinfo  {journal} {Phys. Rev. Lett.}\
  }\textbf {\bibinfo {volume} {100}},\ \bibinfo {pages} {110404} (\bibinfo
  {year} {2008})}\BibitemShut {NoStop}%
\bibitem [{\citenamefont {Sherson}\ \emph {et~al.}(2010)\citenamefont
  {Sherson}, \citenamefont {Weitenberg}, \citenamefont {Endres}, \citenamefont
  {Cheneau}, \citenamefont {Bloch},\ and\ \citenamefont
  {Kuhr}}]{sherson2010single}%
  \BibitemOpen
  \bibfield  {author} {\bibinfo {author} {\bibfnamefont {J.~F.}\ \bibnamefont
  {Sherson}}, \bibinfo {author} {\bibfnamefont {C.}~\bibnamefont {Weitenberg}},
  \bibinfo {author} {\bibfnamefont {M.}~\bibnamefont {Endres}}, \bibinfo
  {author} {\bibfnamefont {M.}~\bibnamefont {Cheneau}}, \bibinfo {author}
  {\bibfnamefont {I.}~\bibnamefont {Bloch}},\ and\ \bibinfo {author}
  {\bibfnamefont {S.}~\bibnamefont {Kuhr}},\ }\bibfield  {title} {\bibinfo
  {title} {Single-atom-resolved fluorescence imaging of an atomic mott
  insulator},\ }\href@noop {} {\bibfield  {journal} {\bibinfo  {journal}
  {Nature}\ }\textbf {\bibinfo {volume} {467}},\ \bibinfo {pages} {68}
  (\bibinfo {year} {2010})}\BibitemShut {NoStop}%
\bibitem [{\citenamefont {Tokuno}\ \emph {et~al.}(2008)\citenamefont {Tokuno},
  \citenamefont {Oshikawa},\ and\ \citenamefont {Demler}}]{tokuno2008dynamics}%
  \BibitemOpen
  \bibfield  {author} {\bibinfo {author} {\bibfnamefont {A.}~\bibnamefont
  {Tokuno}}, \bibinfo {author} {\bibfnamefont {M.}~\bibnamefont {Oshikawa}},\
  and\ \bibinfo {author} {\bibfnamefont {E.}~\bibnamefont {Demler}},\
  }\bibfield  {title} {\bibinfo {title} {Dynamics of one-dimensional bose
  liquids: Andreev-like reflection at y junctions and the absence of the
  aharonov-bohm effect},\ }\href@noop {} {\bibfield  {journal} {\bibinfo
  {journal} {Physical review letters}\ }\textbf {\bibinfo {volume} {100}},\
  \bibinfo {pages} {140402} (\bibinfo {year} {2008})}\BibitemShut {NoStop}%
\bibitem [{\citenamefont {Pon}(1961)}]{pon1961hybrid}%
  \BibitemOpen
  \bibfield  {author} {\bibinfo {author} {\bibfnamefont {C.~Y.}\ \bibnamefont
  {Pon}},\ }\bibfield  {title} {\bibinfo {title} {Hybrid-ring directional
  coupler for arbitrary power divisions},\ }\href@noop {} {\bibfield  {journal}
  {\bibinfo  {journal} {IRE Transactions on Microwave Theory and Techniques}\
  }\textbf {\bibinfo {volume} {9}},\ \bibinfo {pages} {529} (\bibinfo {year}
  {1961})}\BibitemShut {NoStop}%
\bibitem [{\citenamefont {Yariv}(2002)}]{yariv2002critical}%
  \BibitemOpen
  \bibfield  {author} {\bibinfo {author} {\bibfnamefont {A.}~\bibnamefont
  {Yariv}},\ }\bibfield  {title} {\bibinfo {title} {Critical coupling and its
  control in optical waveguide-ring resonator systems},\ }\href@noop {}
  {\bibfield  {journal} {\bibinfo  {journal} {IEEE Photonics Technology
  Letters}\ }\textbf {\bibinfo {volume} {14}},\ \bibinfo {pages} {483}
  (\bibinfo {year} {2002})}\BibitemShut {NoStop}%
\bibitem [{\citenamefont {Dong}\ \emph {et~al.}(2010)\citenamefont {Dong},
  \citenamefont {Feng}, \citenamefont {Feng}, \citenamefont {Qian},
  \citenamefont {Liang}, \citenamefont {Lee}, \citenamefont {Luff},
  \citenamefont {Banwell}, \citenamefont {Agarwal}, \citenamefont {Toliver}
  \emph {et~al.}}]{dong2010ghz}%
  \BibitemOpen
  \bibfield  {author} {\bibinfo {author} {\bibfnamefont {P.}~\bibnamefont
  {Dong}}, \bibinfo {author} {\bibfnamefont {N.-N.}\ \bibnamefont {Feng}},
  \bibinfo {author} {\bibfnamefont {D.}~\bibnamefont {Feng}}, \bibinfo {author}
  {\bibfnamefont {W.}~\bibnamefont {Qian}}, \bibinfo {author} {\bibfnamefont
  {H.}~\bibnamefont {Liang}}, \bibinfo {author} {\bibfnamefont {D.~C.}\
  \bibnamefont {Lee}}, \bibinfo {author} {\bibfnamefont {B.}~\bibnamefont
  {Luff}}, \bibinfo {author} {\bibfnamefont {T.}~\bibnamefont {Banwell}},
  \bibinfo {author} {\bibfnamefont {A.}~\bibnamefont {Agarwal}}, \bibinfo
  {author} {\bibfnamefont {P.}~\bibnamefont {Toliver}}, \emph {et~al.},\
  }\bibfield  {title} {\bibinfo {title} {Ghz-bandwidth optical filters based on
  high-order silicon ring resonators},\ }\href@noop {} {\bibfield  {journal}
  {\bibinfo  {journal} {Optics express}\ }\textbf {\bibinfo {volume} {18}},\
  \bibinfo {pages} {23784} (\bibinfo {year} {2010})}\BibitemShut {NoStop}%
\bibitem [{\citenamefont {Landauer}(1957)}]{landauer1957spatial}%
  \BibitemOpen
  \bibfield  {author} {\bibinfo {author} {\bibfnamefont {R.}~\bibnamefont
  {Landauer}},\ }\bibfield  {title} {\bibinfo {title} {Spatial variation of
  currents and fields due to localized scatterers in metallic conduction},\
  }\href@noop {} {\bibfield  {journal} {\bibinfo  {journal} {IBM Journal of
  research and development}\ }\textbf {\bibinfo {volume} {1}},\ \bibinfo
  {pages} {223} (\bibinfo {year} {1957})}\BibitemShut {NoStop}%
\bibitem [{\citenamefont {Landauer}(1970)}]{landauer1970electrical}%
  \BibitemOpen
  \bibfield  {author} {\bibinfo {author} {\bibfnamefont {R.}~\bibnamefont
  {Landauer}},\ }\bibfield  {title} {\bibinfo {title} {Electrical resistance of
  disordered one-dimensional lattices},\ }\href@noop {} {\bibfield  {journal}
  {\bibinfo  {journal} {Philosophical magazine}\ }\textbf {\bibinfo {volume}
  {21}},\ \bibinfo {pages} {863} (\bibinfo {year} {1970})}\BibitemShut
  {NoStop}%
\bibitem [{\citenamefont {Landauer}(1987)}]{landauer1987electrical}%
  \BibitemOpen
  \bibfield  {author} {\bibinfo {author} {\bibfnamefont {R.}~\bibnamefont
  {Landauer}},\ }\bibfield  {title} {\bibinfo {title} {Electrical transport in
  open and closed systems},\ }\href@noop {} {\bibfield  {journal} {\bibinfo
  {journal} {Zeitschrift f{\"u}r Physik B Condensed Matter}\ }\textbf {\bibinfo
  {volume} {68}},\ \bibinfo {pages} {217} (\bibinfo {year} {1987})}\BibitemShut
  {NoStop}%
\bibitem [{\citenamefont {Kramer}\ \emph {et~al.}(2012)\citenamefont {Kramer},
  \citenamefont {Bergmann},\ and\ \citenamefont
  {Bruynseraede}}]{kramer2012localization}%
  \BibitemOpen
  \bibfield  {author} {\bibinfo {author} {\bibfnamefont {B.}~\bibnamefont
  {Kramer}}, \bibinfo {author} {\bibfnamefont {G.}~\bibnamefont {Bergmann}},\
  and\ \bibinfo {author} {\bibfnamefont {Y.}~\bibnamefont {Bruynseraede}},\
  }\href@noop {} {\emph {\bibinfo {title} {Localization, Interaction, and
  Transport Phenomena: Proceedings of the International Conference, August
  23--28, 1984 Braunschweig, Fed. Rep. of Germany}}},\ Vol.~\bibinfo {volume}
  {61}\ (\bibinfo  {publisher} {Springer Science \& Business Media},\ \bibinfo
  {year} {2012})\BibitemShut {NoStop}%
\bibitem [{\citenamefont {Shapiro}(1983)}]{shapiro1983quantum}%
  \BibitemOpen
  \bibfield  {author} {\bibinfo {author} {\bibfnamefont {B.}~\bibnamefont
  {Shapiro}},\ }\bibfield  {title} {\bibinfo {title} {Quantum conduction on a
  cayley tree},\ }\href@noop {} {\bibfield  {journal} {\bibinfo  {journal}
  {Physical Review Letters}\ }\textbf {\bibinfo {volume} {50}},\ \bibinfo
  {pages} {747} (\bibinfo {year} {1983})}\BibitemShut {NoStop}%
\end{thebibliography}%

\onecolumngrid
%\appendix
\newpage 

\appendix
\setcounter{secnumdepth}{2}
\setcounter{equation}{0}
\setcounter{figure}{0}
\renewcommand{\thetable}{S\arabic{table}}
\renewcommand{\theequation}{S\arabic{equation}}
\renewcommand{\thefigure}{S\arabic{figure}}

\clearpage
\begin{center}
	\textbf{\large Supplemental Material}
\end{center}
\setcounter{equation}{0}
\setcounter{figure}{0}
\setcounter{table}{0}
\makeatletter
\renewcommand{\theequation}{S\arabic{equation}}
\renewcommand{\thefigure}{S\arabic{figure}}
\renewcommand{\bibnumfmt}[1]{[S#1]}
We provide additional technical details and data supporting the claims in the main text.

% \paragraph{Landauer Formulation.---}
\section{Landauer formalism}\label{sec:landauer}
\begin{figure}
    \centering
    \includegraphics[width=0.40\textwidth]{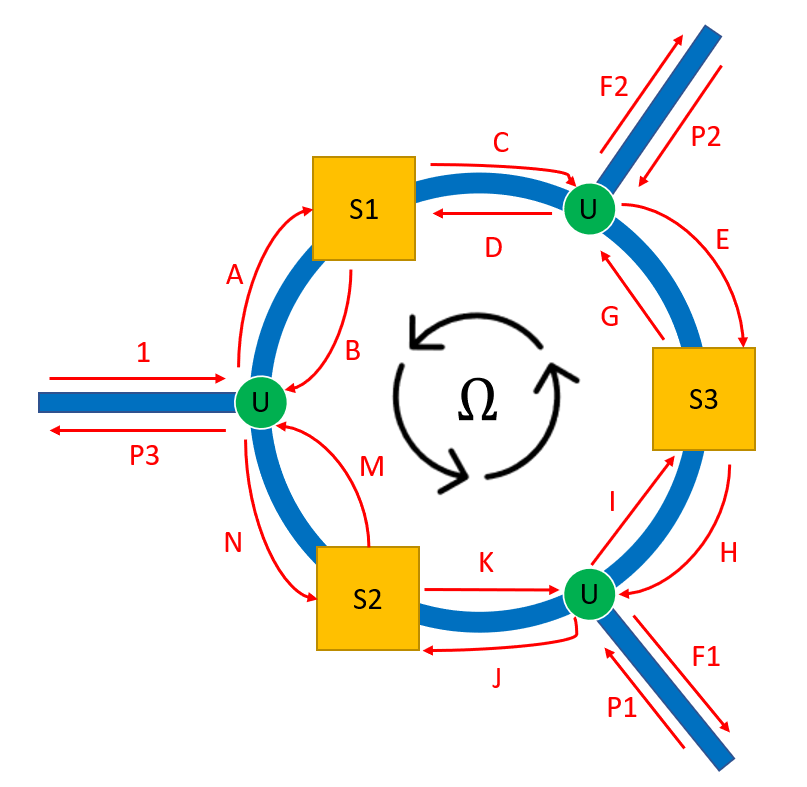}
    \caption{Three-terminal ring device described with Landauer formula. The left channel represents the path towards the source lead, while the top and bottom right channels represent the path towards the two drains. The arms of the ring are represented by scattering matrices $\mathcal{S}_1$, $\mathcal{S}_2$ and $\mathcal{S}_3$, and the flux is absorbed into their parameters via the Peierls substitution. The junctions connecting ring to source and drains are represented by the scattering matrix $\mathcal{U}$. }
    \label{fig:landauer_1}
\end{figure}

The Landauer formula~\cite{landauer1957spatial,landauer1970electrical,landauer1987electrical,kramer2012localization} is used in mesoscopic systems to express the conductance in terms of the scattering properties of the geometry of the problem. It is a useful tool to study transport properties and obtain exact expressions for the transport properties of quantum conductors. It has the advantage over semiclassical transport theories that it captures interference phenomena in the conductance. It has been used to calculate the conductance for rings subject to a flux~\cite{gefen1984quantum,buttiker1984quantum}. The formalism expresses branches and junctions in terms of scattering matrices. The geometry of the system we are considering is shown in Fig.~\ref{fig:landauer_1}.

We model the ring as single-channel conductors connecting the branches between the three junctions to the source and drain leads. Each such branch is described as a single scatterer connected to an ideal one-dimensional channel. The scattering can be described with a $2\times2$ scattering matrix
\begin{gather}
\mathcal{S}_j =
    \begin{bmatrix}
    r_j&t_j\\
    t_j^\prime & r_j^\prime
    \end{bmatrix}.
\end{gather}
All phases and scattering effects that occur within each branch are absorbed into the parameters of the scattering matrix. In the absence of flux through the ring, time-reversal and current-conservation (demanding unitarity) impose the following constraints on the scattering matrix: $t_j = t_j^\prime$ and $-t_j/(t_j^\prime)^*=r_j/(r_j^\prime)^*$. To account for the flux which breaks time-reversal symmetry, we use the Peierls substitution $t_j\rightarrow{}t_j e^{-i2\pi\Omega/3},t_j^\prime\rightarrow{}t_j e^{i2\pi\Omega/3},r_j\rightarrow{}r_j,r_j^\prime\rightarrow{}r_j^\prime$. The prior constraints on $t_j,r_j$ and $r_j^\prime$ remain the same such that we obtain valid scattering matrices. For the $\mathcal{S}_1$ arm, we get a system of equations that describe the scattering in this arm
\begin{gather}
\begin{bmatrix}
B\\C
\end{bmatrix} =
    \begin{bmatrix}
    r_1&t_1 e^{-i2\pi\Omega/3}\\
    t_1 e^{i2\pi\Omega/3} & r_1^\prime
    \end{bmatrix}
    \begin{bmatrix}
    A\\D
    \end{bmatrix}.
\end{gather}
We can model the junctions that connect to the leads with a $3\times 3$ scattering matrix $\mathcal{U}$~\cite{shapiro1983quantum}
\begin{gather}
\mathcal{U} =
    \begin{bmatrix}
    0&-\frac{1}{\sqrt{2}}&-\frac{1}{\sqrt{2}}\\
    -\frac{1}{\sqrt{2}} & \frac{1}{2} & -\frac{1}{2}\\
    -\frac{1}{\sqrt{2}}&-\frac{1}{2}&\frac{1}{2}
    \end{bmatrix}.
\end{gather}
The first channel is the path leading out of the ring to one of the leads. $\mathcal{U}$ describes a scenario where there is no instantaneous reflection for particles entering the ring from the channel and we can scatter in both directions in the ring with equal probability. While the scattering matrix $\mathcal{U}$ that fulfils these conditions is not unique, the results do not qualitatively depend on the specific choice of $\mathcal{U}$.

We solve the equations by writing down the relationships between the various amplitudes at the junctions and scatterers. We demand that the channel of the source lead into the ring is unitary and setting that $P_1$ and $P_2$ is zero, which means that particles that exit to the drains cannot return. As a result, we get a linear system of 15 equations and 15 unknowns. We solve it for $F_1$ and $F_2$ which are the complex amplitudes of the transmission into the respective drains. Then, we insert the values for all the scattering coefficients $t_j,r_j$ and $r_j^\prime$. 

The final result depends on how we choose these scattering coefficients as they determine the transport properties of the ring. From physical consideration, we demand that transmission and reflection occur with equal probability and we demand that clockwise and anti-clockwise reflection is the same $r_j=r_j^\prime$. These considerations lead us to the choice $r_j=r_j^\prime=\sqrt{1/2}$ and $t_j=i\sqrt{1/2}$. 

The transmission $G_\alpha=|F_\alpha |^2$ into the respective drains $\alpha\in\{0,1\}$ is then given by
\begin{equation}
G_\alpha=16\left\vert\frac{1-\sqrt{2}+2i(2\sqrt{2}-3)\exp(-i\pi(2\Omega+\alpha))}{62-46\sqrt{2}+2i\cos(\pi(2\Omega+\alpha))}\right\vert^2\,.
\end{equation}
As seen in the main text, the transmission strongly depends on the flux $\Omega$ and can be tuned to maximize transmission into a particular drain.

\section{Flux dependence of dynamics close to ground state}\label{sec:gs_sup}
\begin{figure*}[htbp]
	\centering	
	\subfigimg[width=0.9\textwidth]{}{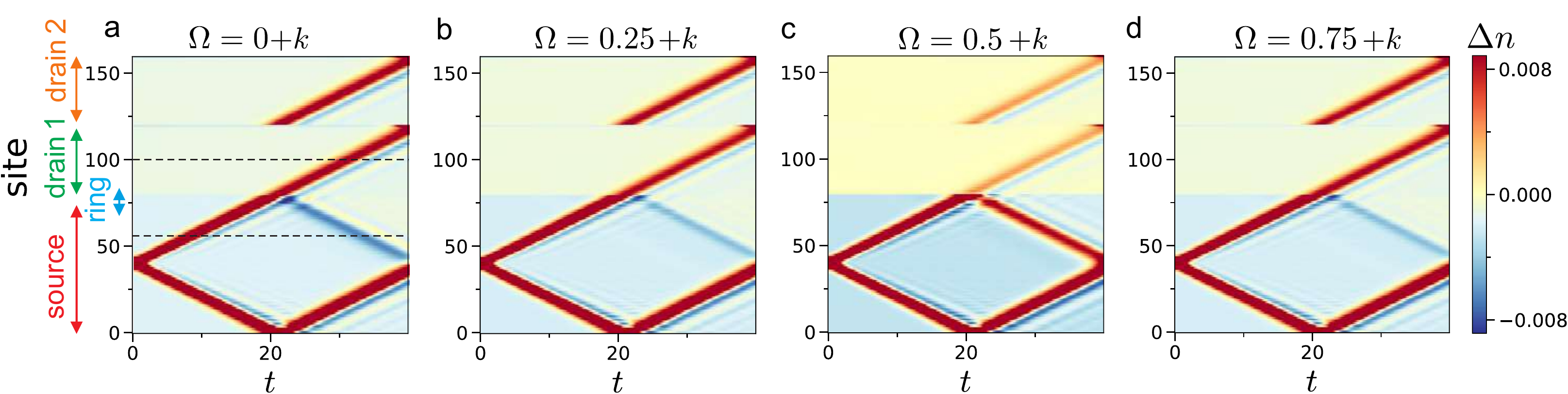}
	\caption{Dynamics of density wave in three-terminal device. \idg{a} Change of density relative to average density $\Delta n(t)=\langle n(t)\rangle-n_0$ as function of time and sites of source, ring, drain 1 and 2. We show \idg{a} $\Omega=k$ \idg{b} $\Omega=0.25+k$, \idg{c} $\Omega=0.5+k$ and \idg{d} $\Omega=0.75+k$, where $k$ is an integer. Perturbation starts in source, and moves through ring into drains. Note that the dynamics in both drains is symmetric for all values of flux. We have hard-core bosons with $J=1$, $K=0.5$ and in total $L=160$ lattice sites and $N=80$ bosons ($L=3$, $L_s=79$, $L_b=L_c=39$).
	}
	\label{fig:fulldynamics_sup}
\end{figure*}

Here, we show further results on the dynamics in the three terminal device close to the ground state. We prepare a density perturbation in the source, then quench the Hamiltonian and create a propagating density wave. The dynamics of the change of density in time for different values of flux is shown in Fig.\ref{fig:fulldynamics_sup}. The perturbation starts from the source and moves both forward and backward in sites. The backward propagating wave stays in the source within our simulation time, and is simply reflected at the other end of the source. It has no influence on the rest of the system and can be ignored. The forward moving density wave propagates through the ring, where it is both transmitted into the drains, as well as reflected back into the source. For $\Omega=0$, we find a negative Andreev reflection, whereas for $\Omega=0.5$, we find a clear positive reflection.

The dynamics is periodic with $\Omega\rightarrow\Omega+k$, where $k$ is an integer. Note that $\Omega=0.25+k$ and $\Omega=0.75+k$ produces nearly the same dynamics, demonstrating a reflection symmetry $\Omega\rightarrow-\Omega$ for the dynamics close to the ground state. For any value of flux, we observe that the current into drain 1 and drain 2 is nearly identical. Both the reflection symmetry and the symmetry in the drains is a low-energy effect that is absent for the strongly perturbed dynamics. 

\begin{figure}[htbp]
	\centering	
	\subfigimg[width=0.28\textwidth]{}{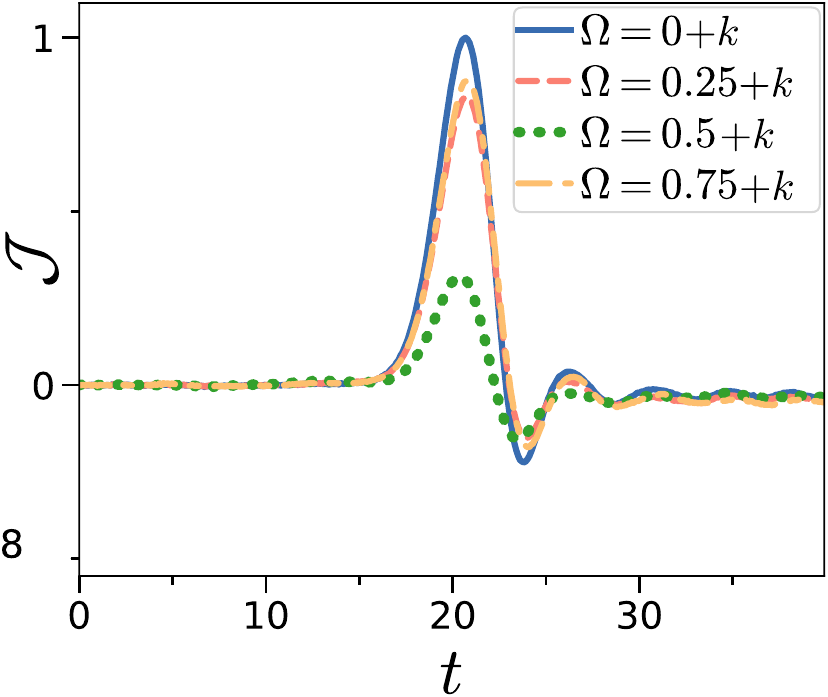}
	\caption{Current $\mathcal{J}$ into the first drain against time and $\Omega$ for low-energy dynamics. We use same parameters as in Fig.\ref{fig:fulldynamics_sup}.
	}
	\label{fig:currentLow_sup}
\end{figure}

In Fig.\ref{fig:currentLow_sup}, we show the current into the first drain for different $\Omega$ against time $t$. 

\section{Non-equilibrium dynamics}\label{sec:nonequ}
\begin{figure*}[htbp]
	\centering	
		\subfigimg[width=0.3\textwidth]{a}{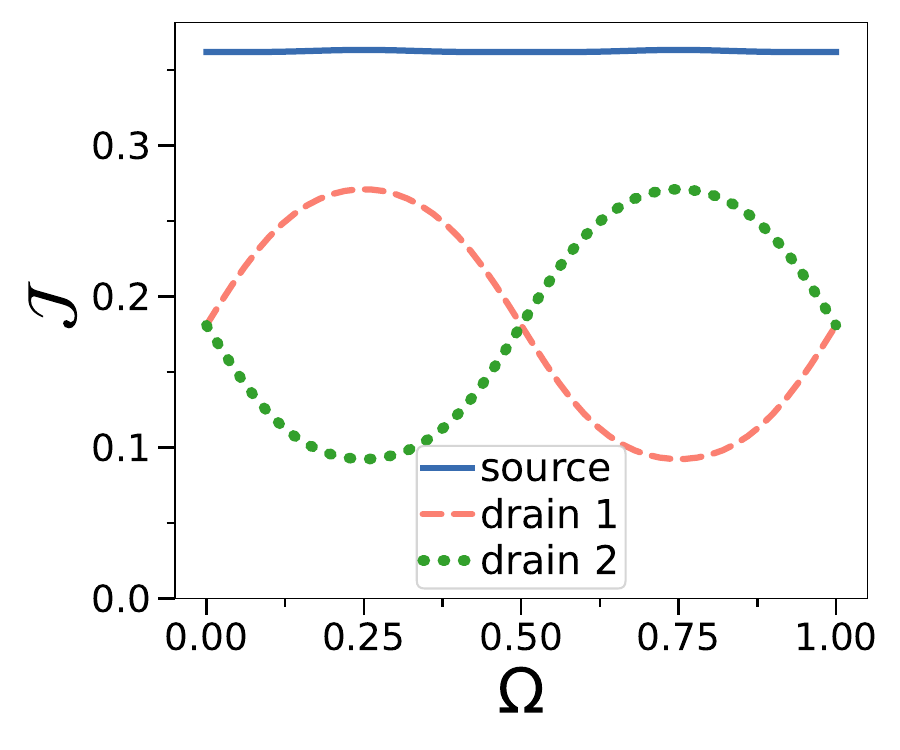}
	\subfigimg[width=0.3\textwidth]{b}{currentsteadySD_BHMDrainL3N4m3J1U0g1u0M1s1p51fluxT10g1G1r0R0.pdf}
		\subfigimg[width=0.3\textwidth]{c}{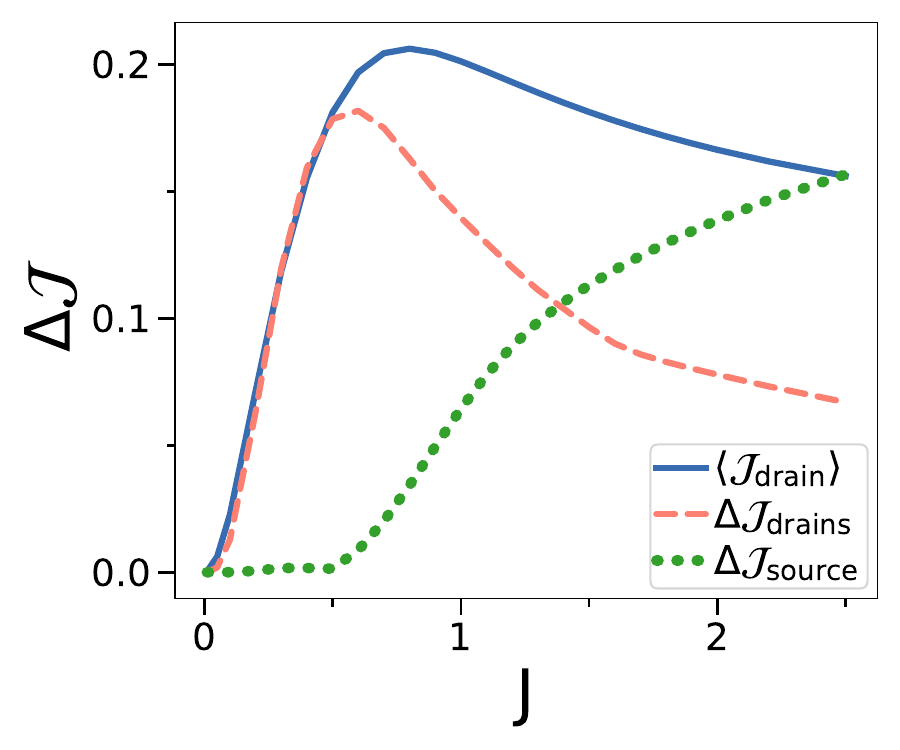}
	\caption{Steady-state current $\mathcal{J}$ for hard-core bosons as function of flux $\Omega$ for \idg{a} $J=0.5$ \idg{b} $J=1$. \idg{c} Mean value of drain current $\langle \mathcal{J}_\text{drain}\rangle$ taken over $\Omega$,  maximal difference between drain 1 and drain 2 current $\Delta \mathcal{J}_\text{drains}=\text{max}_\Omega(\vert\mathcal{J}_\text{drain 1}(\Omega)-\mathcal{J}_\text{drain 2}(\Omega)\vert)$ and amplitude of source current $\Delta \mathcal{J}_\text{source}=\text{max}_\Omega\mathcal{J}_\text{source}-\text{min}_\Omega\mathcal{J}_\text{source}$  as function of intra-ring coupling $J$.
	We set $L=3$, $K=1$ and $B_\text{s}=B_\text{d}=1$.
	}
	\label{fig:currsteady}
\end{figure*}

Here we provide further results on the non-equilibrium dynamics.
In Fig.\ref{fig:currsteady}a,b we show the steady-state current through source and drains as function of flux $\Omega$. The current in drain 1 and 2 varies strongly with $\Omega$, allowing for directional control into either drain. For the source current, we find for $J=0.5$ in Fig.\ref{fig:currsteady}a no variation with $\Omega$, whereas for $J=1$ the current in source varies strongly (Fig.\ref{fig:currsteady}b). In Fig.\ref{fig:currsteady}c, we investigate the current as function of intra-ring coupling $J$. The solid blue line shows the mean drain current $\langle \mathcal{J}_\text{drain}\rangle$ averaged over $\Omega$. The dashed orange line is the maximal difference between the two drain currents $\Delta \mathcal{J}_\text{drains}=\text{max}_\Omega(\vert\mathcal{J}_\text{drain 1}(\Omega)-\mathcal{J}_\text{drain 2}(\Omega)\vert)$. The dotted green curve is the amplitude of the source current $\Delta \mathcal{J}_\text{source}=\text{max}_\Omega\mathcal{J}_\text{source}-\text{min}_\Omega\mathcal{J}_\text{source}$. We find a transition from $\Delta \mathcal{J}_\text{source}\approx0$ to $\Delta \mathcal{J}_\text{source}\gtrsim 0$ for $J>0.5$. Further, the amplitude and mean value of the drain current shows a peak at the transition $J\approx0.5$. This is indicates that our system features a non-equilibrium transition in the flux dependence of the current.

\section{Driving flux linearly in time}\label{sec:drive_sup}
\begin{figure*}[htbp]
	\centering	
	\subfigimg[width=0.3\textwidth]{a}{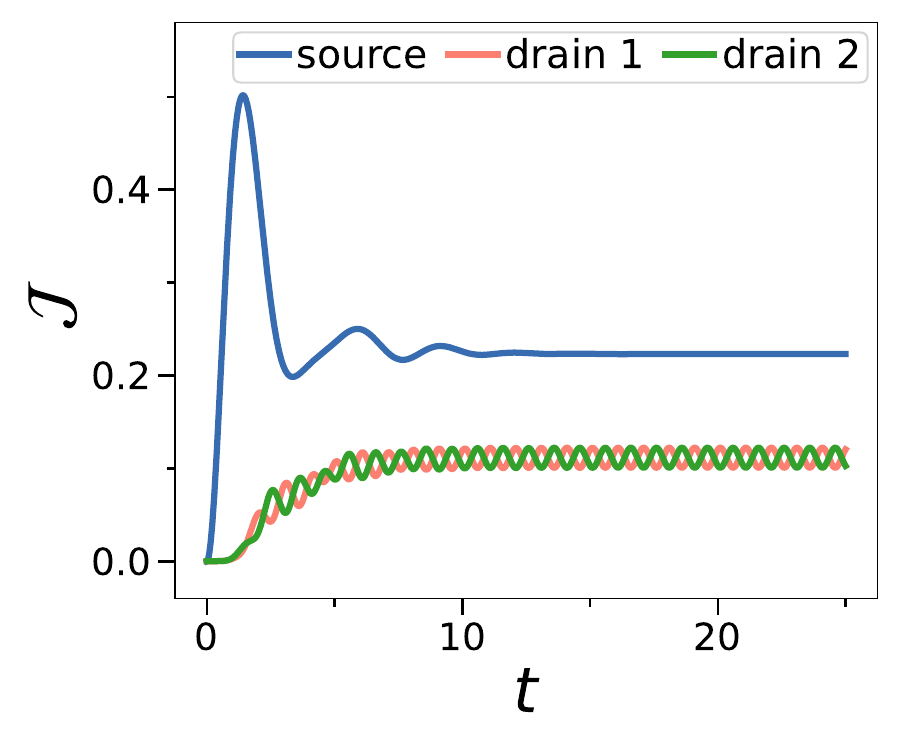}	\subfigimg[width=0.3\textwidth]{b}{currentSDEval_BHMDrainL3N4m3J0_5U0g1u0M1s1p1periodT2_8g1G1r0R0.pdf}
	\subfigimg[width=0.3\textwidth]{c}{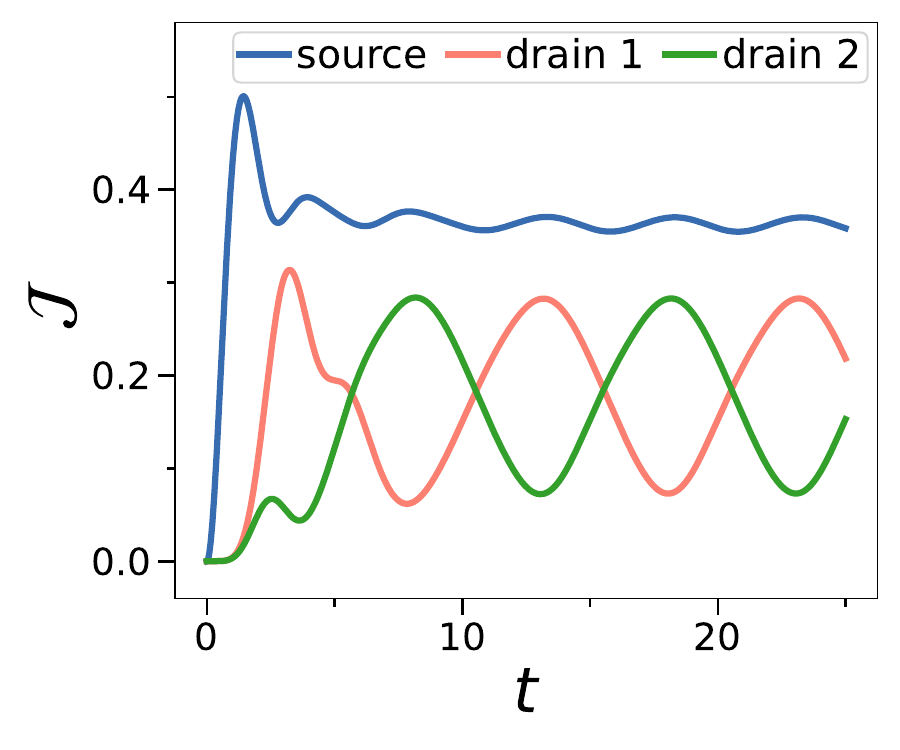}
	\caption{Current for linear driving $\Omega(t)=t/T$ of flux within ring. We show $\mathcal{J}(t)$ in time $t$ for various driving periods $T$ for \idg{a} $T=1$, \idg{b} $T=2.8$ and \idg{c} $T=10$. We have $L=3$, $K=1$, $J=0.5$ and $B_\text{s}=B_\text{d}=1$.
	}
	\label{fig:dynamicflux_sup}
\end{figure*}

We show results on driving the flux of the ring linearly in time with 
\begin{equation}
\Omega(t)=t/T
\end{equation}
in Fig.\ref{fig:dynamicflux_sup}.
For very fast driving $T=1$ in Fig.\ref{fig:dynamicflux_sup}a, we find that the current shows barely any oscillation as the driving is much faster than then system dynamics. 
For intermediate driving $T=2.8$ in Fig.\ref{fig:dynamicflux_sup}b, we find strong oscillations. The amplitude of the drain currents is nearly the same as the source current, implying that most of the incoming direct current is converted into an \revA{AC modulated} current.
For very slow driving $T\gg1$ in Fig.\ref{fig:dynamicflux_sup}c, the system dynamics is much faster than the driving. Thus, the system has sufficient time to settle into the instantaneous steady state of the flux. The current then simply takes the same value as the current of the steady state of the system $\rho_\text{SS}(\Omega(t))$ at a given flux $\Omega(t)$.

\section{Oscillating flux in time}\label{sec:drive_osc_sup}
\begin{figure*}[htbp]
	\centering	
	\subfigimg[width=0.3\textwidth]{a}{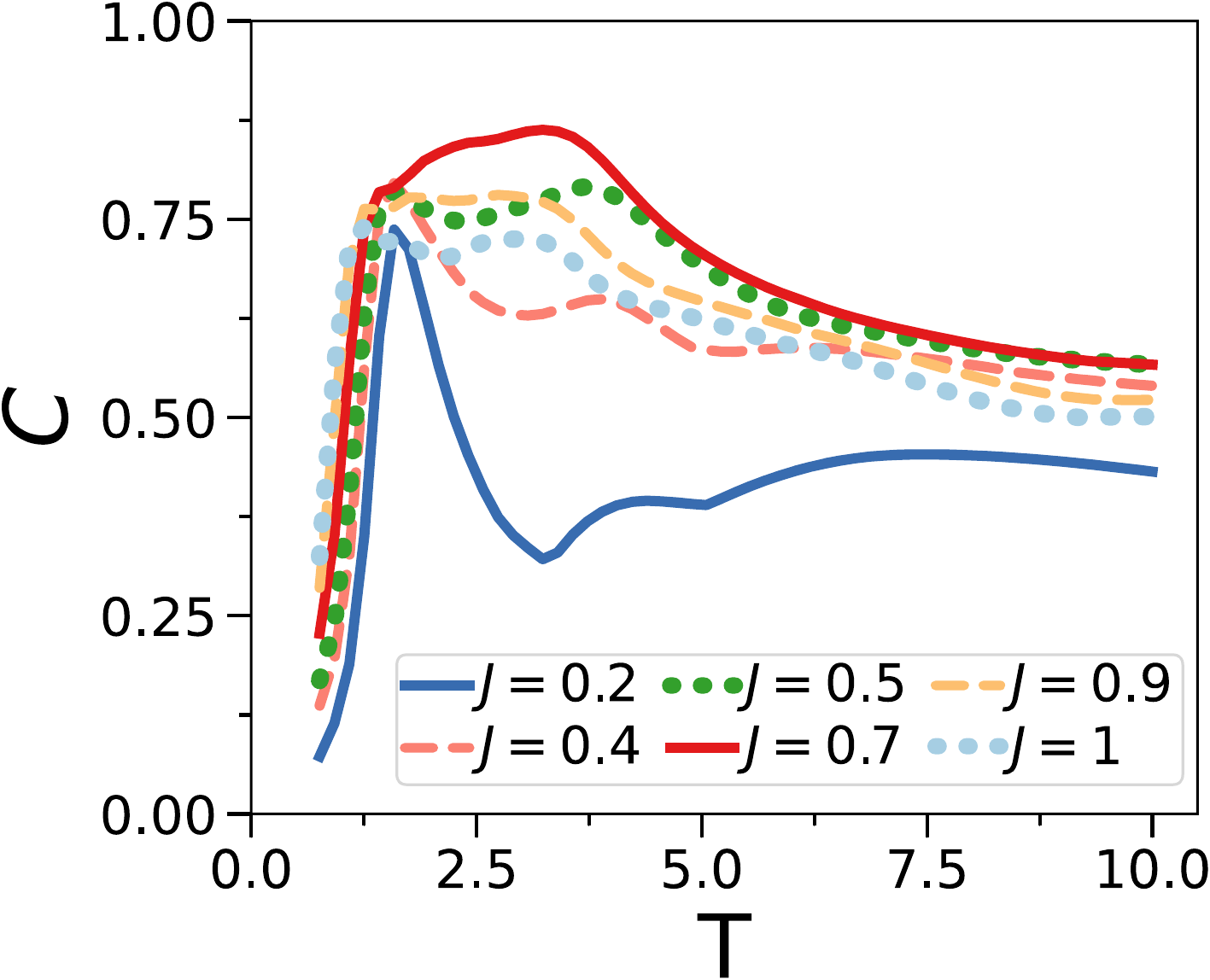}	\subfigimg[width=0.3\textwidth]{b}{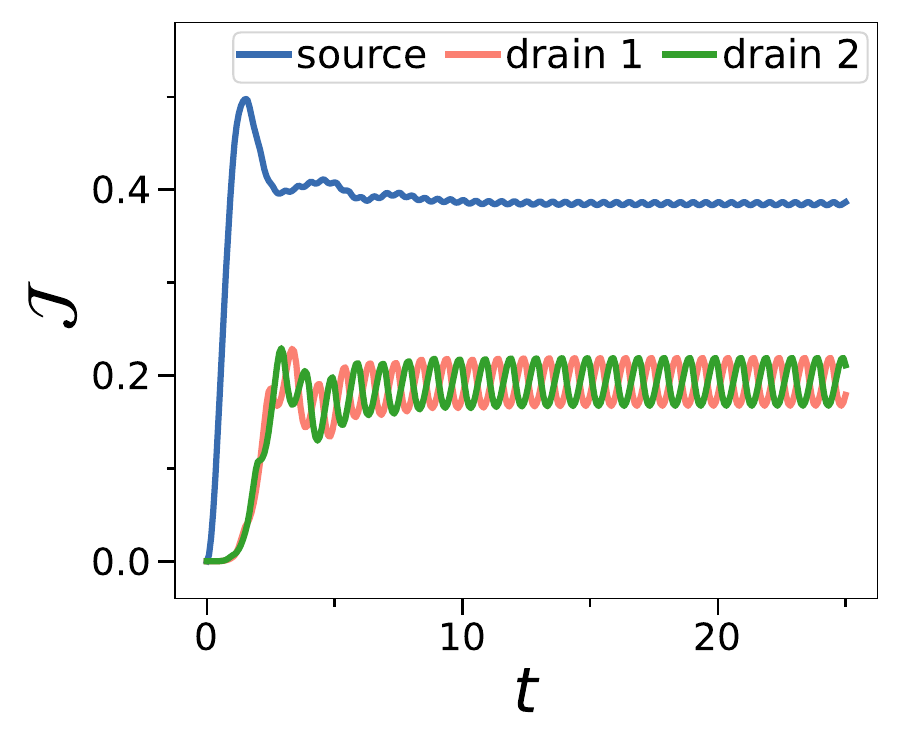}
	\subfigimg[width=0.3\textwidth]{c}{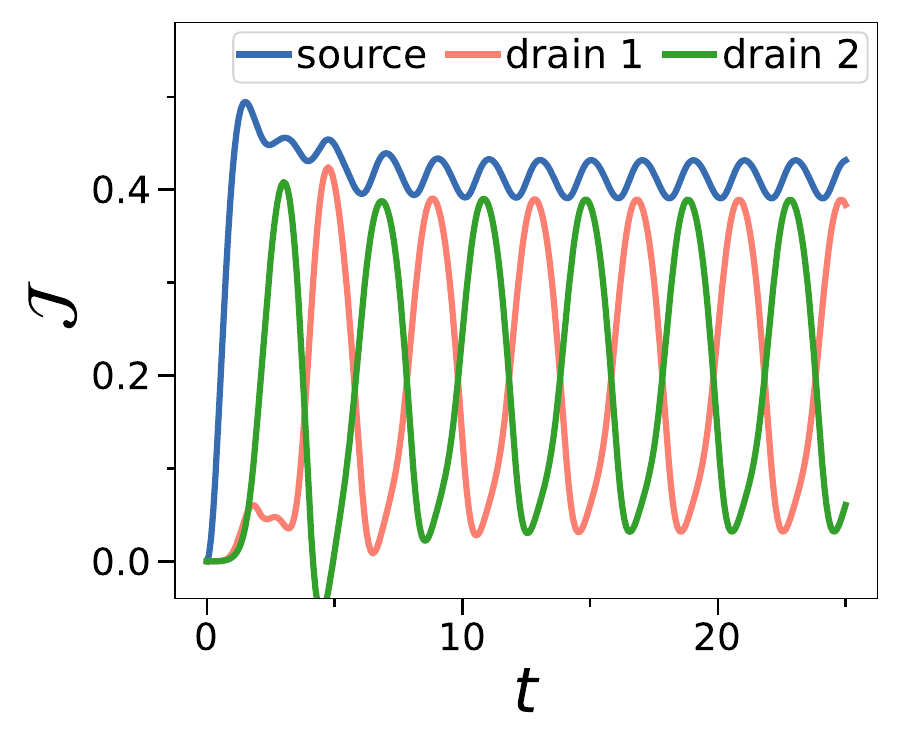}
 	\subfigimg[width=0.3\textwidth]{d}{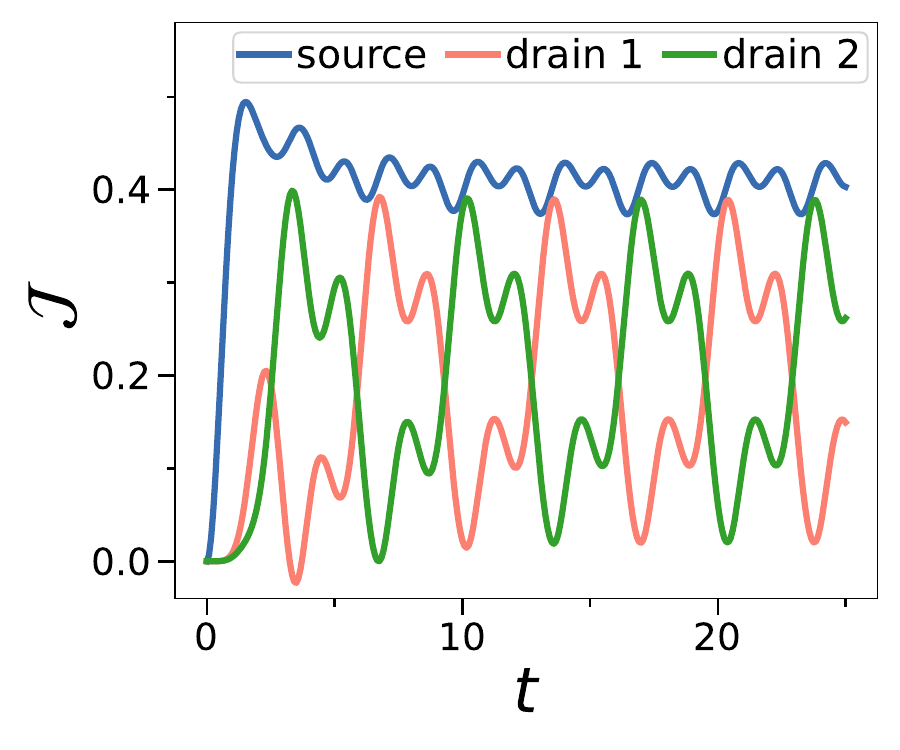}
 	\subfigimg[width=0.3\textwidth]{e}{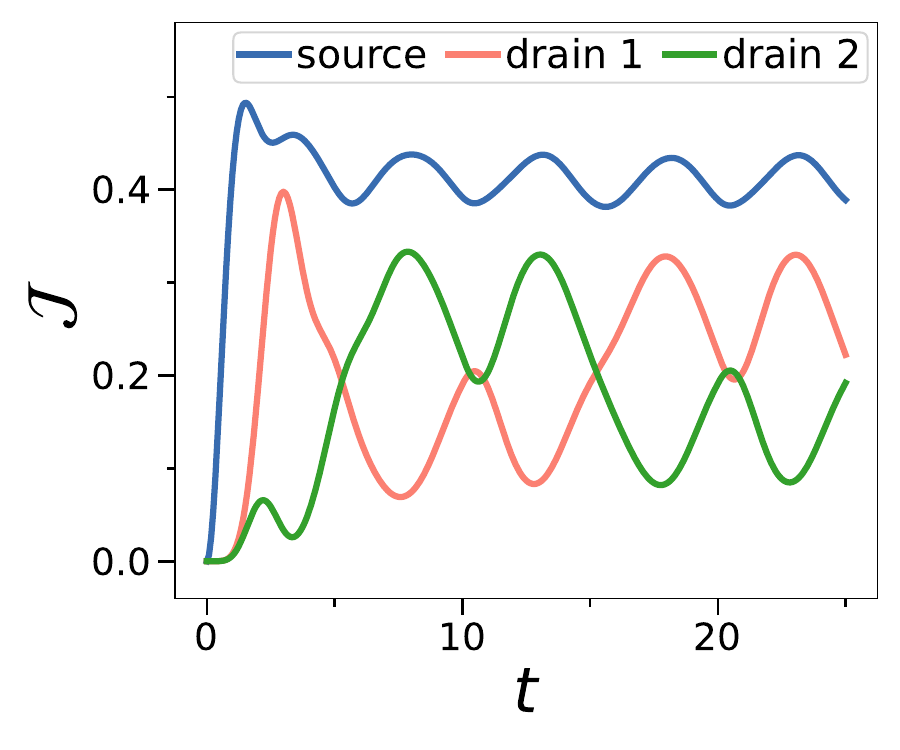}
	\caption{Current for triangle flux driving protocol~\eqref{eq:flux_triangle_sup}.
 \idg{a} Conversion efficiency $C$.
 \idg{b-e} $\mathcal{J}(t)$ in time $t$ for various driving periods $T$ of the flux.  \idg{b} $T=0.5$, \idg{c} $T=2$, \idg{d} $T=3.4$ and \idg{e} $T=10$.
	We have $L=3$, $K=1$, $J=0.7$ and $B_\text{s}=B_\text{d}=1$.
	}
	\label{fig:dynamictriangle_sup}
\end{figure*}

\revA{We now study the driving of the flux of the ring with a triangle function
\begin{equation}\label{eq:flux_triangle_sup}
    \Omega(t)=2\left\vert \frac{t}{2T}-\left\lfloor \frac{t}{2T}-\frac{1}{2} \right\rfloor\right\vert\,.
\end{equation}
This driving with a period of $2T$ corresponds to a linear increase of $\Omega(t)$ from $\Omega=0$ to $\Omega=1$ within a time $T$, followed by a linear decrease from $\Omega=1$ to $\Omega=0$ within $T$. These steps are repeated periodically.

We show the conversion efficiency $C$ and currents for various $T$ in Fig.\ref{fig:dynamictriangle_sup}. 
We find a maximal conversion efficiency $C_\text{max}\approx0.86$ for $J\approx0.7$ and $T=3.4$. We find that the conversion efficiency is large for $J=0.7$ over a larger range of $T$ compared to the linear driving. For smaller $T\le2$, we find that the driving is very close to a sine curve. For larger $T$ we note that the driving becomes less sine-like compared to the linear flux driving.}

\section{Experimental details}\label{sec:experiment}
Here we give details on the experimental demonstration of the Bose-Einstein condensate in the potential of a three-terminal ring as shown in Fig.1b. 

The experiment starts with a Magneto-Optical Trap (MOT) which traps up to $7\times 10^8$ Rubidium $87$ atoms with a temperatures of 20µK for 5ms. Magnetic field biases with an optical pump beam pumps the atoms into the $F=2, m_f=+2$ state. A magnetic field gradient of 220G/cm is then ramped up in 150ms to trap up to $3\times 10^8$ atoms. 
A 2 second translation stage brings the atoms to a second magnetic trap with lower vacuum pressure. 
Due to heating processes during this step, a Radio-Frequency (RF) evaporative cooling step is then further performed to condense the atoms back to $20\mu K$. An atom cloud of $1 \times 10^7$ atoms remains after this step. At the same time, a 1064nm laser beam for optical trapping of the atoms is ramped up to 5W, and subsequently a ramp down of the MOT is performed, leaving behind a pure optical trap.

Next, an optical evaporation step is performed. 
The laser power is first lowered from 9.0V to 6.0V in 2s, where the atom temperature reaches $1\mu K$ and the atom number is $1 \times 10^6$. 
The second ramp brings the voltage from 6.0V to 5.0V in 3s with a temperature of 250nK and atom number $3.5 \times 10^5$. 

The final ramp occurs in 2s from 5.0V to 4.8V where we get a Bose-Einstein condensate (BEC) of $1.8 \times 10^5$ atoms with a phase space density of 3.0.
The optical setup for creating arbitrary traps in 2D comprises a Digital Micromirror Device (DMD) setup and an optical sheet. The DMD (DLP9500) has dimensions of 20.7mm x 11.7mm with pitch of $10.8\mu m$. A controller board (V4395) from Texas Instruments controls the operation of the DMD. We draw the potential of the ring with the three leads by electronically controlling the mirror array of the DMD. The drawn ring has a outer radius of $R_1=15\mu m$, an inner radius of $R_0=9\mu m$, and the center of the ring has a radius of $R=12\mu m$.
A blue-detuned (532nm) 10mm beam diameter is reflected off the DMD with 40\% efficiency in the first order with all mirrors on. This beam is first de-magnified 4 times with a 300mm-75mm lens configuration, and later 20 times with a 200mm-20x objective (Mitutoyo M Plan APO NIR 20x, NA = 0.40) for a total of 80x de-magnification. The 1st order beam at the atom plane has a maximum 190mW of beam power at a beam radius of $62.5\mu m (1/e^2)$ giving a maximum trap depth of $1.8\mu K$. Utilising the same objective lens, a dichronic mirror reflects the absorption image of the atoms through a 200mm lens and imaged onto an ANDOR EMCCD. The resonant imaging beam is delivered from below the atoms with a pulse time of 10µs to minimise the force on the atoms.
The optical sheet for 2D trapping is generated via intersecting 2 beams using a single lens. The interference of the 2 beams creates optical sheets superimposed on the beam profile. Our beam is asymmetric at the focus with 50µm radius in the vertical direction and 150 µm in the horizontal direction. This minimises the probability of loading into multiple sheets as the interfering beams are static. The interfering beams are distanced such that the interference sheets are spaced 7.5µm apart.

Loading of a BEC into the combined blue-detuned trap starts with a BEC at 50\% purity in the red-detuned crossed dipole trap. The desired trap is first projected on the DMD, and within 100ms both the DMD and optical sheet are ramped up while keeping the red-detuned trap on. The DMD beam is ramped up to half power while the sheet is ramped up to 63mW per interfering beam. The red-detuned trap is ramped down to a negligible trap depth in 10ms to complete the transfer. The transfer efficiency of the atoms heavily depends on the size of the DMD trap as compared to the initial BEC size. A DMD square trap of side 40µm gives a loading efficiency of 50\%. The condensation of the BEC in the DMD is verified by observing in the time-of-flight images an asymmetric cloud in the direction perpendicular to the sheet. 

\section{2D GPE dynamics}\label{sec:2dgpe}

We show further results on the simulation of the continuous ring-lead system with 2D GPE. The potential is shown in Fig.1c in the main text. The ring has a diameter of $R=30\mu \text{m}$ with a width of $5\mu\text{m}$. The leads have the same width as the ring and a length of $12\mu\text{m}$. The evolution is performed in the $x$-$y$ plane, and we assume that the atoms are confined in $z$-direction with a Gaussian atomic density profile of $10\mu$\text{m}. We use the scattering length and mass of Rubidium atoms with approximation of $z$-confinement~\cite{bao2003numerical}. The simulation is performed on a grid of dimension $256\times256$ using the Trotter-Suzuki library~\cite{wittek2015extended}. The flux $\Omega$ is imparted by rotating the full system around the center of the ring with a rotation frequency $\omega=\frac{\Omega \hbar}{m R^2}$, where $R$ is the center radius of the ring. The rotation results in a Coriolis flux, which induced the flux, as well as a centrifugal force that pushes the atoms away from the rotation center. To compensate the centrifugal force, we add a potential $V_{\text{cent}}=-\frac{1}{2}\omega^2 R^2 m$. $N$ atoms are initialised in the ground state of the source lead only. At $t>0$, the ring and drain potential is switched on, and the atoms evolve into the ring and drains. 

\begin{figure*}[htbp]
	\centering	
	\subfigimg[width=0.28\textwidth]{a}{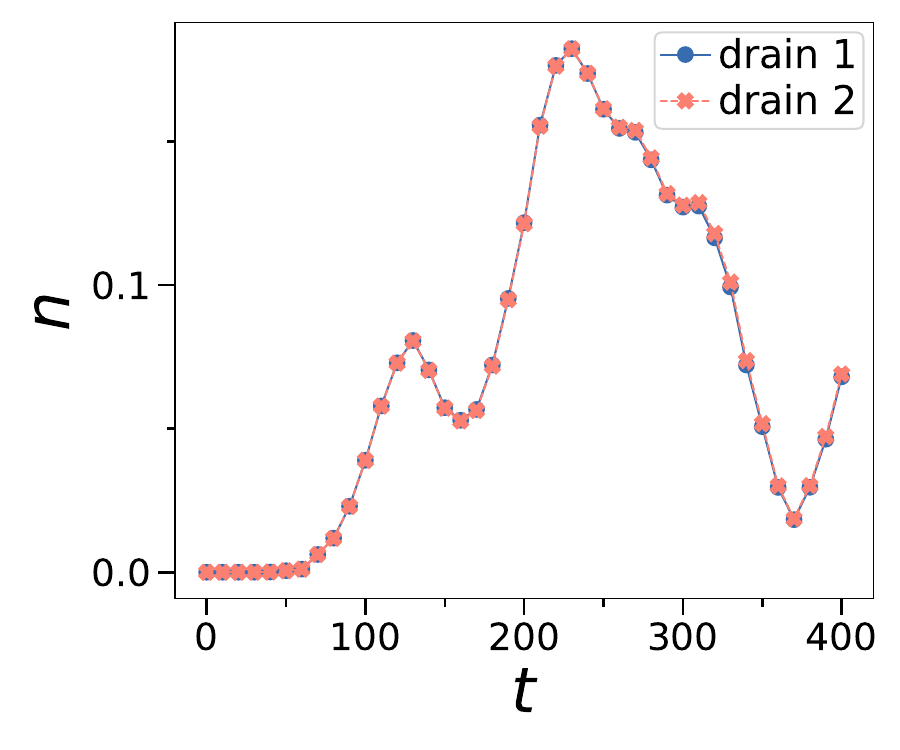}	\subfigimg[width=0.28\textwidth]{b}{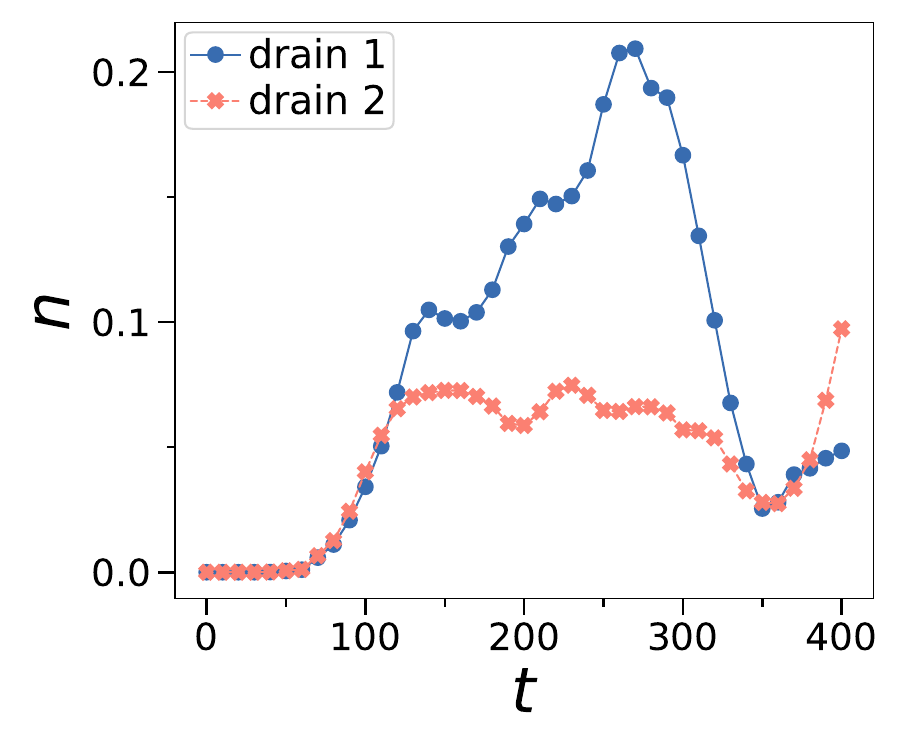}
\subfigimg[width=0.28\textwidth]{c}{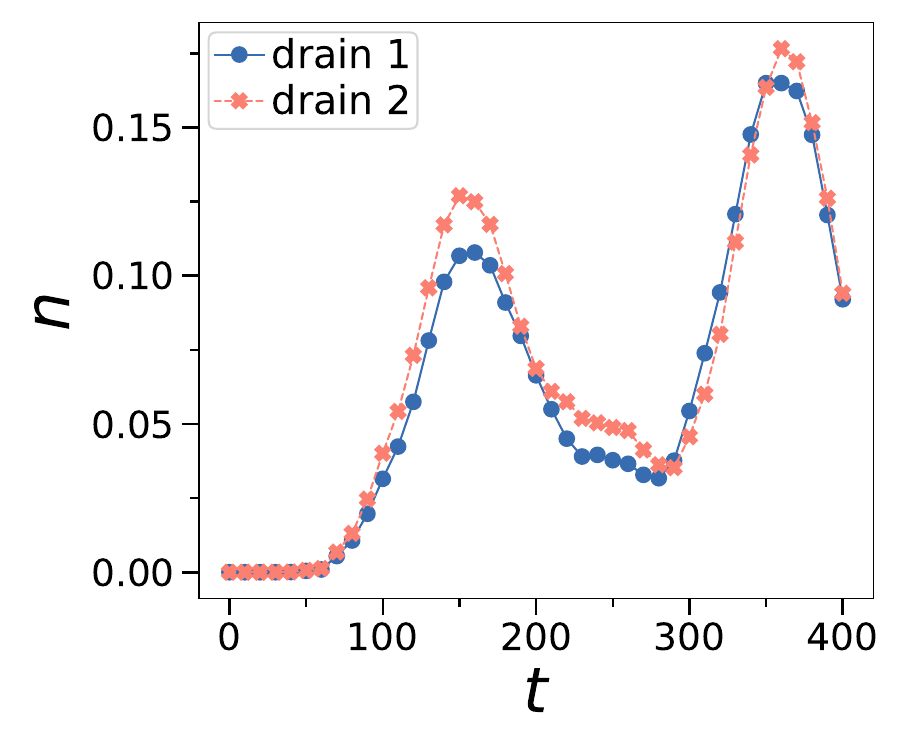}
\subfigimg[width=0.28\textwidth]{d}{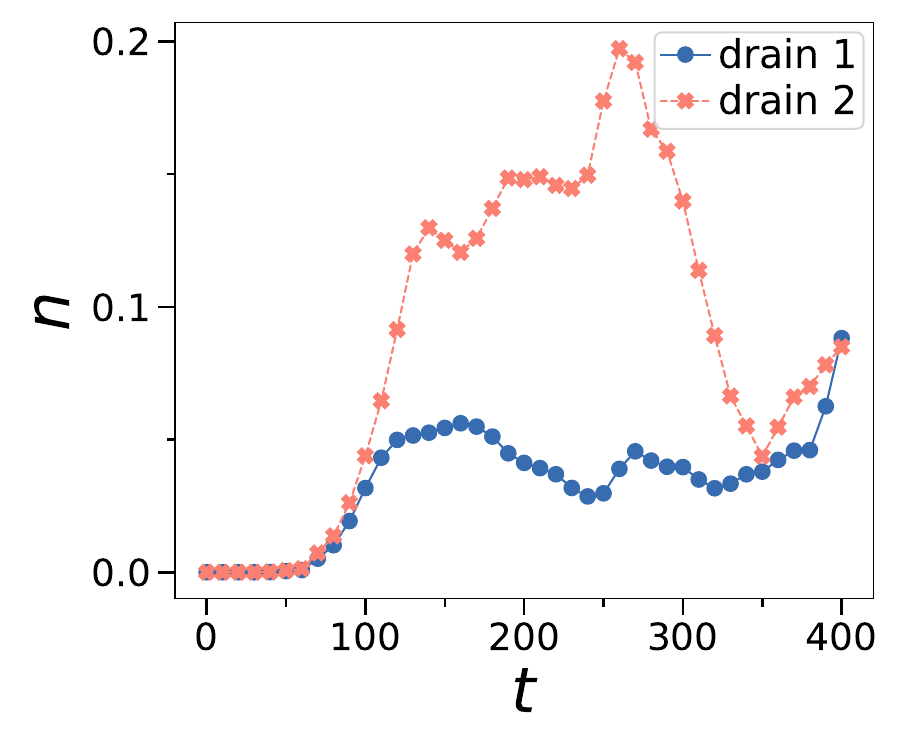}
\subfigimg[width=0.28\textwidth]{e}{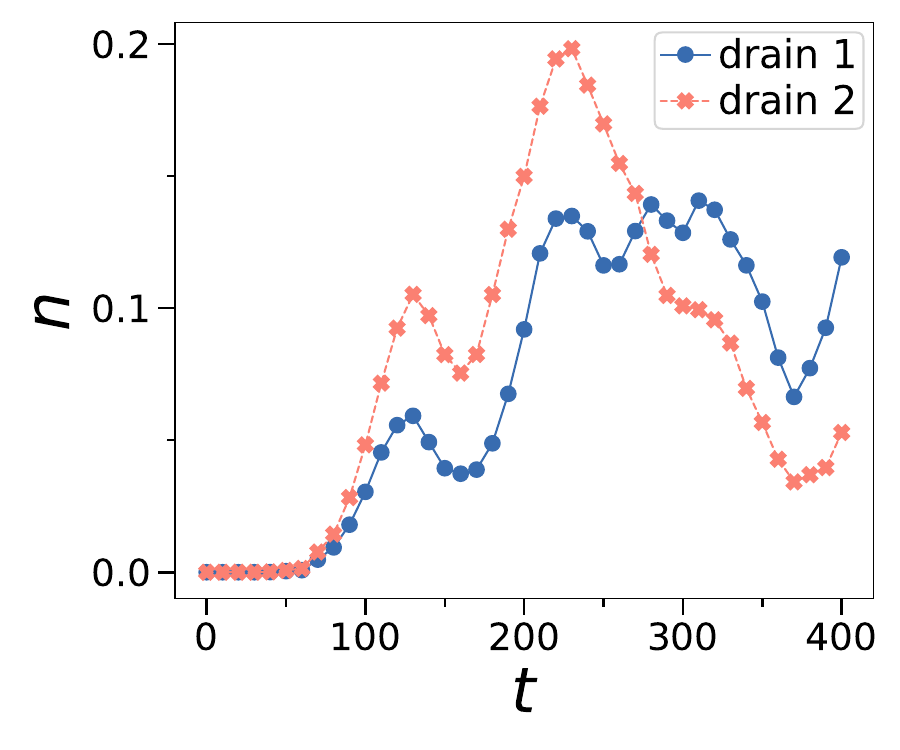}
	\caption{Density fraction in drains for 2D GPE simulation of ring-lead system for $N=2000$ atoms. We show number of atoms in drains divided by total atom number $N$ as function of time $t$ in units of milliseconds. We show values of flux \idg{a} $\Omega=0$, \idg{b} $\Omega=0.25$, \idg{c} $\Omega=0.5$, \idg{d} $\Omega=0.75$ and \idg{b} $\Omega=1$.
	}
	\label{fig:2dgpe_den}
\end{figure*}

\begin{figure*}[htbp]
	\centering	
	\subfigimg[width=0.28\textwidth]{a}{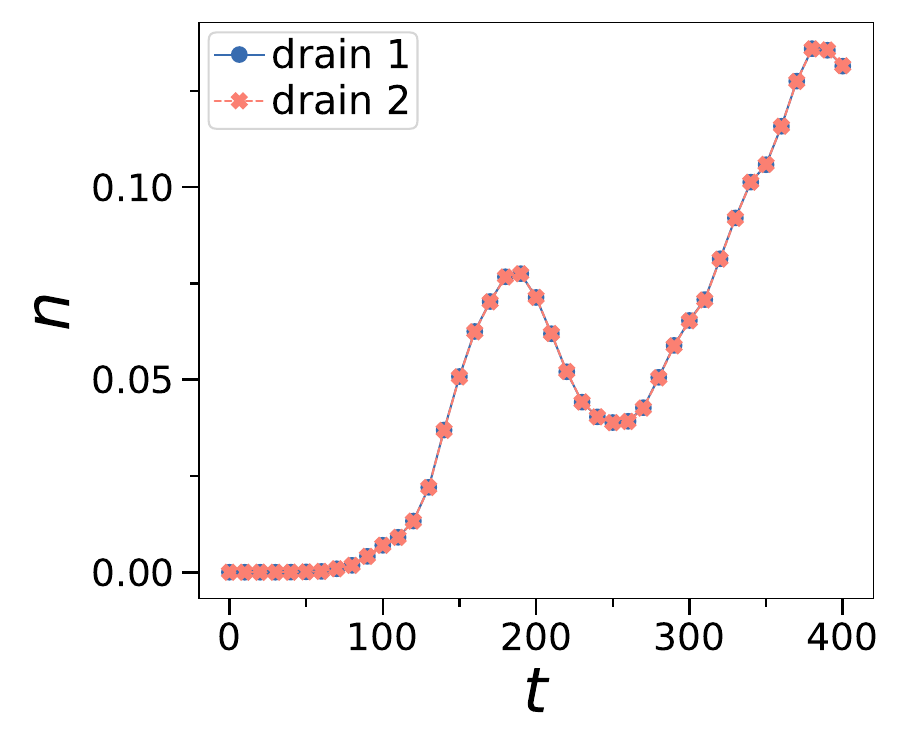}	\subfigimg[width=0.28\textwidth]{b}{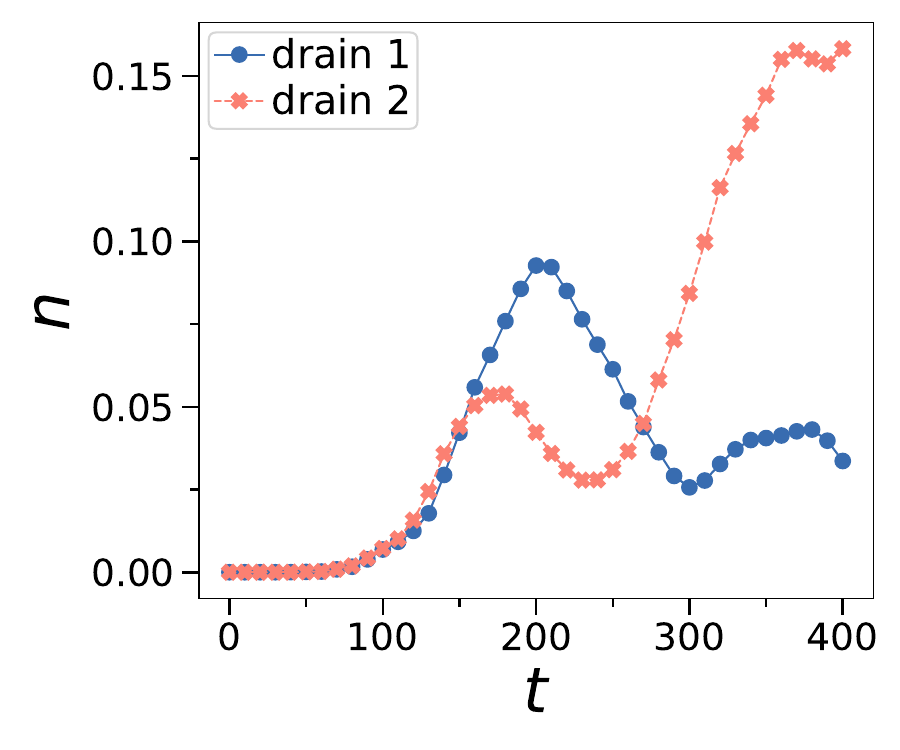}
\subfigimg[width=0.28\textwidth]{c}{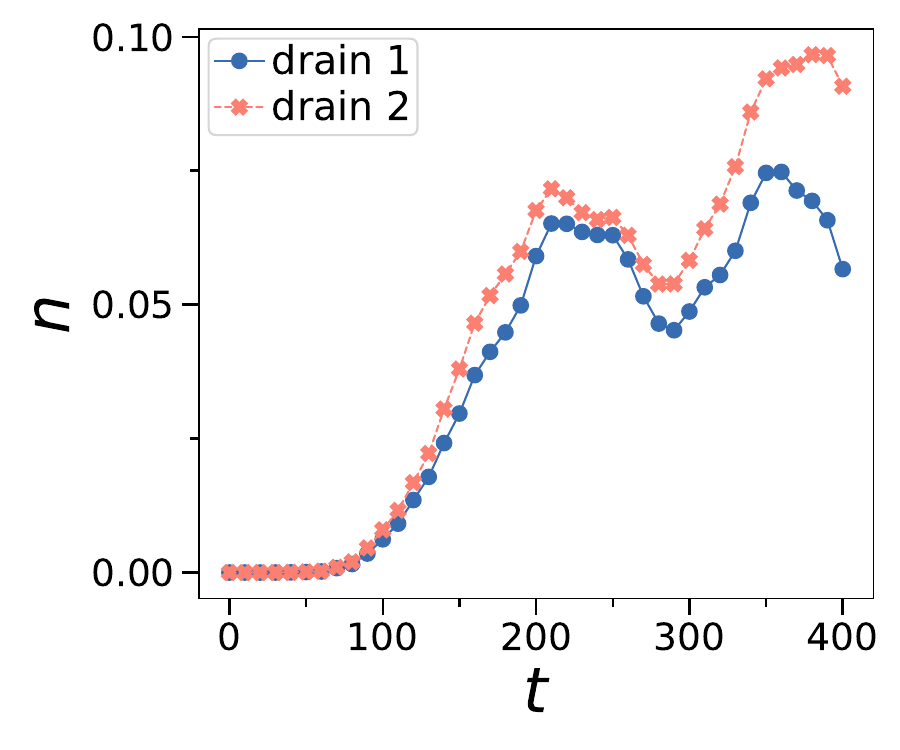}
\subfigimg[width=0.28\textwidth]{d}{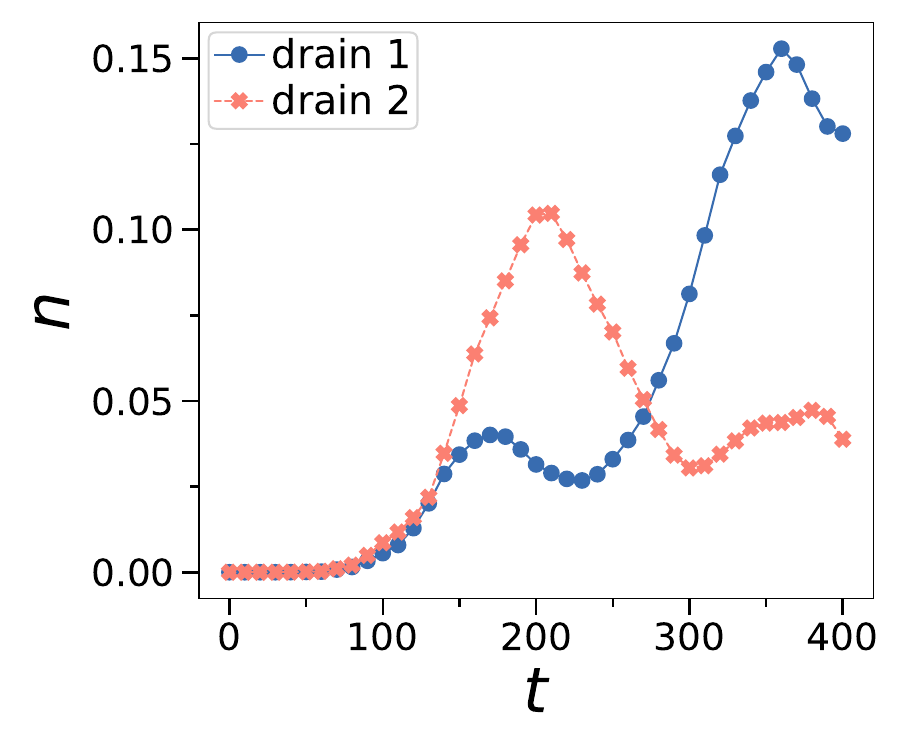}
\subfigimg[width=0.28\textwidth]{e}{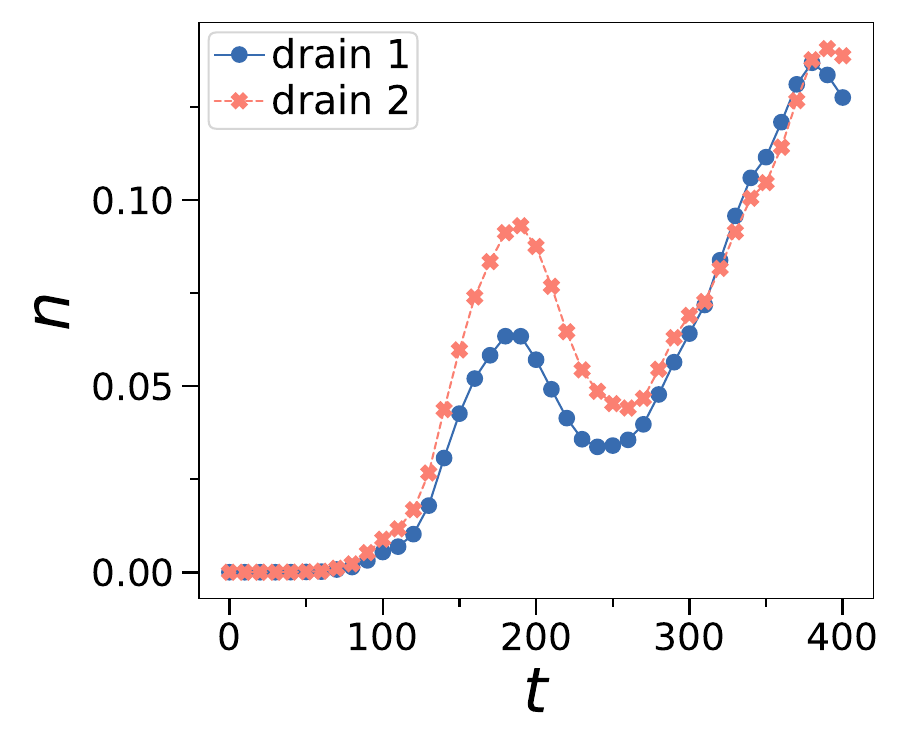}
	\caption{\revB{Density fraction in drains for 2D GPE simulation of ring-lead system for $N=500$ atoms. We show number of atoms in drains divided by total atom number $N$ as function of time $t$ in units of milliseconds. We show values of flux \idg{a} $\Omega=0$, \idg{b} $\Omega=0.25$, \idg{c} $\Omega=0.5$, \idg{d} $\Omega=0.75$ and \idg{b} $\Omega=1$.}
	}
	\label{fig:2dgpe_den500}
\end{figure*}

In Fig.\ref{fig:2dgpe_den}, we show the fraction of atoms in the drains at time $t$ for $N=2000$ atoms, as well as for $N=500$ atoms in Fig.\ref{fig:2dgpe_den500}. 
We find different evolution of drain density as function of flux. For $\Omega=0$, both drains are equally populated.
For $\Omega=1/4$ drain 1 is populated more strongly compared to drain 2 for initial times, i.e. up to $t=300\text{ms}$, and for $\Omega=3/4$ vice versa. We find that with time, the atom density increases initially equally for both drains. Then, for $t\approx 100\text{ms}$, the flux dependence emerges. This is when the atoms propagating through the two arms of  the ring start to overlap and interfere. Then, an interference pattern in the atomic density emerges, which is highly flux dependent and controls the current into the drains. 
We find that the dynamics changes with $N$. We find stronger imbalance for $N=2000$ compared to $N=500$. Further, for $N=500$ we find for $t>250\text{ms}$ the population imbalance between drain 1 and 2 switches. We believe that this is due to reflections from the finite sized leads as well as effects of finite width of the ring and leads.

\begin{figure*}[htbp]
	\centering	
\subfigimg[width=0.28\textwidth]{a}{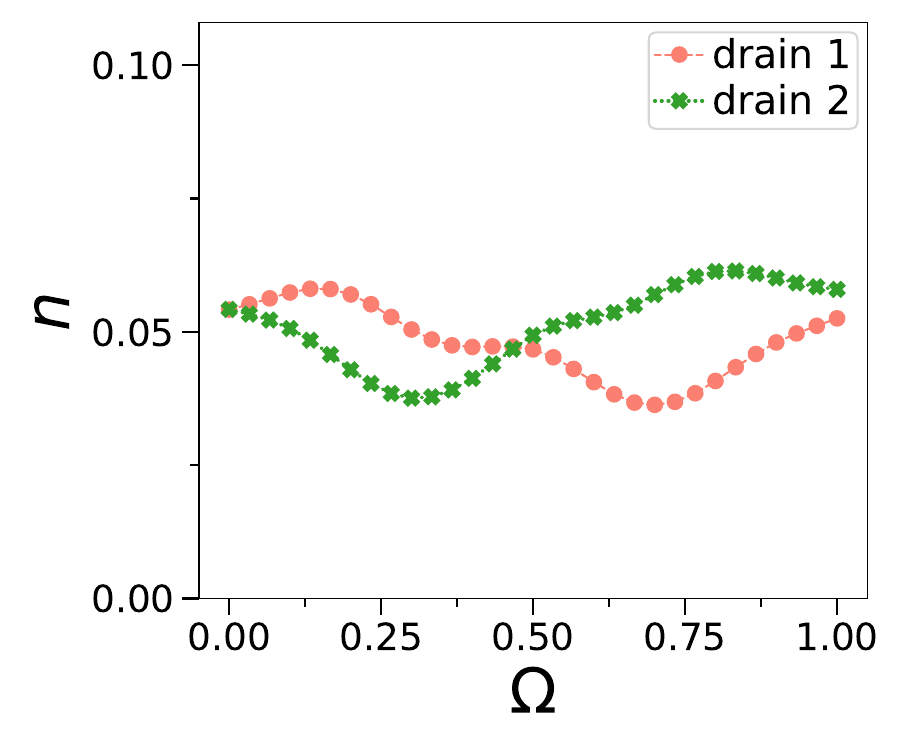}	\subfigimg[width=0.28\textwidth]{b}{GPEN2000flux.pdf}
\subfigimg[width=0.28\textwidth]{c}{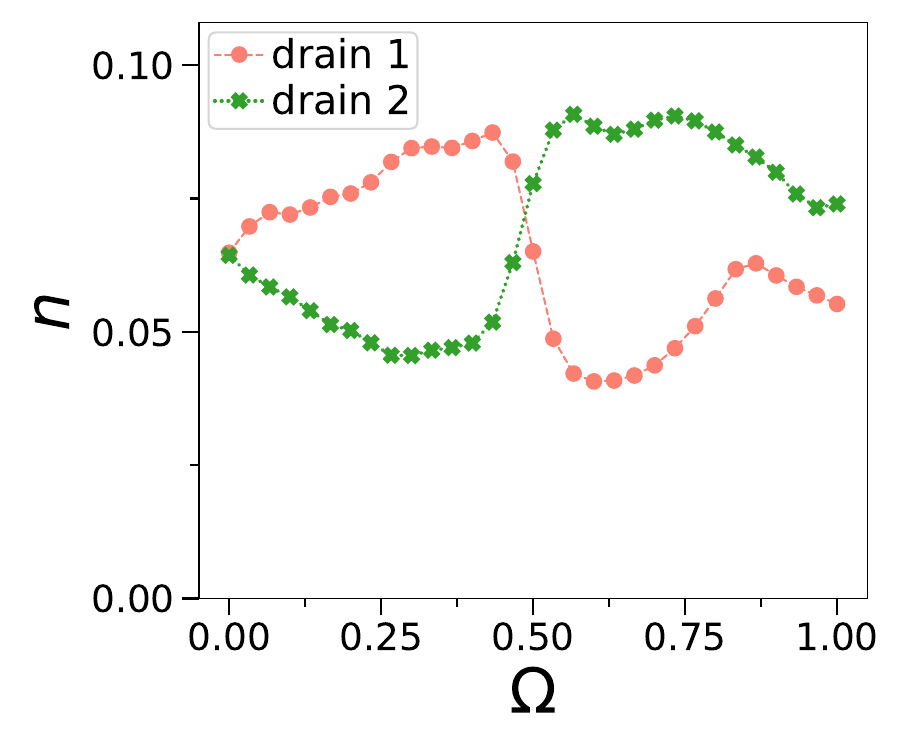}
\subfigimg[width=0.28\textwidth]{d}{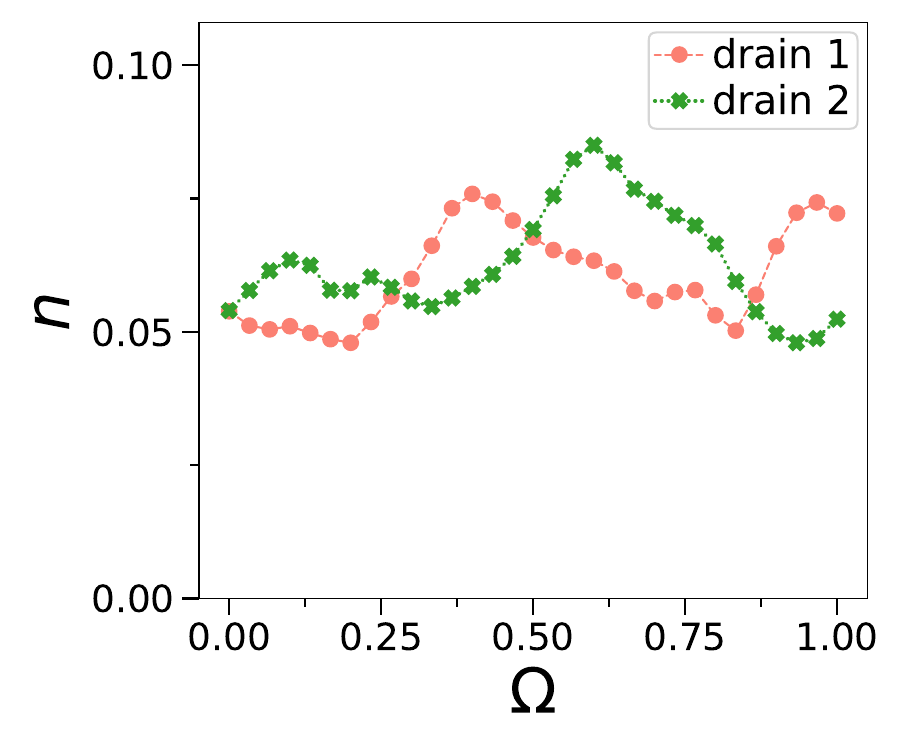}
	\caption{Atom fraction in drains averaged over time $t=0$ to $t=400\text{ms}$ for 2D GPE simulation of ring-lead system for varying flux and atom number $N$.  We show \idg{a} $N=1000$, \idg{b} $N=2000$, \idg{c} $N=3000$ and \idg{d} $N=4000$.
	}
	\label{fig:fluxGPE}
\end{figure*}

In Fig.\ref{fig:fluxGPE}, we show the fraction of atoms in the drains as function of flux $\Omega$ for different atom number $N$, where we integrate the fraction of atoms in the drain between $t=0$ and $t=400\text{ms}$.
We find clear modulation with $\Omega$, with the atomic density being directed into either drain for $\Omega\approx1/4$ and $\Omega\approx3/4$. 
We find largest atom density oscillation with flux for $N=2000$ and $N=3000$. For larger $N$, we find that the oscillation amplitudes decrease with flux.

\begin{figure*}[htbp]
	\centering	
\subfigimg[width=0.66\textwidth]{}{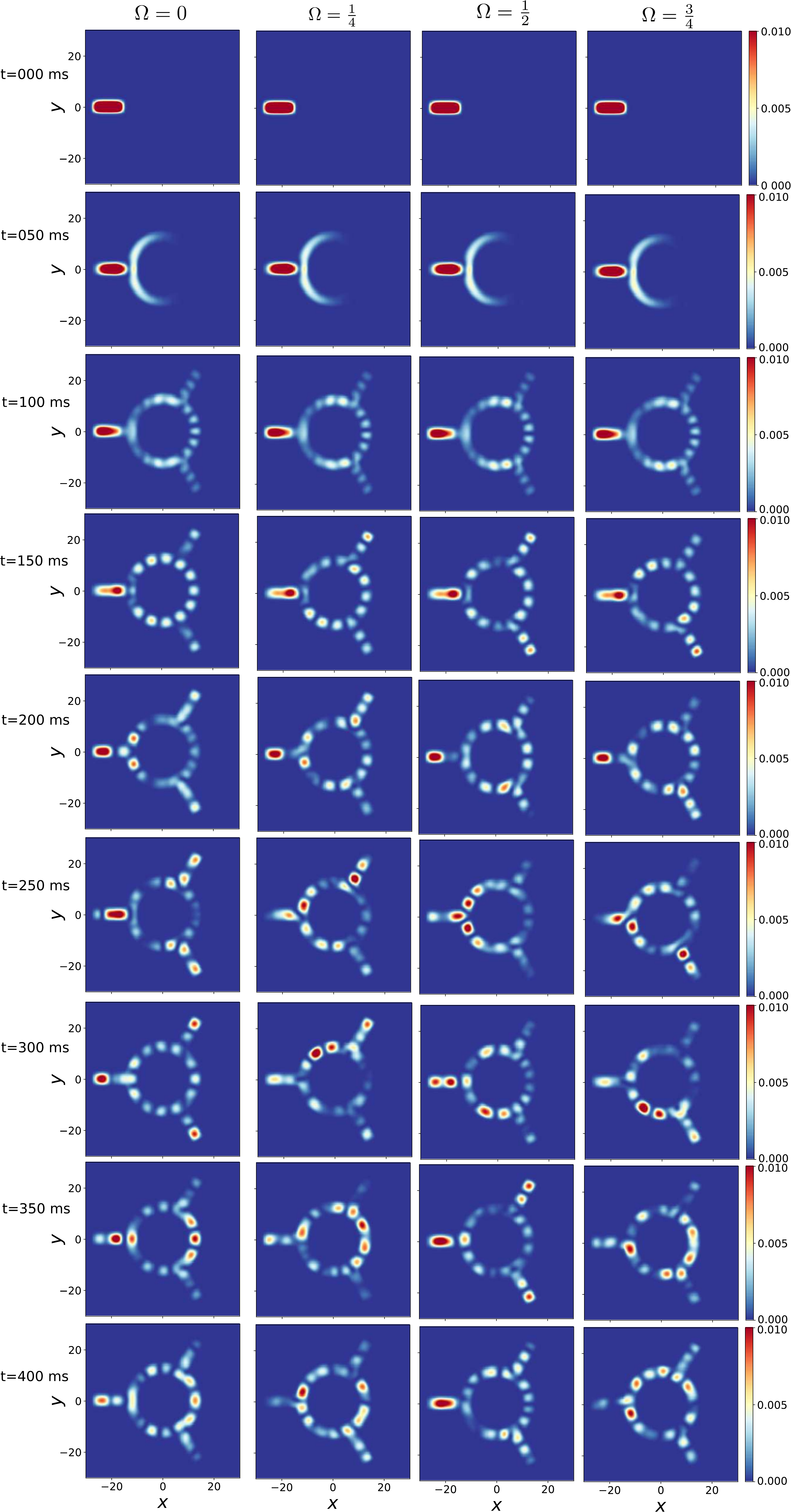}	
	\caption{Density in time $t$ of GPE with different $\Omega$ for $N=2000$ atoms.
	}
	\label{fig:GPEall}
\end{figure*}

\begin{figure*}[htbp]
	\centering	
\subfigimg[width=0.95\textwidth]{}{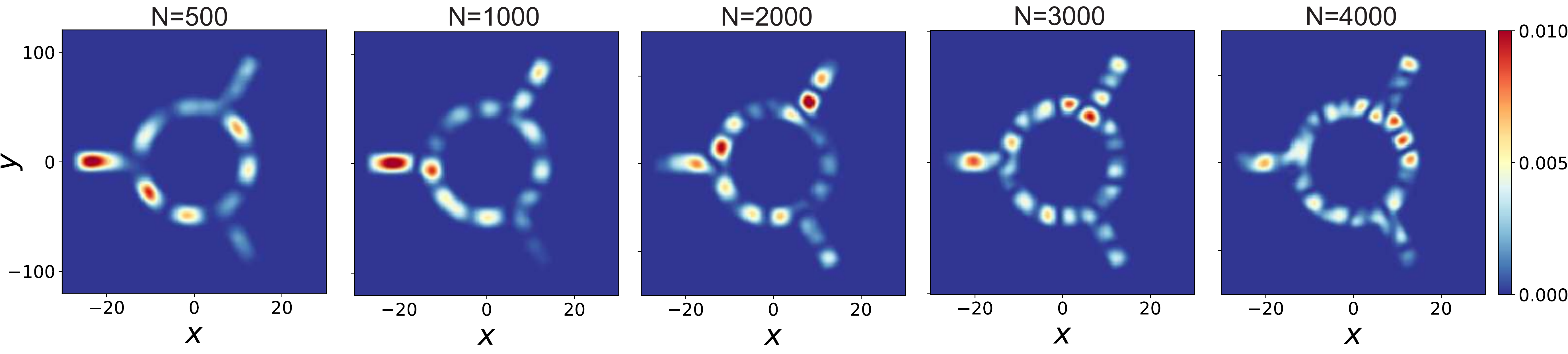}	
	\caption{Density at time $t=250$ms of GPE for $\Omega=1/4$ for different number $N$ of atoms.
	}
	\label{fig:GPEN}
\end{figure*}

\revB{We study the dynamics of the GPE in time $t$ in Fig.~\ref{fig:GPEall} for different flux $\Omega$. Here, we see that the origin of the flux dependence are interference patterns arising from the atoms traveling opposite directions in the ring. These interference pattern emerge at about $t=100$ms once the waves from both directions overlap. 
The flux shifts these interference patterns in the ring, which can be seen well for $t\ge150$ms. While the pattern can dynamically fluctuate in position, they have a pronounced effect on the drain current. 
For $\Omega=1/4$ and $\Omega=3/4$, the symmetry between the leads is broken. For $\Omega=1/4$,  destructive interference pattern are more pronounced at the bottom lead and constructive at the top lead, leading to a net increased current into the top lead. For $\Omega=3/4$, we observe the opposite, leading to an increased density on average in the bottom lead.

Now, we study the effect on the number $N$ of atoms on the interference pattern in Fig.~\ref{fig:GPEN}.
For low $N$ (or equivalently low interaction), the interference pattern emerging is very broad, much larger than the size of the leads.
By tuning the position of destructive interference in respect to the leads using the flux, one can control the current into the drain.
With increasing interaction, the wavelength decreases and the pattern shrinks. For $N=2000$, we observe that the pattern is on the same order as the size of the leads. Here, we find the best control over the lead current.
Further increasing $N$, we find that the interference pattern is smaller than the width of the leads. Here, the flux cannot control the drain current well anymore. In Fig.~\ref{fig:fluxGPE}d for $N=4000$, we observe that the  period of the density modulation with $\Omega$ is only half of that of $N=2000$. This may be explained by the size of the interference pattern for $N=4000$, which is only half of that of $N=2000$. Thus, by variying the flux twice the amount of destructive interference patterns pass across the leads and modulate the current.   }

\end{document}